\documentclass[a4paper, 11pt]{article}
\usepackage{jheppub}
\usepackage{graphicx} 
\numberwithin{equation}{section}
\usepackage{subcaption}
\usepackage{multirow}
\usepackage{bm}
\usepackage{booktabs}
\usepackage{xspace}
\usepackage[nolist,nohyperlinks]{acronym}
\usepackage{hyperref}
\usepackage{siunitx}
\usepackage[table]{xcolor}
\usepackage{enumerate}

\newcommand{\BtoDpiDstar}{B \to D^{*}(\to D\pi)\ell \bar{\nu}_{\ell}}
\newcommand{\BtoDstar}{B \to D^{*}\ell \bar{\nu}_{\ell}}
\newcommand{\BstoDstar}{B_s \to D^{*}_s\ell \bar{\nu}_{\ell}}
\newcommand{\BotoDstar}{\bar{B}^0 \to D^{*+}\ell \bar{\nu}_{\ell}}
\newcommand{\BptoDstar}{B^+ \to \bar{D}^{*0}\ell^+ \nu_{\ell}}
\newcommand{\fref}[1]{Figure~\ref{#1}}
\newcommand{\tref}[1]{Table~\ref{#1}}
\newcommand{\FFs}{\{f, \mathcal{F}_1, \mathcal{F}_2, g\}}
\usepackage{letltxmacro}
\usepackage[normalem]{ulem}
\LetLtxMacro{\originaleqref}{\eqref}
\renewcommand{\eqref}{Equation~\originaleqref}
\usepackage[noabbrev,capitalize]{cleveref}

\preprint{CERN-TH-2026-126, MITP-26-026}
\title{Angular Analysis of \boldmath$\BtoDstar$ from  Lattice and Experiment: \boldmath$|V_{cb}|$ and New Physics Constraints}
\allowdisplaybreaks

\author{Marzia Bordone$^{1}$, Ollie Heald$^{2,3}$, Andreas J\"uttner$^{2,3,4}$}
\affiliation{$^1$ \textit{PRISMA}$^{++}$ Cluster of Excellence \& Mainz Institute for Theoretical Physics
Johannes Gutenberg University, Staudingerweg 9, D-55128 Mainz, Germany}
\affiliation{$^2$ School of Physics and Astronomy, University of Southampton, Southampton, SO17 1BJ, United Kingdom}
\affiliation{$^3$ STAG Research Center, University of Southampton, Southampton, SO17 1BJ, United Kingdom}
\affiliation{$^4$ Theoretical Physics Department, CERN, Geneva, Switzerland}

\emailAdd{marzia.bordone@uni-mainz.de}
\emailAdd{ojh1n20@soton.ac.uk}
\emailAdd{andreas.juttner@cern.ch}

\abstract{We present a combined angular analysis within and beyond the Standard Model (SM) of experimental measurements for the $\BtoDstar$ angular coefficients provided by the Belle collaboration, together with lattice-calculated hadronic form-factor data from the HPQCD, JLQCD, and FNAL/MILC collaborations. We focus on determining the CKM matrix element $|V_{cb}|$ and constraining a set of Wilson coefficients associated with new physics (NP) mediated by scalar and tensor currents. SM predictions for the angular coefficients are obtained using form-factor parameterisations based on the Boyd-Grinstein-Lebed (BGL) ansatz, with unitarity constraints imposed as Bayesian priors. Experimental and theoretical data are analysed jointly by considering the cases $\ell = e, \mu$ separately and comparing with the massless approximation. For the latter, we determine $|V_{cb}| = 0.03997(71)$, with no resolution of the exclusive-inclusive puzzle. Using the full expressions for the angular coefficients in the presence of scalar, vector, and tensor currents, the corresponding Wilson coefficients are constrained through a joint Bayesian fit to lattice and experimental data. By including the renormalisation group evolution of the Wilson coefficients in the SM effective field theory (SMEFT), these constraints translate into bounds on the effective scale of potential heavy NP at the TeV scale. We find, at the 68\% confidence level, that NP mediated by a scalar leptoquark and a vector leptoquark/colourless scalar boson are excluded at the effective scales 1.0 and 2.5 TeV, respectively.
}

\begin{document}

\maketitle

\acrodef{LQCD}{lattice QCD}
\newcommand{\LQCD}{\ac{LQCD}\xspace}
\acrodef{HQET}{Heavy Quark Effective Theory}
\newcommand{\HQET}{\ac{HQET}\xspace}
\acrodef{SM}{Standard Model}
\newcommand{\SM}{\ac{SM}\xspace}
\acrodef{BSM}{Beyond the Standard Model}
\newcommand{\BSM}{\ac{BSM}\xspace}
\acrodef{SMEFT}{Standard Model Effective Field Theory}
\newcommand{\SMEFT}{\ac{SMEFT}\xspace}
\acrodef{WET}{Weak Effective Theory}
\newcommand{\WET}{\ac{WET}\xspace}
\acrodef{CKM}{Cabibbo-Kobayashi-Maskawa}
\newcommand{\CKM}{\ac{CKM}\xspace}
\acrodef{NP}{new physics}
\newcommand{\NP}{\ac{NP}\xspace}
\acrodef{BGL}{Boyd-Grinstein-Lebed}
\newcommand{\BGL}{\ac{BGL}\xspace}
\acrodef{LFU}{Lepton Flavour Universality}
\newcommand{\LFU}{\ac{LFU}\xspace}
\acrodef{RGE}{renormalisation group evolution}
\newcommand{\RGE}{\ac{RGE}\xspace}
\acrodef{AIC}{Akaike information criterion}
\newcommand{\AIC}{\ac{AIC}\xspace}
\acrodef{IGLS}{Iterative Generalised Least Squares}
\newcommand{\IGLS}{\ac{IGLS}\xspace}
\acrodef{INS}{Importance nested sampling}
\newcommand{\INS}{\ac{INS}\xspace}

\section{Introduction}

Within the \SM of particle physics, the \CKM matrix parameterises the relation between weak and mass eigenstates of quarks~\cite{Cabibbo:1963yz,Kobayashi:1973fv}. The CKM matrix elements are free parameters, however, and therefore must be determined experimentally. For instance, the \CKM matrix element $|V_{cb}|$ may be measured from semileptonic decays of $B$ mesons to a charmed final state for either inclusive decays $B \to X_c \ell\bar{\nu}_\ell$, where $X_c$ is a charmed hadronic final state, or an exclusive mode such as $\BtoDstar$. In both cases, the quark-level transition $b \to c\ell\bar{\nu}_\ell$ is mediated by a tree-level W-boson exchange. The importance of the exclusive semileptonic $\BtoDstar$ decay channel lies in two currently unsolved inconsistencies between \SM predictions and experimental measurements. The first is the so-called ``$|V_{cb}|$ puzzle'', a long-standing discrepancy between the exclusive and inclusive determinations of $|V_{cb}|$~\cite{FlavourLatticeAveragingGroupFLAG:2024oxs,ParticleDataGroup:2024cfk,HeavyFlavorAveragingGroupHFLAV:2024ctg}. This \CKM parameter enters various \NP-sensitive \SM predictions, for instance, $\epsilon_K$ of neutral kaon mixing~\cite{FlavourLatticeAveragingGroupFLAG:2024oxs,Jwa:2025fon}, which is and will remain one of the measured \SM processes with sensitivity to the highest energy scales~\cite{deBlas:2025gyz}. The importance and potential impact of resolving the puzzle can therefore not be underestimated, in particular since $\epsilon_K$ depends on  $|V_{cb}|$ to the fourth power. The second lies in testing possible extensions of the \SM. In the $B\to D^* \tau\bar\nu_\tau$ channel there are some long-standing deviations between experimental measurements and theory predictions, in particular in the \LFU ratio $R(D^{(*)})$~\cite{BaBar:2012obs,BaBar:2013mob,Belle:2016dyj,Belle:2019rba,LHCb:2023zxo,LHCb:2023uiv,LHCb:2024jll,Belle-II:2025yjp,Tsaklidis:2025BelleIILFU,Hitlin:2025babarRD}, which motivated measurements of angular observables~\cite{Belle:2016dyj,LHCb:2023ssl}, as well as many \NP analyses~\cite{Fajfer:2012vx,Duraisamy:2013pia,Ivanov:2016qtw,Colangelo:2018cnj,Jung:2018lfu,Blanke:2018yud,Blanke:2019qrx,Mandal:2020htr,Fedele:2022iib,Datta:2022czw,Ray:2023xjn,Bhattacharya:2024zog,Huang:2025kof}. We also note several efforts concentrating on the light-lepton channels, with measurements of \LFU between the $\ell = e$ and $\ell=\mu$ cases, as well as of the differential decay width~\cite{Belle:2018ezy,Belle:2023bwv,Belle-II:2023okj}.

On the experimental side, the Belle and Belle-II collaborations are measuring the exclusive decay~\cite{Belle:2017rcc,Belle:2018ezy,Belle:2023bwv,Belle-II:2023okj,Belle:2023xgj,Belle-II:2023svm}, and recently also LHCb~\cite{LHCb-CONF-2026-001}. Any interpretation of the experimental data in terms of \SM and \BSM physics, however, also requires theory input in the form of \LQCD-calculated form factors, which parameterise the non-perturbative hadronic matrix elements that enter the \SM expression. To this end, the HPQCD, JLQCD, and FNAL/MILC collaborations~\cite{PhysRevD.109.094515,PhysRevD.109.074503,Bazavov_2022} have recently published results, for the first time at non-zero recoil. Owing to the analyticity of the form factors, model-independent parameterisations enable the extrapolation from a small number of kinematic points, for which \LQCD provides data, to the entire semileptonic region~\cite{Boyd:1994tt,Boyd:1995cf,Boyd:1995sq,Boyd_1997,Gubernari:2020eft,DiCarlo:2021dzg,Flynn:2023qmi,Simula:2025fft,Gopal:2024mgb}. The parameterisations can be determined either from fits to \LQCD data alone, or jointly by simultaneous fits to \LQCD and experimental data. While only the former cleanly separates \SM and experiment during the analysis, the latter is still valuable, since it allows for consistency checks and the derivation of constraints on potential \BSM-physics contributions. In both cases, physical observables, such as $|V_{cb}|$, $R(D^\ast)$, or polarisation and asymmetry variables derived from differential shapes may then be computed from these parameterisations~\cite{Martinelli:2021onb,Martinelli:2021myh,Martinelli:2022xir,Fedele:2023ewe,Martinelli:2023fwm,Martinelli:2024bov,Bigi:2017njr,Jaiswal:2020wer,Iguro:2020cpg,Bordone:2024weh,Fang:2026hru}.

In this work, we present a similar analysis, employing the experimental datasets of Refs.~\cite{Belle:2023xgj,hepdata.153299}, which for the first time provide measurements of the $\BtoDstar$ decay width in terms of partially-integrated, normalised angular coefficients. As opposed to the partially-integrated, normalised differential decay rates measured in Refs.~\cite{Belle:2017rcc,Belle:2018ezy,Belle:2023bwv,Belle-II:2023okj}, angular coefficients allow a more granular study of the decay kinematics. Moreover, since some of the possible angular coefficients vanish in the \SM, powerful null tests can be devised. The published datasets also distinguish between charged and neutral initial $B$ mesons, as well as between electrons and muons in the final state. This enables us to assess the impact of each dataset on the extraction of $|V_{cb}|$, and to investigate possible \NP effects in these decay modes. The sensitivity of $\BtoDstar$ decays to \NP has already been studied in the literature, both using earlier experimental results~\cite{Colangelo:2018cnj,Bobeth:2021lya,Huang:2021fuc,Bernlochner:2024xiz,Fang:2026hru} and using the dataset considered here~\cite{Fedele:2023ewe,Colangelo:2024mxe}. In contrast to the latter study, we use the full Belle dataset~\cite{Belle:2023xgj,hepdata.153299}, distinguishing between electrons and muons in the final state, and, in contrast to Ref.~\cite{Colangelo:2024mxe}, taking into account the full experimental correlation.

We now briefly anticipate the main results of this study. With  $|V_{cb}|=0.03997(71)$ we confirm our previous determination $|V_{cb}|=0.04042(71)$ in Ref.~\cite{Bordone:2024weh}, which was based on the same lattice input, but with experimental results for the differential decay rate~\cite{Belle:2023bwv,Belle-II:2023okj,hepdata.137767,hepdata.145129}, and also other previous $|V_{cb}|$ determinations~\cite{Martinelli:2021onb,Martinelli:2021myh,Martinelli:2022xir,Martinelli:2023fwm,Fedele:2023ewe,Fang:2026hru}. While our result in this way confirms the $|V_{cb}|$ puzzle, a correlated-difference analysis finds a significant tension between the results analysed individually for the $B^0$ and $B^+$ decay channels, with the latter being less precise, but much better compatible with the result from the inclusive analysis~\cite{HeavyFlavorAveragingGroupHFLAV:2024ctg}. We find two contributions dominating the error budget of $|V_{cb}|$ by approximately similar amounts, namely \LQCD and the experimentally determined total branching fraction for $\BtoDstar$ decays. The joint analysis of \LQCD and experimental data provides constraints on \SM null tests and, thereby, on the effective energy scales of select extensions of new scalar physics above 1 TeV. However, the way in which \LQCD and experimental data, and their respective correlations, are currently made available to the community assumes Gaussian statistics (central values and covariances). This leads to inconsistencies at various stages of our analysis. We therefore call for future experimental and theoretical results to be made available in terms of distributions (histograms, likelihoods) or correlated data samples, such that the true statistical nature of the data can be carried through the entire analysis chain.

This paper is structured as follows. The angular formalism of exclusive $\BtoDpiDstar$ decays is outlined in \cref{Sec:ang_observables}, with further details and expressions provided in \cref{App:ang_coeff_defs}. Following this is a discussion of the frequentist and Bayesian analysis frameworks in \cref{Sec:analysis_framework}, along with definitions of the various fits and experimental datasets we consider. Descriptions of the \LQCD and experimental datasets are listed in \cref{Appendix:datasets}. In this work, hadronic transition form factors are parameterised by the \BGL~\cite{Boyd:1994tt} ansatz with unitarity imposed in terms of a Bayesian prior~\cite{Flynn:2023qmi}, which is explained in \cref{Appendix:FFbasis}. Results for the \BGL coefficients and $|V_{cb}|$ for \SM fits to \LQCD and experimental data are presented in \cref{Sec:SM_BGL_Vcb_results}. Finally, constraints on potential \BSM contributions and predictions for the expected effective scales for various \NP scenarios are discussed in \cref{Sec:NPinBtoDstar}. 

\section{Angular coefficients in \texorpdfstring{\boldmath$\BtoDpiDstar$}{B -> D*(->Dpi)lnu} Decays}\label{Sec:ang_observables}

The angular distributions of $\BtoDpiDstar$ decays have been comprehensively examined in the literature, in the SM~\cite{Korner:1989qb,Korner:1989ve,Gilman:1989uy} and in NP searches~\cite{Tanaka:1994ay,BaBar:2006taf,Fajfer:2012vx,Tanaka:2012nw,Biancofiore:2013ki,Duraisamy:2013pia,Duraisamy:2014sna}; we summarise here the formalism and introduce notation, following closely that of Ref.~\cite{Martinelli:2024vde}. In full generality, the fourfold differential angular distribution of $B\to D^*(\to D\pi)\ell\bar{\nu}_\ell$ cascade decays can be parameterised as
\begin{equation}
    \begin{aligned}
    \frac{\textrm{d}^{4}\Gamma(\BtoDstar)}{\textrm{d}w\,\textrm{d}\cos\theta_{v}\,\textrm{d}\cos\theta_{\ell}\,\textrm{d}\chi} &= \frac{3}{16\pi}\Gamma_{0}[ J_{1s}(w)\sin^{2}\theta_{v} + J_{1c}(w)\cos^{2}\theta_{v} \\
    &\quad + J_{2s}(w)\sin^{2}\theta_{v}\cos2\theta_{\ell} + J_{2c}(w)\cos^{2}\theta_{v}\cos2\theta_{\ell} \\
    &\quad + J_{3}(w)\sin^{2}\theta_{v}\sin^{2}\theta_{\ell}\cos2\chi + J_{4}(w)\sin2\theta_{v}\sin2\theta_{\ell}\cos\chi \\
    &\quad + J_{5}(w)\sin2\theta_{v}\sin\theta_{\ell}\cos\chi + J_{6s}(w)\sin^{2}\theta_{v}\cos\theta_{\ell} \\
    &\quad + J_{6c}(w)\cos^{2}\theta_{v}\cos\theta_{\ell} + J_{7}(w)\sin2\theta_{v}\sin\theta_{\ell}\sin\chi \\
    &\quad + J_{8}(w)\sin2\theta_{v}\sin2\theta_{\ell}\sin\chi + J_{9}(w)\sin^{2}\theta_{v}\sin^{2}\theta_{\ell}\sin2\chi],
    \end{aligned}
    \label{eq:ang_distribution}
\end{equation}
where $\Gamma_0 = \frac{\eta_{\textrm{EW}}^2 m_B m_{D^*}^2}{(4\pi)^3}G_\textrm{F}^2|V_{cb}|^2$, $G_\textrm{F}$ is Fermi's constant, and $\eta_\textrm{EW} \approx 1.0066$ accounts for electroweak corrections~\cite{Sirlin:1981ie}. For the kinematic variables: $\theta_\ell$ is the helicity angle of the charged lepton in the lepton-neutrino rest frame, $\theta_v$ is the helicity angle of the $D$ meson in the $D\pi$ rest frame, $\chi$ is the azimuthal angle between the two planes defined by the $\ell\bar{\nu}_\ell$ and $D\pi$ directions, respectively, and the dimensionless recoil variable $w$ is defined in terms of the invariant squared mass of the lepton-neutrino pair:
\begin{equation}\label{eq:w(q^2)}
    w(q^2) \equiv \frac{m_B^2 + m_{D^*}^2 - q^2}{2m_Bm_{D^*}} = \frac{p_B \cdot p_{D^*}}{m_B m_{D^*}},
\end{equation}
where $q = p_B - p_{D^\ast} = p_\ell + p_{\bar{\nu}_\ell}$ is the momentum transfer between the initial $B$ and final $D^\ast$ meson. The quantities $J_i(w)$ are known as angular coefficients, which may be written in terms of helicity amplitudes~\cite{Hagiwara:1989gza} -- see \cref{App:ang_coeff_defs} for the details. In the \SM, $J_7 = J_8 = J_9 = 0$, with $J_{6c}$ also vanishing in the massless-lepton limit $m_\ell \to 0$. Compared to the normalised differential decay rates measured in Refs.~\cite{Belle:2018ezy,Belle:2023bwv}, 
\eqref{eq:ang_distribution} parameterises the full angular distribution of the decay, thus providing a more comprehensive set of observables. For comparison, the normalised rates $\frac{1}{\Gamma}\frac{\textrm{d}\Gamma}{\textrm{d}x}$, $x \in \{w, \cos\theta_\ell, \cos\theta_v, \chi\}$, are only sensitive to eight of the twelve angular coefficients present in the full fourfold differential distribution in \eqref{eq:ang_distribution} -- see Equations (16)-(18) of Ref.~\cite{Martinelli:2024vde}. This additional information is valuable in the search for \NP, which could itself induce non-\SM angular structure~\cite{Huang:2021fuc}.

In the recent experimental analysis by the Belle collaboration~\cite{Belle:2023xgj,hepdata.153299}, binned measurements of the normalised, partially-integrated angular coefficients, defined as
\begin{equation}\label{eq:JwithN}
    \hat{J}_i(w_n) / \mathcal{N}, \quad \mathrm{with}\quad \mathcal{N} = \frac{8\pi}{9}(6\bar{J}_{1s} + 3\bar{J}_{1c} - 2\bar{J}_{2s} - \bar{J}_{2c}),
\end{equation}
where
\begin{equation}\label{eq:JhatJbar}
    \hat{J}_i(w_n) = \int_{\Delta w_n}J_i(w)\,\textrm{d}w, \quad \mathrm{and} \quad \bar{J}_i = \int_{1}^{w_\textrm{max}^{(\ell)}}J_i(w)\,\textrm{d}w,
\end{equation}
are provided in four bins of $w$ with widths $\Delta w_n$. The maximum value of the recoil variable $w$, and hence the upper limit of integration in the last bin, depends on the species of lepton -- see \eqref{eq:w_max^ell}. In what follows, we denote $w_\textrm{max}^{(0)}$ the upper limit of $w$ for massless leptons, and $w_\textrm{max}^{(\ell)}$ for $\ell = e,\mu$. The experimental and \LQCD datasets used in this angular analysis are described in \cref{Appendix:datasets}. A similar analysis is also available from Belle II~\cite{Belle-II:2023svm}. However, given the more limited statistical power, it presents measurements of angular observables only in two $w$ bins, with large errors. Therefore, we do not include it here. The LHCb experiment has also presented a first measurement of $B\to D^*\mu\nu_\mu$ decays~\cite{LHCb-CONF-2026-001}. However, no data in terms of one-dimensional distributions or binned measurements of angular observables have been released as of yet.

The total decay rate for $\BtoDstar$ can be found by integrating \eqref{eq:ang_distribution} over the three angles and over $w$:
\begin{equation}\label{eq:Gamma}
    \Gamma(\BtoDstar) = \frac{\Gamma_0}{6}[6\bar{J}_{1s} + 3\bar{J}_{1c} - 2\bar{J}_{2s} - \bar{J}_{2c}]\,.
\end{equation}
The normalisation factor $\mathcal{N}$ is thus proportional to the integrated width. Within the \SM, the angular coefficients and the total decay rate can be expressed in terms of the four hadronic transition form factors $\FFs$ that parameterise the matrix elements of the $V - A$ current that mediates the $B \to D^*$ transition (see \cref{Appendix:FFbasis} for the Lorentz decomposition of these matrix elements). The form factors have been computed using \LQCD by the HPQCD~\cite{PhysRevD.109.094515}, JLQCD~\cite{PhysRevD.109.074503}, and FNAL/MILC~\cite{Bazavov_2022} collaborations. As explained in detail later, we obtain a parameterisation of each form factor over the full semileptonic kinematic range in terms of  \BGL ans\"atze~\cite{Boyd:1994tt}. The details of the \BGL parameterisations are discussed in \cref{Subsection:BGL}, along with the necessary input parameters. The helicity amplitudes are then computed in terms of the form-factor parameterisations to construct the eight (nine, in the massive lepton case) non-vanishing angular coefficients in the \SM. Integrating these over the bin widths $\Delta w_n$ and normalising provides the lattice predictions for the quantities in \eqref{eq:JwithN} measured by the Belle experiment~\cite{Belle:2023xgj,hepdata.153299}. In the following, we define nomenclature for the various types of fits considered to establish the notation used here and within the \texttt{BFF} library~\cite{BFF_code}.

\section{Analysis Framework}\label{Sec:analysis_framework}

The analysis problem consists of determining the coefficients $a_{F,n}$ of truncated \BGL form-factor parameterisations 
\begin{equation}\label{eq:FF_B_phi_series_trunc}
    F(z) = \frac{1}{B_{J^P}(z)\phi_F(z)}\sum_{n=0}^{K_F-1}a_{F,n} z^n,
\end{equation}
for form factors $F \in \FFs$ with spin-parity $J^P$, to either only theory, or theory and experimental data (termed, as in Ref.~\cite{Bordone:2024weh}, Type A and Type B fits, respectively). As detailed in \cref{Appendix:FFbasis}, $B_{J^P}(z)$ and $\phi_F(z)$ are Blaschke factors and outer functions, respectively, and $z(q^2)$ is a conformal map of the real $q^2$ axis onto the complex unit disk. As a further ingredient, we impose \SM unitarity in terms of the non-linear constraints on the coefficients of the \BGL ansatz,
\begin{align}\label{eq:BGL_unitarity_vector}
    \sum_{n = 0}^\infty |a_{f,n}|^2 + |a_{\mathcal{F}_1,n}|^2 \leq 1, && \sum_{n = 0}^\infty |a_{\mathcal{F}_2,n}|^2 \leq 1, && \sum_{n = 0}^\infty |a_{g,n}|^2 \leq 1,
\end{align}
and similarly for tensor form factors in \eqref{eq:BGL_unitarity_tensor}, following the Bayesian-inference framework of Ref.~\cite{Flynn:2023qmi} as implemented in the Python library \texttt{BFF}~\cite{BFF_code}. Joint fits to \LQCD and experimental data are implemented using
\texttt{PyMultiNest}~\cite{Buchner_2014}, a wrapper for the \texttt{MultiNest} nested sampling algorithm~\cite{Skilling:2004pqw,Feroz_2008,Feroz_2009,Feroz_2019}. This approach has been utilised previously in Refs.~\cite{Flynn:2023qmi,Flynn:2023eok} for the analysis of $B_s \to K\ell\bar{\nu}_\ell$ decay and in Ref.~\cite{Bordone:2024weh} for $\BtoDstar$. By virtue of the unitarity constraints, which act as a regulator, the truncation order $K_F - 1$ in a fit to $N_F$ data points can be freely varied in a Bayesian fit~\cite{Flynn:2023qmi,Flynn:2023eok,Bordone:2024weh}. The number of free parameters must, on the other hand, not exceed the number of data points $N_F$ for a frequentist fit with a meaningful quality-of-fit statistic. 

Whilst the Bayesian framework lacks any absolute goodness-of-fit metric such as the $\chi^2$ or $p$-value in the frequentist picture, relative model comparisons (model selections) may be carried out following an interpretation of the Bayesian evidence provided by the so-called Jeffreys' scale~\cite{JeffreysScale}. The ratio of the Bayesian evidence of two competing models -- the Bayes factor $\mathcal{K}$ -- quantifies the preference, if at all, for one model over another. First, we define the Bayes factor $\mathcal{K}$ as the difference in the natural logarithm of the \INS evidence $\ln\mathcal{Z}_{\textrm{INS}}$ (calculated by \texttt{MultiNest}) for models 1 and 2:
\begin{equation}\label{eq:Bayes_factor_12}
    \mathcal{K} = \ln\mathcal{Z}_{\textrm{INS}}^{(1)} - \ln\mathcal{Z}_{\textrm{INS}}^{(2)}.
\end{equation}
If $\mathcal{K} > 2.3$ there is strong evidence for model 1 over model 2, and if $\mathcal{K} > 4.6$ the evidence is decisive. For negative values of $\mathcal{K}$, the converse is true, suggesting support for model 2. The Bayes factor $\mathcal{K}$ will be reported in \cref{Sec:SM_BGL_Vcb_results,Sec:NPinBtoDstar} to compare fits to $|V_{cb}|$ and \NP parameters for the various experimental and \LQCD datasets considered. 

\subsection{Fits to Form Factors and Angular Coefficients}\label{Subsection:fit_nomenclature}

We broadly distinguish two types of fit strategies:
\begin{itemize}
    \item[--] \textbf{Type A}: fitting only to lattice data with the generalised least-squares kernel:
    \begin{equation}\label{eq:chi^2_lat}
        \chi^2_\textrm{A}(\mathbf{a}) = \big[ \mathbf{f}^{\textrm{lat}} - \mathbf{f}^{\textrm{BGL}}(\mathbf{a}) \big]^\intercal\mathsf{C}^{-1}_{\textrm{lat}}\big[ \mathbf{f}^{\textrm{lat}} - \mathbf{f}^{\textrm{BGL}}(\mathbf{a}) \big],
    \end{equation}
    where $\mathbf{f}^{\textrm{lat}}$ is a vector of form-factor data in each $q^2$ bin with covariance matrix $\mathsf{C}_{\textrm{lat}}$, and $\mathbf{f}^{\textrm{BGL}}(\mathbf{a})$ the corresponding theory prediction for the set of \BGL coefficients $\mathbf{a}$. The \BGL coefficients $\mathbf{a}$ may be determined in both the frequentist and Bayesian frameworks -- see Ref.~\cite{Flynn:2023qmi} for a description of the Bayesian-inference sampling algorithm. Depending on the truncation order $K$ of the \BGL series in \eqref{eq:FF_B_phi_series_trunc}, a frequentist fit is possible provided that the number of data points exceeds the number of free parameters, so that the number of degrees of freedom $N_\textrm{dof} \geq 1$. In this case, the fit is performed by minimising \eqref{eq:chi^2_lat}, with $\mathbf{f}^{\textrm{BGL}}(\mathbf{a}) = \mathsf{Z}\mathbf{a}$ by virtue of linearity, where
    $\mathsf{Z}$ is the design matrix containing the appropriate Blaschke factors, outer functions, and kinematic-endpoint constraints in \eqref{aligneq:fF1F2F1_constraints}. Details of the design matrix and its construction are given in \cref{App:Z_constraint}. In this case, \eqref{eq:chi^2_lat} is minimised by
    \begin{equation}\label{eq:freq_a_cov}
        \mathbf{a} = (\mathsf{Z}^\intercal\mathsf{C}_{\textrm{lat}}^{-1}\mathsf{Z})^{-1}\mathsf{Z}^\intercal \mathsf{C}_{\textrm{lat}}^{-1}\mathbf{f}.
    \end{equation}
    with covariance matrix $\mathsf{C}_{\mathbf{a}} = (\mathsf{Z}^\intercal\mathsf{C}_{\textrm{lat}}^{-1}\mathsf{Z})^{-1}$.  
    \item[--] \textbf{Type B}: simultaneous fit to both lattice and experimental data, with
    \begin{equation}\label{eq:chi^2_latexp}
        \chi^2_\textrm{B}(\mathbf{a}) = \chi^2_\textrm{A}(\mathbf{a}) + \big[ \bm{\mathcal{J}}^{\textrm{exp}} - \bm{\mathcal{J}}^{\textrm{lat}}(\mathbf{a}) \big]^\intercal\mathsf{C}^{-1}_{\textrm{exp}}\big[ \bm{\mathcal{J}}^{\textrm{exp}} - \bm{\mathcal{J}}^{\textrm{lat}}(\mathbf{a}) \big],
    \end{equation}
    where $\bm{\mathcal{J}}^{\textrm{exp}}$ is a vector of experimentally-measured partially-integrated, normalised angular coefficients (i.e. $\hat{J} / \mathcal{N}$) with covariance matrix $\mathsf{C}_{\textrm{exp}}$, and $\bm{\mathcal{J}}^{\textrm{lat}}(\mathbf{a})$ is the corresponding \LQCD prediction.  
\end{itemize}
Type-A fits are linear in the BGL parameters, and the unitarity constraints in \cref{eq:BGL_unitarity_vector,eq:BGL_unitarity_tensor} can therefore be implemented as a restriction of the integration range in the Monte-Carlo integration of the likelihood integral (see Ref.~\cite{Flynn:2023qmi} for details). The angular coefficients entering the $\chi^2$ in Type-B fits are, on the other hand, quadratic in the BGL coefficients. In this case, the Monte-Carlo integration 
is carried out using \texttt{PyMultiNest}~\cite{Buchner_2014,Skilling:2004pqw,Feroz_2008,Feroz_2009,Feroz_2019}, which is forced to comply with the unitarity constraints by introducing a window function for each spin-parity channel that yields a negative contribution to the log-likelihood as the unitarity sum saturates, acting as a penalty to the fit.

\subsubsection[{Determination of \texorpdfstring{$|V_{cb}|$}{|Vcb|}}]{Determination of \texorpdfstring{\boldmath$|V_{cb}|$}{|Vcb|}}\label{Subsubsection:Vcb_determination}

Regarding the determination of the \CKM matrix element $|V_{cb}|$, we use different strategies for Type-A and Type-B fits, respectively:
In the case of Type-A fits, $|V_{cb}|$ can be extracted by comparing the lattice-calculated and experimentally measured decay rates and angular coefficients bin-by-bin in $w$. In this approach, \SM unitarity is imposed only on the \LQCD results for the form factors, in this way cleanly separating \SM from experiment, which could contain \NP that is incompatible with unitarity. 

What is computable just from fits to \LQCD form-factor data is the decay rate modulo $|V_{cb}|^2$:
\begin{equation}\label{eq:Gamma_lat}
    \widetilde{\Gamma}^{\textrm{lat}}(\mathbf{a}) = \frac{\Gamma^{\textrm{lat}}(\mathbf{a})}{|V_{cb}|^2},
\end{equation}
which is proportional to the normalisation $\mathcal{N}$ of the angular coefficients -- see \cref{eq:JwithN,eq:Gamma}. Hence, as in Ref.~\cite{Bordone:2024weh}, the ratio of experimental to \LQCD quantities may be employed as a predictor of $|V_{cb}|$:
\begin{equation}\label{eq:chi2_Vcb_ratio}
    \chi^2_{\textrm{ratio}}(|V_{cb}|^2) = \left( \bm{\mathcal{R}}^{\textrm{exp/lat}} - |V_{cb}|^2 \right)^\intercal \mathsf{C}^{-1}_{\textrm{ratio}}\left( \bm{\mathcal{R}}^{\textrm{exp/lat}} - |V_{cb}|^2 \right),
\end{equation}
where
\begin{equation}\label{eq:Vcb_binned_quotient}
    \bm{\mathcal{R}}^{\textrm{exp/lat}} = \frac{\Gamma^\mathrm{exp}\bm{\mathcal{J}}^{\textrm{exp}}}{\widetilde{\Gamma}^\mathrm{lat}\bm{\mathcal{J}}^{\textrm{lat}}}\,,
\end{equation}
predicts $|V_{cb}|^2$ in each $w$ bin of each angular coefficient. Contrary to expectations, this led to inconsistent results: the value of $|V_{cb}|$ obtained from the combination of \LQCD datasets (\texttt{Combined}) exceeded the results from individual \LQCD predictions by up to $\sim 1\sigma$. The ratios in \eqref{eq:Vcb_binned_quotient} calculated by resampling led to many $|V_{cb}|^2 < 0$ samples, particularly for bins where the experimental measurements are close to zero with large relative errors. A possible culprit could be the covariance matrix $\mathsf{C}^{-1}_{\textrm{ratio}}$ of the ratios in \eqref{eq:Vcb_binned_quotient}, which was computed by error propagation via resampling. This approximates the ratio of two Gaussians as a Gaussian, which could explain the observed artefacts. To mitigate this sign issue, we avoid computing ratios for the fit and instead consider the following reformulation of the likelihood:
\begin{equation}\label{eq:chi2_Vcb_diff}
    \chi^2_{\textrm{diff}}(|V_{cb}|^2) = \left( \Gamma^{\textrm{exp}}\bm{\mathcal{J}}^{\textrm{exp}} - |V_{cb}|^2\widetilde{\Gamma}^{\textrm{lat}}\bm{\mathcal{J}}^{\textrm{lat}} \right)^\intercal \mathsf{C}^{-1}_{\textrm{diff}}(|V_{cb}|^2) \left( \Gamma^{\textrm{exp}}\bm{\mathcal{J}}^{\textrm{exp}} - |V_{cb}|^2\widetilde{\Gamma}^{\textrm{lat}}\bm{\mathcal{J}}^{\textrm{lat}}  \right),
\end{equation}
where 
\begin{equation}\label{eq:exp_lat_cov}
    \mathsf{C}_{\textrm{diff}}(|V_{cb}|^2) = \textrm{Cov}(\Gamma^{\textrm{exp}}\bm{\mathcal{J}}^{\textrm{exp}}) + (|V_{cb}|^2)^2\textrm{Cov}(\widetilde{\Gamma}^{\textrm{lat}}\bm{\mathcal{J}}^{\textrm{lat}})\,,
\end{equation}
is now a parameter-dependent covariance matrix. The expression in \eqref{eq:exp_lat_cov} assumes the statistical independence of experimental and lattice-calculated quantities. Employing \eqref{eq:chi2_Vcb_diff} avoids having to compute the covariance matrix of the ratios in \eqref{eq:Vcb_binned_quotient}, but introduces the further complication of having a parameter-dependent covariance matrix $\mathsf{C}_{\textrm{diff}}(|V_{cb}|^2)$, which sacrifices an analytic solution for the minimiser. To proceed, \eqref{eq:chi2_Vcb_ratio} is minimised iteratively using \IGLS, which estimates both the fit parameter and covariance matrix at each iteration until convergence. Convergence of the fit parameter is expected to be quick, since the smallness of $(|V_{cb}|^2)^2$ results in a small perturbation from $\textrm{Cov}(\Gamma^{\textrm{exp}}\bm{\mathcal{J}}^{\textrm{exp}})$ in \eqref{eq:exp_lat_cov}. This is then cross-checked with the result obtained by numerically minimising the Gaussian log-likelihood function
\begin{equation}
    \chi^2_{\textrm{opt}}(|V_{cb}|^2) = \chi^2_{\textrm{diff}}(|V_{cb}|^2) + \ln\det[\mathsf{C}_{\textrm{diff}}(|V_{cb}|^2)].
\end{equation}
While this alternative approach largely rectifies the unexpected tensions between the $|V_{cb}|$ predictions from the \texttt{Combined} and individual \LQCD datasets, they persist (albeit less severely) for $B = B^0$ in both Type-A and Type-B analyses -- this is discussed in more detail in \cref{Sec:SM_BGL_Vcb_results}.

Regarding Type-B fits, the \CKM matrix element $|V_{cb}|$ may be simultaneously determined alongside the \BGL coefficients by adding the following contribution to the Type-B log-likelihood kernel  in \eqref{eq:chi^2_latexp}:
\begin{equation}\label{eq:chi^2_norm}
    \chi^2_\textrm{norm}(\mathbf{a}, |V_{cb}|^{2}) = \frac{\big[\Gamma^{\textrm{exp}} - |V_{cb}|^{2}\widetilde{\Gamma}^{\textrm{lat}}(\mathbf{a}) \big]^{2}}{\sigma_{\Gamma^{\textrm{exp}}}^{2}},
\end{equation}
where $\Gamma^\textrm{exp}$ is the experimentally-determined total decay rate for $\BtoDstar$, and $\sigma_{\Gamma^{\textrm{exp}}}$ its uncertainty. The lattice-calculated rate $\widetilde{\Gamma}^\textrm{lat}(\mathbf{a})$ is computed from the angular coefficients using \eqref{eq:Gamma} (which themselves are calculated from the form factors $f, \mathcal{F}_1, \mathcal{F}_2$, and $g$ -- see \cref{Sec:ang_observables}). 

\subsection{Experimental Input}
The measurements provided by Belle cover the $\BotoDstar$ and $\BptoDstar$ decay channels\footnote{Charge conjugation is implied throughout.}, with $\ell = e, \mu$~\cite{Belle:2023xgj,hepdata.153299}. In this analysis, we consider the raw spectra for each decay channel, as well as averages over the initial-state $B$ meson and final-state lepton $\ell$ provided in the HEPData repository~\cite{hepdata.153299}. We therefore have seven datasets, which are denoted as $\langle x \rangle$, with $x$ being the quantities averaged:
\begin{itemize}
    \item Fully-averaged: $\langle B^0, B^+, e, \mu \rangle$.
    \item Initial-state-averaged: $\langle B^0, B^+, e \rangle$ and $\langle B^0, B^+, \mu \rangle$.
    \item Individual: $\langle B^0, e \rangle$, $\langle B^0, \mu \rangle$, $\langle B^+, e \rangle$, and $\langle B^+, \mu \rangle$.
\end{itemize}
In \cref{Appendix:datasets}, the initial-state-averaged datasets for each lepton mass are shown in \fref{fig:exp_ang_coeffs_ml}, and the lepton-averaged datasets for each $B$ meson are shown in \cref{fig:exp_ang_coeffs_B0B+_e,fig:exp_ang_coeffs_B0B+_mu}. For more details on these datasets, see the accompanying text in \cref{Appendix:datasets}. In the expressions for the angular coefficients in \crefrange{aligneq:J1s}{aligneq:J789}, the lepton mass enters as $m_\ell^2 / q^2$, multiplying the time-like/scalar helicity amplitude $H_t(w) \propto \mathcal{F}_2(w)$. Due to the larger masses of the $B$ and $D^*$ mesons, non-zero lepton masses introduce small corrections, most significant at small $q^2$ or large $w$. For the fully-averaged $\langle B^0, B^+, e, \mu \rangle$ dataset, we set $m_\ell = 0$ as done in the analysis by the Belle collaboration~\cite{Belle:2023xgj}, and for the remaining datasets, we explicitly account for lepton mass effects in the \LQCD calculation of the angular coefficients. In particular, the role of helicity suppression that arises from the small lepton masses on \NP searches is discussed more in \cref{Subsubsection:NP_sensitivity}. 

Common to both methods that calculate $|V_{cb}|$, \cref{eq:chi2_Vcb_ratio,eq:chi^2_norm}, is the reliance on experimental measurements of the branching fraction $\mathcal{B}^{\textrm{exp}}(\BtoDstar)$ and the $B$ meson lifetime $\tau(B)$, which together determine the total integrated decay rate:
\begin{equation}
    \Gamma^{\textrm{exp}} = \frac{\mathcal{B}^{\textrm{exp}}(\BtoDstar)}{\tau(B)}.
\end{equation}
In this work, we employ the current values provided by HFLAV (charge conjugation is implied)~\cite{HeavyFlavorAveragingGroupHFLAV:2024ctg}:
\begin{align}
    \mathcal{B}^{\text{exp}}(\BotoDstar) &= (4.90 \pm 0.01 \pm 0.12)\%, \label{eq:branching_fraction_B0} \\
    \mathcal{B}^{\text{exp}}(\BptoDstar) &= (5.53 \pm 0.07 \pm 0.21)\%, \label{eq:branching_fraction_B-}
\end{align}
where the uncertainties are statistical and systematic, respectively. Assuming strong isospin symmetry, the branching fractions in \cref{eq:branching_fraction_B0,eq:branching_fraction_B-} may be combined into a single average for the $B^0$~\cite{HeavyFlavorAveragingGroupHFLAV:2024ctg}:
\begin{equation}
    \mathcal{B}^{\text{exp}}(\BtoDstar) = (4.90 \pm 0.01 \pm 0.11)\%, \label{eq:branching_fraction_B0_avg}
\end{equation}
which takes into consideration the breaking of isospin symmetry due to the Coulomb effect. For the lifetimes of the $B$ mesons~\cite{HeavyFlavorAveragingGroupHFLAV:2024ctg,ParticleDataGroup:2024cfk}:
\begin{align}
    \tau(B^0) &= 1.517(4) \textrm{ ps}, \label{eq:tau_B0} \\
    \tau(B^+) &= 1.638(4) \textrm{ ps}. \label{eq:tau_B-}
\end{align}
Where a meson-averaged experimental dataset is analysed, it is understood that \cref{eq:branching_fraction_B0_avg,eq:tau_B0} are employed. For individual channels, the appropriate branching fraction and lifetime are used. 

\section{Results for BGL coefficients and \texorpdfstring{\boldmath$|V_{cb}|$}{|Vcb|}}\label{Sec:SM_BGL_Vcb_results} 

We build on the results of Ref.~\cite{Bordone:2024weh}, which studied the truncation dependence of fits to the form factors $F \in \FFs$ for the same \LQCD datasets considered here. The results show that the \BGL coefficients converge to stable values at order $K_F = 4$, with third-order coefficients and higher in the expansion being compatible with zero. In the following, we therefore use $K_F = 3$. Truncating the \BGL series at this order parameterises each form factor as a quadratic polynomial in $z$. We reproduce here the same Type-A fits with updated input values for the $\bar{c}b$ QCD spectrum for the Blaschke factors (detailed in \cref{Subsection:BGL}).

\subsection{Type-A Fit Results}\label{Subsec:Type_A_results}

\begin{table}[t]
    \centering
    \resizebox{\linewidth}{!}{
\begin{tabular}{cS[table-format = -1.5(2)]S[table-format = -1.4(2)]S[table-format = -1.2(2)]S[table-format = -1.6(2)]S[table-format = -1.4(2)]S[table-format = -1.3(2)]}
\toprule
\textbf{Dataset} & $a_{f,0}$ & $a_{f,1}$ & $a_{f,2}$ & $a_{\mathcal{F}_1,0}$ & $a_{\mathcal{F}_1,1}$ & $a_{\mathcal{F}_1,2}$ \\
\midrule
\texttt{HPQCD 23} & 0.01232(20) & 0.015(15) & -0.20(43) & 0.002065(34) & -0.0083(48) & -0.03(12) \\
\texttt{JLQCD 23} & 0.01209(19) & 0.017(10) & -0.04(42) & 0.002026(32) & 0.0005(38) & 0.011(49) \\
\texttt{FNAL/MILC 21} & 0.01239(23) & 0.002(11) & -0.33(35) & 0.002077(38) & -0.0051(23) & -0.081(45) \\
\texttt{Combined} & 0.01221(12) & 0.0135(66) & -0.16(26) & 0.002047(19) & -0.0038(16) & -0.018(34) \\
\toprule
\textbf{Dataset} & $a_{\mathcal{F}_2,0}$ & $a_{\mathcal{F}_2,1}$ & $a_{\mathcal{F}_2,2}$ & $a_{g,0}$ & $a_{g,1}$ & $a_{g,2}$ \\
\midrule
\texttt{HPQCD 23} & 0.0466(34) & -0.24(14) & -0.07(55) & 0.0314(24) & -0.110(94) & 0.02(57) \\
\texttt{JLQCD 23} & 0.0500(17) & -0.081(78) & -0.01(57) & 0.0289(18) & -0.055(36) & -0.05(56) \\
\texttt{FNAL/MILC 21} & 0.0527(16) & -0.346(68) & 0.04(54) & 0.0328(12) & -0.155(53) & -0.04(56) \\
\texttt{Combined} & 0.05046(95) & -0.210(42) & 0.00(55) & 0.03101(85) & -0.061(26) & -0.12(55) \\
\bottomrule
\end{tabular}}
    \caption{Type-A Bayesian fit results at the truncation order $K = 3$ to \LQCD form factor data from the HPQCD~\cite{PhysRevD.109.094515}, JLQCD~\cite{PhysRevD.109.074503}, and FNAL/MILC~\cite{Bazavov_2022} collaborations.}
    \label{tab:TypeA_fit_3_3_3_3}
\end{table}

In \tref{tab:TypeA_fit_3_3_3_3} we show the results of a Type-A fit with $K = 3$ to each \LQCD dataset, and their collective combination (\texttt{Combined}). These results are consistent with the analogous results reported in Ref.~\cite{Bordone:2024weh}, with minor shifts in central values that arise due to the differing input parameters for the masses of $\bar{c}b$ bound states in the $B_c$ spectrum (compare \tref{tab:cb_masses} with Table 9 of Ref.~\cite{Bordone:2024weh}). We observe the same $2.5 \sigma$ tension for $a_{\mathcal{F}_2,1}$ between \texttt{JLQCD 23} and \texttt{FNAL/MILC 21}. Note that the frequentist approach in this work has not implemented unitarity constraints on the \BGL coefficients, leading to results for some coefficients at second order and higher that break the unitarity bounds. While the Type-A frequentist fits are not reproduced here, they are reported in \cref{Subsection:TFF_results} for fits to the tensor form factors from the HPQCD collaboration~\cite{PhysRevD.109.094515}, together with the Bayesian inference fits. 

\begin{figure}[t]
    \centering
    \includegraphics[width=\linewidth]{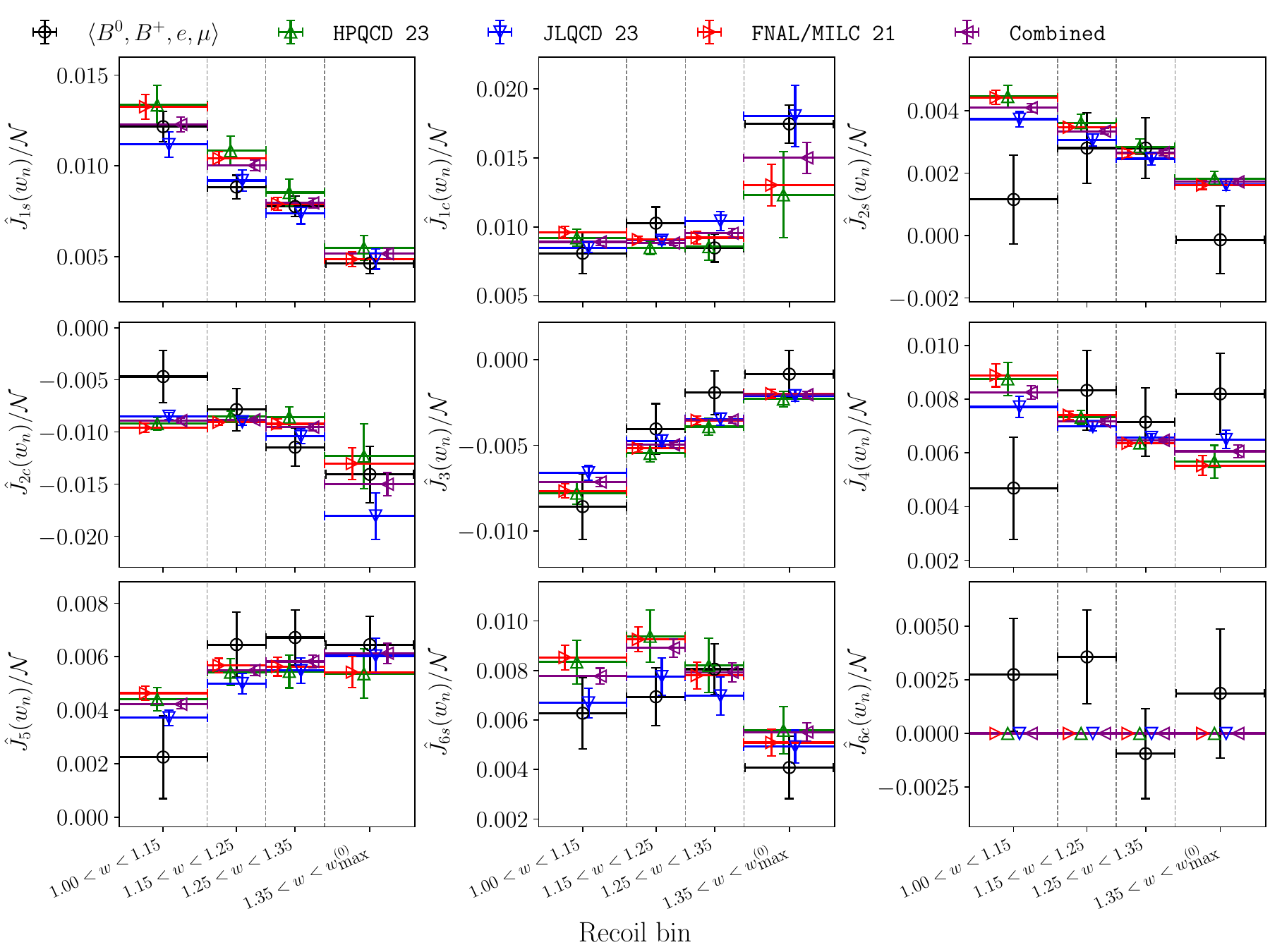}
    \caption{Partially-integrated, normalised angular coefficients from the \texttt{Belle 23 angular} $\langle B^0, B^+, e, \mu \rangle$ fully-averaged experimental dataset~\cite{Belle:2023xgj,hepdata.153299}, and \LQCD predictions from a $m_\ell = 0$, $K = 3$ Type-A fit to form-factor data from the HPQCD~\cite{PhysRevD.109.094515}, JLQCD~\cite{PhysRevD.109.074503}, and FNAL/MILC~\cite{Bazavov_2022} collaborations that follow from the results in \tref{tab:TypeA_fit_3_3_3_3}. Note that the \LQCD markers have been horizontally displaced to improve visibility when data points overlap.}
    \label{fig:angcoeffs}
\end{figure}

Using the \BGL coefficients in \tref{tab:TypeA_fit_3_3_3_3}, the angular coefficients in \crefrange{aligneq:J1s}{aligneq:J789} are calculated from the helicity amplitudes in \crefrange{aligneq:Hpm}{aligneq:Ht}. These are computed for each posterior sample from the Type-A Bayesian fit, and the central values and errors are determined by statistical bootstrapping the resulting samples for $\hat{J}_i(w_n) / \mathcal{N}$. \fref{fig:angcoeffs} plots the partially-integrated, normalised angular coefficients for each \LQCD dataset (calculated with $m_\ell = 0$), as well as the experimental $\langle B^0, B^+, e, \mu \rangle$ fully-averaged dataset. Overall, the \LQCD predictions are consistent with one another, with a few small tensions $(\sim 1.7\sigma)$ between the \texttt{JLQCD 23} results and those from \texttt{HPQCD 23} and \texttt{FNAL/MILC 21}. Similarly, this is also seen with the \LQCD predictions and those measured experimentally, most notably for the first and fourth bin of $\hat{J}_{2s} / \mathcal{N}$ (top right plot). As discussed in \cref{Subsection:fit_nomenclature}, a mismatch in signs from \LQCD and experiment for a given bin can cause complications in the binned extraction of $|V_{cb}|^2$, which is a strictly non-negative quantity. This is discussed more in the following section. 

\subsubsection[Results for \texorpdfstring{$|V_{cb}|$}{|Vcb|}]{Results for \texorpdfstring{\boldmath$|V_{cb}|$}{|Vcb|}}\label{Subsec:binned_Vcb}

The Type-A \BGL coefficients in \tref{tab:TypeA_fit_3_3_3_3} parameterise the form factors $\FFs$ over the full kinematic range according to \eqref{eq:FF_B_phi_series_trunc}, providing \LQCD predictions for the angular coefficients as a function of $w$ that are independent of the experimental data. Note that, as in \eqref{eq:Belle_w_bins}, the endpoint bin in $w$ from the experimental data is quoted as $1.35 < w < 1.50$~\cite{Belle:2023xgj}, whereas the lattice angular coefficients are integrated up to $w_\textrm{max}^{(\ell)}$. As a reminder, the angular coefficient $J_{6c} = 0$ for massless leptons, and only obtains a noticeable deviation from zero for $\ell = \mu$ at large $w$, or $\ell = \tau$. As such, this leaves eight angular coefficients suitable for a binned extraction of $|V_{cb}|$. 

\begin{figure}[t]
    \centering
    \includegraphics[width=\linewidth]{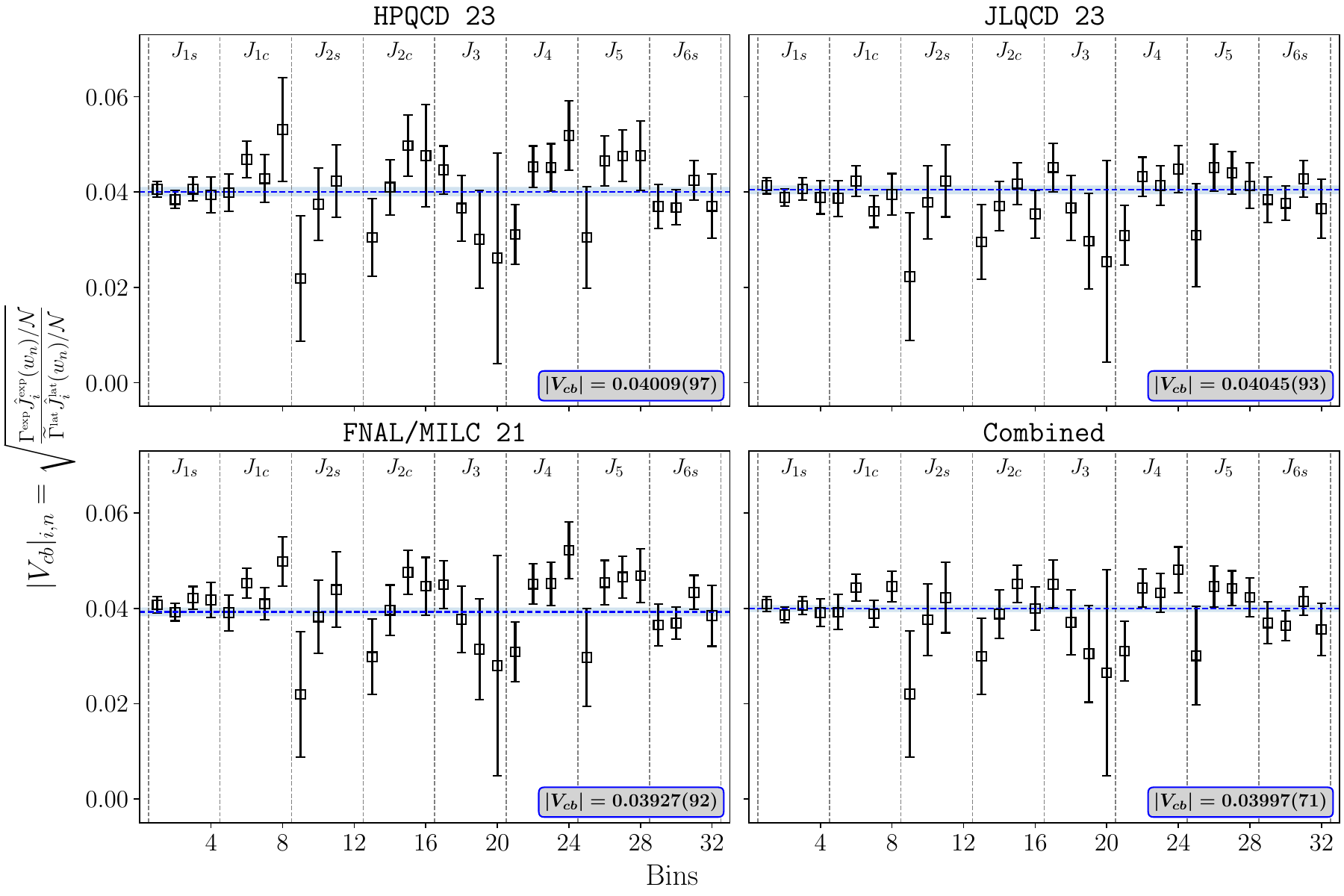}
    \caption{Plots of binned $|V_{cb}|$ values calculated from \eqref{eq:Vcb_binned_quotient} using the \texttt{Belle 23 angular} $\langle B^0, B^+, e, \mu \rangle$ experimental dataset~\cite{Belle:2023xgj,hepdata.153299}, together with the \texttt{HPQCD 23}~\cite{PhysRevD.109.094515}, \texttt{JLQCD 23}~\cite{PhysRevD.109.074503} and \texttt{FNAL/MILC 21}~\cite{Bazavov_2022} \LQCD datasets with $m_\ell = 0$ -- the quantities in \fref{fig:angcoeffs}. The \IGLS fit value for $|V_{cb}|$ is quoted in the bottom right of each plot. Note that the final point for $\hat{J}_{2s} / \mathcal{N}$ is absent from the plot as it predicts $|V_{cb}|^2 < 0$, but it still contributes to the overall fit to $|V_{cb}|^2$.}
    \label{fig:Vcb_fit}
\end{figure}

It is evident from \fref{fig:angcoeffs} that the experimental and lattice measurements may differ in sign, for instance, the first and fourth bins of $\hat{J}_{2s} / \mathcal{N}$. However, the non-negativity of $|V_{cb}|^2$ requires that the experimental and lattice angular coefficients are of the same sign in a given bin. As explained above, the generalised least-squares kernel in \eqref{eq:chi2_Vcb_diff} is employed with central values in the constant correlated fit, rather than the ratios in \eqref{eq:chi2_Vcb_ratio} determined by resampling. One may wonder whether the negative $|V_{cb}|^2$ values are a result of incomplete experimental data representations. For instance, while the angular coefficient $\hat J_{2s} / \mathcal{N}$ defined in \eqref{aligneq:J2s} should be manifestly positive, its representation by a (even negative) central value and covariance matrix in Refs.~\cite{Belle:2023xgj,hepdata.153299}, which assumes a normal distribution, is likely not optimal. A more versatile data representation, such as histogram distributions, bootstrap bins, or likelihood functions, should be employed and made publicly available for both experimental and \LQCD data. In this way, more detailed analyses are possible through resampling, which a priori excludes unphysical results from the analysis chain, such as negative values of $|V_{cb}|^2$ encountered here.

\begin{table}[t] 
    \centering
    \resizebox{.75\linewidth}{!}{
\begin{tabular}{ccS[table-format = 1.5(2)]cc}
\toprule
\textbf{Experimental Dataset} & \textbf{LQCD Dataset} & \multicolumn{1}{c}{\boldmath$|V_{cb}|$} & \multicolumn{1}{c}{\boldmath$\chi^2/N_{\mathrm{dof}}$} & \multicolumn{1}{c}{\boldmath$p$-\textbf{value}} \\
\midrule
\multirow{4}{*}{$\langle B^0, B^+, e, \mu \rangle$} & \texttt{HPQCD 23} & 0.04009(97) & 29.2/31 & 0.56 \\
& \texttt{JLQCD 23} & 0.04045(93) & 28.4/31 & 0.60 \\
& \texttt{FNAL/MILC 21} & 0.03927(92) & 35.1/31 & 0.28 \\
& \texttt{Combined} & 0.03997(71) & 29.9/31 & 0.52 \\
\midrule
\multirow{4}{*}{$\langle B^0, B^+, e \rangle$} & \texttt{HPQCD 23} & 0.0399(10) & 28.5/31 & 0.59 \\
& \texttt{JLQCD 23} & 0.0402(10) & 27.0/31 & 0.67 \\
& \texttt{FNAL/MILC 21} & 0.03940(99) & 33.4/31 & 0.35 \\
& \texttt{Combined} & 0.04000(76) & 28.6/31 & 0.59 \\
\midrule
\multirow{4}{*}{$\langle B^0, B^+, \mu \rangle$} & \texttt{HPQCD 23} & 0.0402(10) & 29.0/31 & 0.57 \\
& \texttt{JLQCD 23} & 0.04046(99) & 27.9/31 & 0.63 \\
& \texttt{FNAL/MILC 21} & 0.03964(99) & 33.4/31 & 0.35 \\
& \texttt{Combined} & 0.04017(76) & 29.4/31 & 0.55 \\
\midrule
\multirow{4}{*}{$\langle B^0, e \rangle$} & \texttt{HPQCD 23} & 0.0396(12) & 29.7/31 & 0.53 \\
& \texttt{JLQCD 23} & 0.0396(12) & 27.8/31 & 0.63 \\
& \texttt{FNAL/MILC 21} & 0.0397(12) & 32.6/31 & 0.39 \\
& \texttt{Combined} & 0.04007(87) & 28.6/31 & 0.59 \\
\midrule
\multirow{4}{*}{$\langle B^0, \mu \rangle$} & \texttt{HPQCD 23} & 0.0393(12) & 26.8/31 & 0.68 \\
& \texttt{JLQCD 23} & 0.0392(12) & 24.3/31 & 0.80 \\
& \texttt{FNAL/MILC 21} & 0.0393(11) & 32.2/31 & 0.41 \\
& \texttt{Combined} & 0.03983(87) & 27.2/31 & 0.66 \\
\midrule
\multirow{4}{*}{$\langle B^+, e \rangle$} & \texttt{HPQCD 23} & 0.0406(13) & 26.5/31 & 0.69 \\
& \texttt{JLQCD 23} & 0.0408(13) & 25.1/31 & 0.76 \\
& \texttt{FNAL/MILC 21} & 0.0400(13) & 32.8/31 & 0.38 \\
& \texttt{Combined} & 0.0405(11) & 28.9/31 & 0.57 \\
\midrule
\multirow{4}{*}{$\langle B^+, \mu \rangle$} & \texttt{HPQCD 23} & 0.0412(13) & 28.8/31 & 0.58 \\
& \texttt{JLQCD 23} & 0.0413(13) & 28.2/31 & 0.61 \\
& \texttt{FNAL/MILC 21} & 0.0407(13) & 30.6/31 & 0.49 \\
& \texttt{Combined} & 0.0410(11) & 29.3/31 & 0.55 \\
\bottomrule
\end{tabular}}
    \caption{Frequentist $|V_{cb}|$ fit values (based on the Type-A form-factor fit results in \tref{tab:TypeA_fit_3_3_3_3}) to the \texttt{HPQCD 23}~\cite{PhysRevD.109.094515}, \texttt{JLQCD 23}~\cite{PhysRevD.109.074503}, and \texttt{FNAL/MILC 21}~\cite{Bazavov_2022} \LQCD datasets and their combination together with each experimental \texttt{Belle 23 angular} dataset~\cite{Belle:2023xgj,hepdata.153299}, respectively.
    } 
    \label{tab:Vcb_bin_fit} 
\end{table}
 
\begin{figure}[t]
    \centering
    \includegraphics[width=\linewidth]{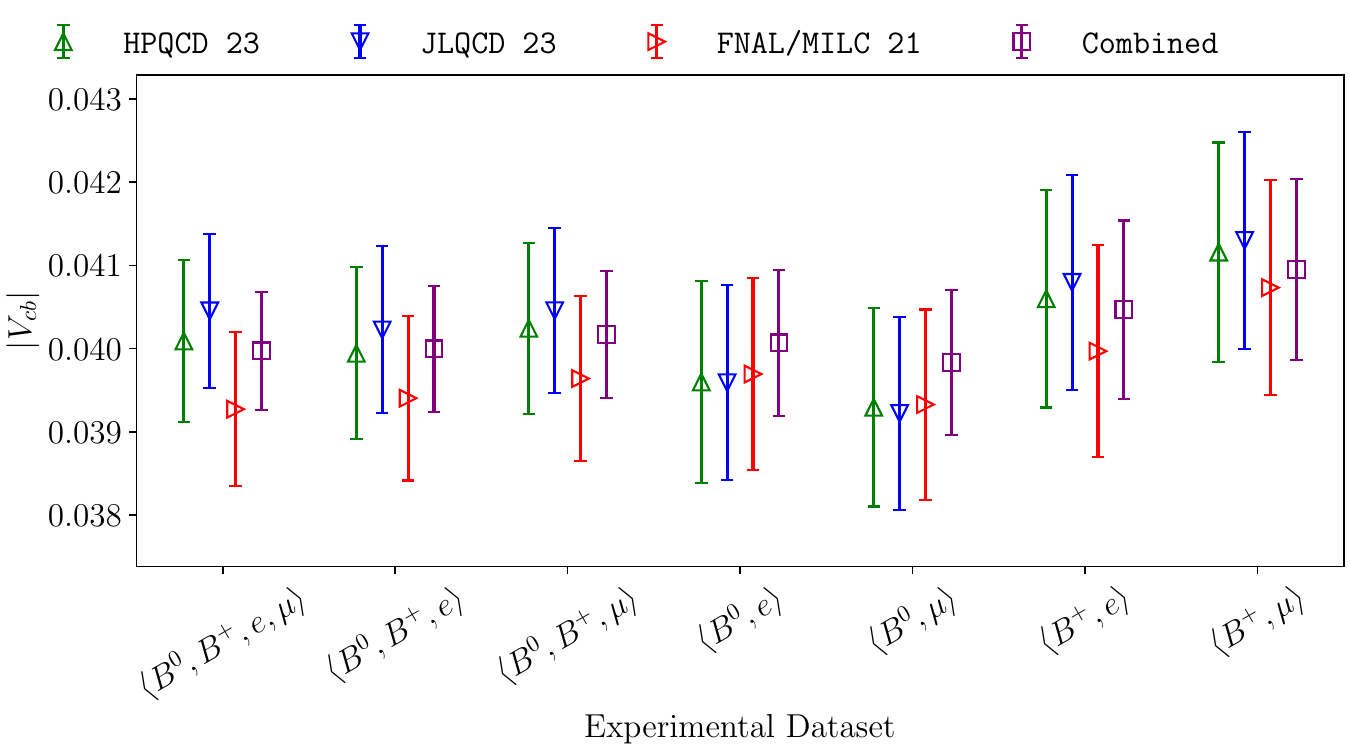}
    \caption{Summary of Type-A frequentist $|V_{cb}|$ fit results for each combination of \texttt{Belle 23 angular} experimental~\cite{Belle:2023xgj,hepdata.153299} and \LQCD datasets~\cite{PhysRevD.109.094515,PhysRevD.109.074503,Bazavov_2022}.}
    \label{fig:Vcb_TypeA_scatter}
\end{figure}

The fits performed with the fully-averaged $\langle B^0, B^+, e, \mu \rangle$ experimental dataset and \LQCD predictions with $m_\ell = 0$ are shown in \fref{fig:Vcb_fit}. For each \LQCD dataset and lepton mass, the fourth bin of $\hat{J}_{2s} / \mathcal{N}$ produces a negative central value for $|V_{cb}|^2$ (which is expected from \fref{fig:angcoeffs}), and is hence excluded from the plot (but not the fit) in \fref{fig:Vcb_fit}. Results for $|V_{cb}|$ using the other experimental datasets and the same $K = 3$ Type-A calculation of the angular coefficients with the relevant lepton masses are tabulated in \tref{tab:Vcb_bin_fit} and plotted in \fref{fig:Vcb_TypeA_scatter}. The reduced chi-squared test statistic, $\chi^2 / N_{\textrm{dof}}$, is reported for each frequentist fit, where $N_{\textrm{dof}} = N_{\textrm{data}} - 1$ is the number of data points contributing minus the number of free parameters (one, a constant) in the fit ansatz. All of the Type-A fit results for $|V_{cb}|$ in \tref{tab:Vcb_bin_fit} are self-consistent and produce acceptable $p$-values, with all but \texttt{FNAL/MILC 21} achieving $\chi^2 / N_{\textrm{dof}} \lesssim 1$. Excluding the $B^0$ experimental datasets, we find that \texttt{FNAL/MILC 21} consistently predicts the smallest value for $|V_{cb}|$, concordant with the results of the binned\footnote{Note that as the bootstrap samples for the total rate $\widetilde{\Gamma}^\textrm{lat}$ are calculated from four fully integrated coefficients as in \eqref{eq:Gamma}, the approach in this work is not a true binned analysis. That is, each angular coefficient bin is multiplied by the same $\widetilde{\Gamma}^{\textrm{lat}}$.} analysis for the $w$ channel in Ref.~\cite{Bordone:2024weh}. However, for the $B^0$ experimental datasets, this is not the case: The trending hierarchy in $|V_{cb}|$ amongst \texttt{HPQCD 23}, \texttt{JLQCD 23}, and \texttt{FNAL/MILC 21} is not present, easing the tension seen in the ratio approach by a factor $\sim 2$. As this reshuffling of the \texttt{Combined} prediction relative to the constituent \LQCD datasets is not the case for the \BGL coefficients nor the angular coefficients (see \tref{tab:TypeA_fit_3_3_3_3} and \fref{fig:angcoeffs}), then it could, as discussed above, be caused by the covariance representation of the experimental data that imposes Gaussianity. Furthermore, the largest and significant tension in $|V_{cb}|$ that we observe is between $\langle B^0, \mu\rangle$ and $\langle B^+, \mu \rangle$ for \texttt{JLQCD 23}. 

A similar binned calculation of $|V_{cb}|$ in the study of differential decay rates in the kinematic variables $\{w, \cos\theta_\ell, \cos\theta_v, \chi\}$ carried out in Ref.~\cite{Bordone:2024weh} did not suffer from the aforementioned negativity issue. However, significant bin-to-bin variations in $|V_{cb}|$ for the angular variables $\{\cos\theta_\ell, \cos\theta_v, \chi\}$ for \texttt{HPQCD 23} and \texttt{FNAL/MILC 21} necessitated an analysis implementing the \AIC~\cite{Akaike:1974vps} to mitigate the effect of the d'Agostini bias~\cite{DAgostini:1993arp}. Such an approach is not required here, however, as the binned predictions for $|V_{cb}|$ are much more compatible -- compare \fref{fig:Vcb_fit} with Figures 5 and 6 of Ref.~\cite{Bordone:2024weh}. 

It is also worth noting that the differences in the $|V_{cb}|$ determinations from the $B^0$ and $B^+$ channels, as visible in \cref{fig:Vcb_TypeA_scatter,fig:Vcb_summary_plot}, although not very significant, raise an interesting question concerning possible isospin-breaking effects. While we refrain from drawing any firm conclusion, a recent analysis in Ref.~\cite{Jung:2026ewj} reports indications of possible isospin-violating effects in the relative production of $B^0$ and $B^+$ mesons in $\Upsilon(4S)$ decays. Such an effect could enter the determination of the $\BotoDstar$ and $\BptoDstar$ branching fractions, and may help alleviate the differences observed in this analysis.

To summarise, the final fit result for the binned average of $|V_{cb}|$ using the fully-averaged $\langle B^0, B^+, e, \mu \rangle$ experimental dataset and the combination of \LQCD datasets neglecting lepton masses is:
\begin{equation}\label{eq:TypeA_Vcb_final}
    |V_{cb}|_\textrm{Type A} = 0.03997(71),
\end{equation}
with $(p, \chi^2 / N_{\textrm{dof}}, N_{\textrm{dof}}) = (0.52, 0.96, 31)$. For comparison, the Type-A analysis in Ref.~\cite{Bordone:2024weh} of the same three \LQCD results with a combination of the \texttt{Belle 23}~\cite{Belle:2023bwv,hepdata.137767} and \texttt{Belle II 23}~\cite{Belle-II:2023okj,hepdata.145129} datasets by HFLAV~\cite{HeavyFlavorAveragingGroupHFLAV:2024ctg,HFLAV_spectrum} for the differential decay rates obtained the fully compatible value $|V_{cb}|=0.04042(71)$. Furthermore, our result is compatible with the analysis in Refs.~\cite{Martinelli:2021myh,Martinelli:2024bov}, which used the same datasets as in Ref.~\cite{Bordone:2024weh}.

\begin{figure}[t]
    \centering
    \includegraphics[width=0.5\linewidth]{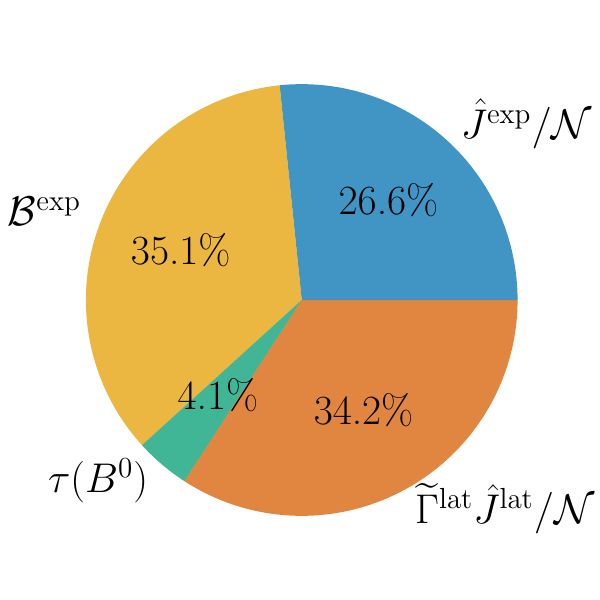}
    \caption{Decomposed error budget for the Type-A determination of $|V_{cb}|$ for the \texttt{Combined} \LQCD dataset and the fully averaged \texttt{Belle 23 angular} $\langle B^0, B^+, e, \mu \rangle$ experimental dataset~\cite{Belle:2023xgj,hepdata.153299}.}
    \label{fig:Vcb_error_budget}
\end{figure}

In \fref{fig:Vcb_error_budget} we decompose the sources of error and their relative contribution in the Type-A determination of $|V_{cb}|$ (see \cref{Subsubsection:Vcb_determination}), for the specific case of the combination of \LQCD inputs and the fully averaged \texttt{Belle 23 angular} $\langle B^0, B^+, e, \mu \rangle$ experimental dataset~\cite{Belle:2023xgj,hepdata.153299}. Recall that for this combination, it is the averaged branching fraction and $B^0$ lifetime in \cref{eq:tau_B0,eq:branching_fraction_B0_avg}, respectively, that are employed. Looking at the individual components of the analysis, the error budget is dominated by the \LQCD form factors and the total branching fraction, with approximately equal contributions. A very similar decomposition was found for the analogous analysis in Ref.~\cite{Bordone:2024weh}. There is therefore scope for further improvement by working on more precise \LQCD computations, as well as dedicated experimental measurements. While there is still some way to go towards sub-percent level precision, it is worthwhile keeping in mind that this will also require further formal developments. In particular, taking QED and strong isospin effects into account more comprehensively. While first ideas have been formulated~\cite{Christ:2024xzj}, a pathway to a full calculation of semileptonic decay rates on the lattice incorporating QCD, QED, and strong isospin still needs to be found. It is also worth mentioning in this context there are new efforts to tackle the $|V_{cb}|$ determination via inclusive $B\to X_c\ell\bar\nu_\ell$ on the lattice~\cite{Bailas:2020qmv,Gambino:2020crt,Barone:2023tbl,Kellermann:2025pzt,DeSantis:2025qbb,DeSantis:2025yfm}. This approach is orthogonal to the established operator product expansion strategy, and hence, could in the future prove decisive in resolving the $|V_{cb}|$ puzzle.

\subsection{Type-B Fit Results}\label{Subsec:Type_B_results}

\begin{table}[t]
    \centering
    \resizebox{\linewidth}{!}{
\begin{tabular}{cS[table-format = 1.5(2)]S[table-format = -1.4(2)]S[table-format = -1.2(2)]S[table-format = 1.6(2)]S[table-format = -1.4(2)]S[table-format = -1.3(2)]}
\toprule
\textbf{Dataset} & \multicolumn{1}{c}{$a_{f,0}$} & \multicolumn{1}{c}{$a_{f,1}$} & \multicolumn{1}{c}{$a_{f,2}$} & \multicolumn{1}{c}{$a_{\mathcal{F}_1,0}$} & \multicolumn{1}{c}{$a_{\mathcal{F}_1,1}$} & \multicolumn{1}{c}{$a_{\mathcal{F}_1,2}$} \\
\midrule
\texttt{HPQCD 23} & 0.01229(20) & 0.013(14) & -0.29(33) & 0.002060(34) & -0.0064(34) & 0.092(67) \\
\texttt{JLQCD 23} & 0.01206(18) & 0.0139(94) & -0.21(29) & 0.002022(31) & -0.0022(22) & 0.015(40) \\
\texttt{FNAL/MILC 21} & 0.01244(23) & 0.0121(95) & -0.21(27) & 0.002084(38) & -0.0022(17) & -0.001(30) \\
\texttt{Combined} & 0.01223(11) & 0.0149(63) & -0.21(21) & 0.002049(19) & -0.0027(13) & 0.013(25) \\
\toprule
\textbf{Dataset} & \multicolumn{1}{c}{$a_{\mathcal{F}_2,0}$} & \multicolumn{1}{c}{$a_{\mathcal{F}_2,1}$} & \multicolumn{1}{c}{$a_{\mathcal{F}_2,2}$} & \multicolumn{1}{c}{$a_{g,0}$} & \multicolumn{1}{c}{$a_{g,1}$} & \multicolumn{1}{c}{$a_{g,2}$} \\
\midrule
\texttt{HPQCD 23} & 0.0481(29) & -0.085(67) & 0.01(53) & 0.0317(20) & -0.095(68) & 0.01(53) \\
\texttt{JLQCD 23} & 0.0498(15) & -0.133(38) & 0.01(52) & 0.0299(11) & -0.046(27) & -0.08(53) \\
\texttt{FNAL/MILC 21} & 0.0531(16) & -0.213(36) & 0.45(38) & 0.0329(11) & -0.103(38) & -0.03(53) \\
\texttt{Combined} & 0.05087(93) & -0.173(30) & 0.32(43) & 0.03110(74) & -0.058(24) & -0.15(51) \\
\bottomrule
\end{tabular}}
    \caption{Type-B Bayesian fit results at the truncation order $K = 3$ to \texttt{HPQCD 23}~\cite{PhysRevD.109.094515}, \texttt{JLQCD 23}~\cite{PhysRevD.109.074503}, and \texttt{FNAL/MILC 21}~\cite{Bazavov_2022} \LQCD datasets 
    and the \texttt{Belle 23 angular} $\langle B^0, B^+, e, \mu \rangle$ experimental dataset~\cite{Belle:2023xgj,hepdata.153299}.}
    \label{tab:TypeB_fit_3_3_3_3}
\end{table}

Recall that fitting to the \BGL coefficients through the angular coefficients forfeits linearity in the BGL parameters. Thus, the Type-B fits are implemented in the \texttt{BFF} library~\cite{BFF_code} using \texttt{PyMultiNest}~\cite{Buchner_2014,Feroz_2008,Feroz_2009,Feroz_2019}. \tref{tab:TypeB_fit_3_3_3_3} lists the results of a Type-B fit with $K = 3$ for the \BGL coefficients and $|V_{cb}|$ using the \texttt{Belle 23 angular} $\langle B^0, B^+, e, \mu \rangle$ fully-averaged experimental dataset~\cite{Belle:2023xgj,hepdata.153299}. Comparing with the Type-A fit results in \tref{tab:TypeA_fit_3_3_3_3}, both are in agreement for the leading-order coefficients, with some small differences arising for the higher-order coefficients. In particular, $a_{\mathcal{F}_1, 1}$ shows a $\sim 1 \sigma$ deviation and $a_{\mathcal{F}_1, 2}$ a $\sim 1.5 \sigma$ deviation for \texttt{FNAL/MILC 21}. While in the Type-A fit, the results for $a_{\mathcal{F}_1, 2}$ and $a_{\mathcal{F}_2, 1}$ for \texttt{HPQCD 23} have a lower precision than other \LQCD datasets, this is restored in the Type-B fits. Inclusion of experimental data can modify the \BGL coefficients of $\mathcal{F}_2$, even for massless leptons, by virtue of the kinematic constraint in \eqref{aligneq:fF1F2F1_constraints} that relates $\mathcal{F}_2$ to $\mathcal{F}_1$ at $w = w_{\textrm{max}}^{(\ell)}$. Despite fitting with non-zero lepton masses provides sensitivity to $H_t(w) \propto \mathcal{F}_2(w)$, no improvement in the precision of the \BGL coefficients was observed as the non-averaged experimental datasets have larger uncertainties than those averaged -- compare \cref{fig:exp_ang_coeffs_ml,fig:exp_ang_coeffs_B0B+_e,fig:exp_ang_coeffs_B0B+_mu}. 

\begin{figure}[t]
    \centering
    \includegraphics[width=\linewidth]{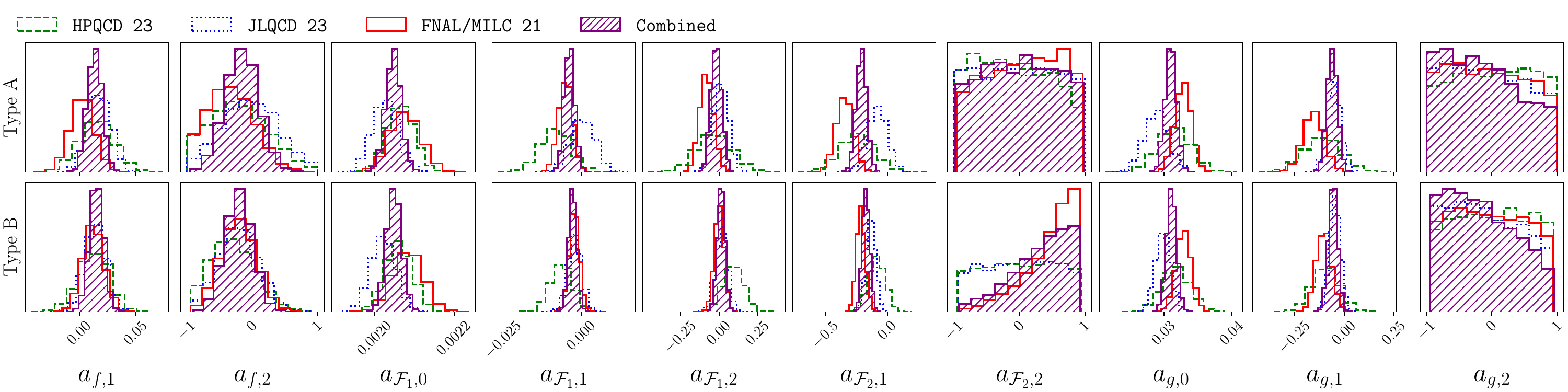}
    \caption{Normalised posterior distributions of the \BGL coefficients that parameterise the form factors $\FFs$ at order $K = 3$ for different cases. Top: Type-A fit to the individual \LQCD datasets \texttt{HPQCD 23}~\cite{PhysRevD.109.094515}, \texttt{JLQCD 23}~\cite{PhysRevD.109.074503}, \texttt{FNAL/MILC 21}~\cite{Bazavov_2022}, and their combination. Bottom: the analogous Type-B fits using the \texttt{Belle 23 angular} $\langle B^0, B^+, e, \mu \rangle$ dataset~\cite{Belle:2023xgj,hepdata.153299}. Note that the coefficients $a_{f,0}$ and $a_{\mathcal{F}_2,0}$ are not shown as they are eliminated by the kinematical constraints in \eqref{aligneq:fF1F2F1_constraints}.}
    \label{fig:TypeAB_hists}
\end{figure}

Histograms of the posterior distributions of the 
Type-A and Type-B fits are compared in \fref{fig:TypeAB_hists}. Whilst in this work we consider experimental data for the angular coefficients $\hat{J}_i(w) / \mathcal{N}$, the fits in Figure 2 of Ref.~\cite{Bordone:2024weh}\footnote{Note that the last four axis labels should read $a_{g,0}, a_{g,1}, a_{g,2}, a_{g,3}$, not $a_{f,0}, a_{f,1}, a_{f,2}, a_{f,3}$.} took as experimental input the statistical combination of the \texttt{Belle 23}~\cite{Belle:2023bwv,hepdata.137767} and \texttt{Belle II 23}~\cite{Belle-II:2023okj,hepdata.145129} datasets (provided by HFLAV~\cite{HFLAV_spectrum}) on normalised differential decay rates of $\BtoDstar$ in the kinematic variables $w$, $\cos\theta_\ell, \cos\theta_v$, and $\chi$. Each dataset contains a similar number of observables, but the relative uncertainties for the angular coefficients $\hat{J}_i(w) / \mathcal{N}$ are roughly an order of magnitude larger than for the normalised differential rates. Overall, the fits show good consistency with those in Ref.~\cite{Bordone:2024weh} across different experimental inputs, with the only exception being the \BGL coefficients $a_{\mathcal{F}_1, 1}$ and $a_{\mathcal{F}_1, 2}$. This is particularly true for the \texttt{HPQCD 23}~\cite{PhysRevD.109.094515} dataset, though the origin is not clear. The higher-precision data provided in the \texttt{Belle II 23} and \texttt{HFLAV 23} datasets better constrain the two \BGL coefficients and, to a lesser extent, $a_{f,2}$. In addition, as in Ref.~\cite{Bordone:2024weh}, we also observe that the inclusion of experimental input with the \texttt{FNAL/MILC 21}~\cite{Bazavov_2022} dataset skews the posterior distribution of $a_{\mathcal{F}_2,2}$ towards +1, which is the upper limit that the unitarity constraint can accommodate. The fact that this persists across all experimental inputs indicates consistent potential incompatibility with unitarity for this particular \LQCD dataset. Note that the unitarity sums are not violated due to the window function (see \cref{Subsection:fit_nomenclature}). What is unclear is the exact cause of this artefact, specific to Type-B fits. The kinematic constraint relating $\mathcal{F}_1$ to $\mathcal{F}_2$ at $w = w_{\textrm{max}}^{(\ell)}$ in \eqref{aligneq:fF1F2F1_constraints} is not implemented exactly by FNAL/MILC as is done by HPQCD~\cite{PhysRevD.109.094515} and JLQCD~\cite{PhysRevD.109.074503}, but is instead checked a posteriori~\cite{Bazavov_2022,Tsang:2023nay}. Without explicitly imposing this constraint, the relevant unitarity sum as computed by FNAL/MILC is satisfied within one standard deviation of the Gaussian posterior~\cite{Bazavov_2022}. As speculated in Ref.~\cite{Bordone:2024weh}, this difference in methodology relative to the other \LQCD collaborations could be responsible for the observed posterior skew of $a_{\mathcal{F}_2, 2}$.

\subsubsection[\texorpdfstring{Results for $|V_{cb}|$}{Results for |Vcb|}]{\texorpdfstring{Results for \boldmath$|V_{cb}|$}{Results for |Vcb|}}

In the Type-B method, the \BGL coefficients and $|V_{cb}|$ are determined simultaneously in the same fit through the addition of \eqref{eq:chi^2_norm} to the likelihood, which relies on the experimental inputs given in \cref{eq:branching_fraction_B0,eq:branching_fraction_B-,eq:branching_fraction_B0_avg,eq:tau_B0,eq:tau_B-}. The Type-B results for $|V_{cb}|$, from the same fits as in \tref{tab:TypeB_fit_3_3_3_3} and in the bottom panel of \fref{fig:TypeAB_hists}, are shown in \tref{tab:TypeB_Vcb_ml}. These results show excellent agreement both across \LQCD datasets for a given experimental dataset and with the analogous Type-A results. The same trends as in the Type-A results carry through, namely \texttt{FNAL/MILC} having the smallest predictions, excluding the case $B = B^0$. Collectively, the largest shifts between the Type-A and Type-B determinations are found for the $\langle B^+, \mu \rangle$ experimental dataset, and they are well below one standard deviation. 

\begin{table}[t]
    \centering
    \begin{tabular}{cS[table-format=1.5(2)]S[table-format=1.5(2)]S[table-format=1.5(2)]S[table-format=1.5(2)]}
\toprule
\multirow{2}{*}{\textbf{Dataset}} & \multicolumn{4}{c}{$|V_{cb}|$} \\
 & \multicolumn{2}{c}{$\langle B^0, B^+, e, \mu \rangle$} & \multicolumn{1}{c}{$\langle B^0, B^+, e \rangle$} & \multicolumn{1}{c}{$\langle B^0, B^+, \mu \rangle$} \\
\midrule
\texttt{HPQCD 23} & \multicolumn{2}{c}{0.04046(97)} & 0.0404(10) & 0.0406(10) \\
\texttt{JLQCD 23} & \multicolumn{2}{c}{0.04073(94)} & 0.0405(10) & 0.0407(10) \\
\texttt{FNAL/MILC 21} & \multicolumn{2}{c}{0.03973(91)} & 0.03984(97) & 0.04008(99) \\
\texttt{Combined} & \multicolumn{2}{c}{0.04029(71)} & 0.04033(77) & 0.04052(76) \\
\midrule
\textbf{Dataset} & \multicolumn{1}{c}{$\langle B^0, e \rangle$} & \multicolumn{1}{c}{$\langle B^0, \mu \rangle$} & \multicolumn{1}{c}{$\langle B^+, e \rangle$} & \multicolumn{1}{c}{$\langle B^+, \mu \rangle$} \\
\midrule
\texttt{HPQCD 23} & 0.0401(12) & 0.0396(12) & 0.0415(13) & 0.0423(13) \\
\texttt{JLQCD 23} & 0.0400(12) & 0.0394(11) & 0.0415(13) & 0.0424(13) \\
\texttt{FNAL/MILC 21} & 0.0401(11) & 0.0396(11) & 0.0410(13) & 0.0418(13) \\
\texttt{Combined} & 0.04042(86) & 0.04008(86) & 0.0414(11) & 0.0420(11) \\
\bottomrule
\end{tabular}

    \caption{Type-B Bayesian fit results for $|V_{cb}|$ at the truncation order $K = 3$ to \texttt{HPQCD 23}~\cite{PhysRevD.109.094515}, \texttt{JLQCD 23}~\cite{PhysRevD.109.074503}, and \texttt{FNAL/MILC 21}~\cite{Bazavov_2022} \LQCD datasets 
    and the \texttt{Belle 23 angular} experimental datasets~\cite{Belle:2023xgj,hepdata.153299}.}
    \label{tab:TypeB_Vcb_ml}
\end{table}

\begin{figure}[t]
    \centering
    \includegraphics[width=\linewidth]{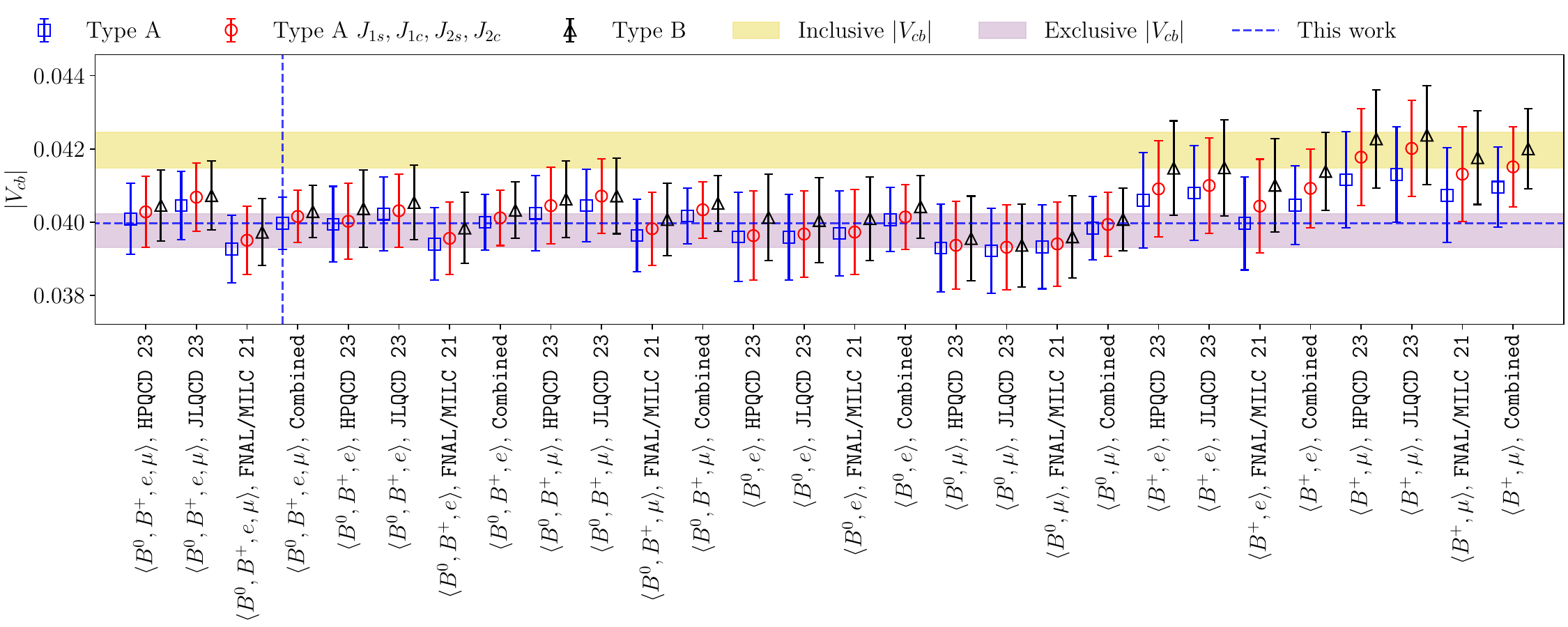}
    \caption{Summary plot of the various $|V_{cb}|$ determinations in this work compared with the $68\%$ confidence bands from inclusive $B$ decays and exclusive $B, B_s$, and $\Lambda_b$ decays~\cite{HeavyFlavorAveragingGroupHFLAV:2024ctg}. We select the fully-averaged $m_\ell = 0$ Type-A result from the \texttt{Belle 23 angular} $\langle B^0, B^+, e, \mu \rangle$ experimental and \texttt{Combined} \LQCD dataset as our nominal value, indicated by the dashed lines.}
    \label{fig:Vcb_summary_plot}
\end{figure}

While the Type-A determination of $|V_{cb}|$ takes into account eight of the angular coefficients, the expression for the total decay rate given in \eqref{eq:Gamma} depends only on the first four angular coefficients ($J_{1s}, J_{1c}, J_{2s}$, and $J_{2c}$). Consequently, there is little to no gain in precision for Type B fits over Type A fits. The Type-A fits to $|V_{cb}|$ were repeated using only these observables for a fairer comparison with the Type-B approach. These have been plotted in \fref{fig:Vcb_summary_plot} alongside the Type-A and Type-B results contained within \cref{tab:Vcb_bin_fit,tab:TypeB_Vcb_ml}. Including only the first four angular coefficients in the Type-A fit pulls up the central value by as much as $0.4\sigma$ for the cases with $B = B^+$, with little to no variation in the error, as many bins with worse experimental precision (producing $|V_{cb}|^2 < 0$) are omitted. Despite this, the Type-B values are consistently larger than the Type-A determinations, while retaining the same uncertainty. This hierarchy persists throughout all combinations of experimental and \LQCD datasets, covering the range spanned by the current inclusive and exclusive $|V_{cb}|$ values reported by HFLAV~\cite{HeavyFlavorAveragingGroupHFLAV:2024ctg} -- see the highlighted bands in \fref{fig:Vcb_summary_plot}. 

It is worth reiterating the following qualitative difference between Type-A and Type-B $|V_{cb}|$ determinations. Type-B fits impose unitarity not only on \LQCD data, but also on experimental data. While the \LQCD simulations assume unitarity within the \SM, this may not be true if there are \NP contributions to the experimental data. Therefore, the Type-A strategy is the conceptually cleaner determination. Despite this, the upward shift in $|V_{cb}|$ when comparing the $B^0$ and $B^+$ cases persists for both Type-A and Type-B calculations. In magnitude, it is consistent with that of the experimental decay rates:
\begin{equation}
    \frac{\Gamma^{\textrm{exp}}(\BptoDstar)}{\Gamma^{\textrm{exp}}(\BotoDstar)} = 1.045 \pm 0.049,
\end{equation}
which follows from \cref{eq:branching_fraction_B-,eq:branching_fraction_B0,eq:tau_B-,eq:tau_B0}. 

\section{New Physics in \texorpdfstring{\boldmath$\BtoDstar$}{B to Dstar}}\label{Sec:NPinBtoDstar}

The parameterisation of the differential angular distribution of $\BtoDstar$ in \eqref{eq:ang_distribution} holds not only in the \SM, but also in any \BSM scenario mediated by heavy \NP. At the low scale $\mu = m_b$ with only left-handed neutrinos, the effective Lagrangian mediating $b\to c\ell\bar{\nu}_\ell$ transitions reads~\cite{Buchmuller:1985jz,Jenkins:2017jig,Grzadkowski:2010es,Jung:2018lfu,Brivio:2017vri}:
\begin{equation}
 \mathcal{L}_\mathrm{eff} = -\frac{4G_\mathrm{F}}{\sqrt{2}} V_{cb} \sum_{i\neq V_L} C^\ell_i \mathcal{O}^\ell_i ,
 \label{eq:eff_hamiltonian}
\end{equation}
with
\begin{equation}\label{eq:SM_BSM_currents}
\begin{aligned}
    \mathcal{O}^\ell_{V_L} &= (\bar{c}_L\gamma^\mu b_L)(\bar{\ell}_L\gamma_\mu \nu^\ell_L)   \,, & \mathcal{O}^\ell_{V_R} &= (\bar{c}_R\gamma^\mu b_R)(\bar{\ell}_L\gamma_\mu \nu^\ell_L)\,, & \mathcal{O}^\ell_{S_R} &= (\bar{c}_L b_R)(\bar{\ell}_R \nu^\ell_L)\,, \\
    \mathcal{O}^\ell_{S_L} &= (\bar{c}_R b_L)(\bar{\ell}_R \nu^\ell_L)\,, & \mathcal{O}^\ell_{T} &= (\bar{c}_R\sigma^{\mu\nu} b_L)(\bar{\ell}_R\sigma_{\mu\nu} \nu^\ell_L)\,,
\end{aligned}
\end{equation}
and, ignoring lepton flavour violating effects, we choose to reabsorb any modification of the \SM-like operator $\mathcal{O}^\ell_{V_L}$. In this case, the \SM is recovered by setting $C^\ell_{V_R} = C^\ell_{S_R} = C^\ell_{S_L} = C^\ell_{T} = 0$ and $C^\ell_{V_L} = 1$. Expressions for the twelve angular coefficients for $\BtoDstar$ in and beyond the \SM for the currents in \eqref{eq:SM_BSM_currents} are available in~\cite{Altmannshofer:2008dz,Duraisamy:2014sna,Colangelo:2018cnj,Bhattacharya:2019olg,Colangelo:2024mxe,Kapoor:2024ufg}. While the \NP contributions associated to the operators $\mathcal{O}_{V_R}^\ell$, 
$\mathcal{O}^\ell_{S_L}$ and $\mathcal{O}^\ell_{S_R}$ are mediated by the same hadronic form factors as in the \SM, this is not the case for $\mathcal{O}^\ell_T$. For $\BtoDstar$ decays, assuming the presence of a non-zero contribution from $\mathcal{O}^\ell_T$ requires the introduction of three tensor form factors: $T_1$, $T_2$, and $T_3$. Therefore, in the following, we extend the analysis of Ref.~\cite{Bordone:2024weh} to include them. The relevant decompositions of hadronic matrix elements for vector, axial, and tensor currents in terms of form factors are given in \cref{Appendix:FFbasis}. Note that in the helicity basis, the form factor $T_3$ is replaced with $T_{23}$, given in \eqref{eq:T23}. In what follows, we drop the superscript $\ell$ on the Wilson coefficients.

\subsection{Tensor Form Factor Results}\label{Subsection:TFF_results}

\begin{figure}[t]
    \centering
    \includegraphics[width=\linewidth]{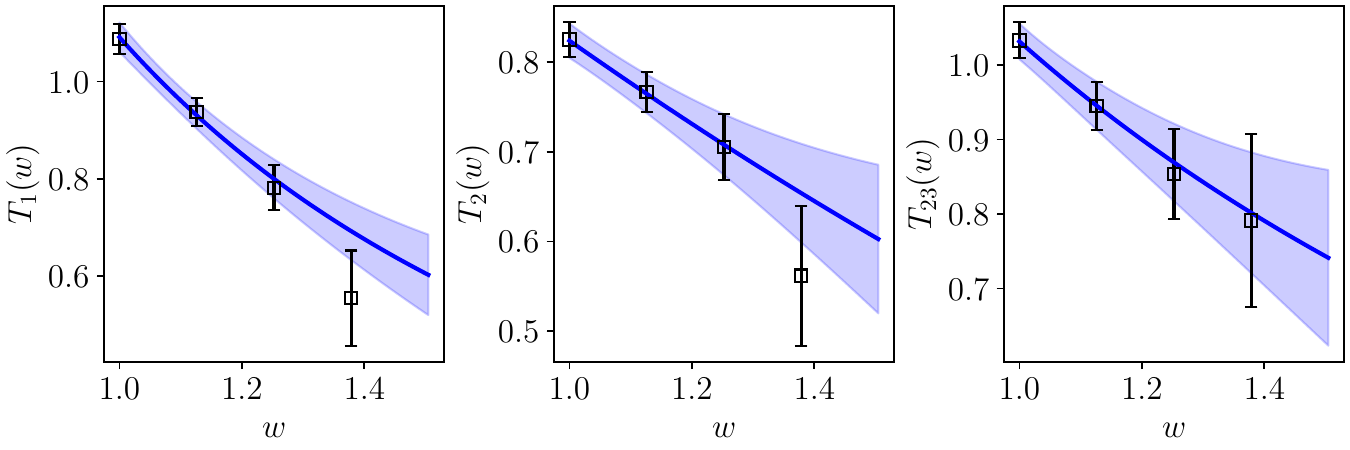}
    \caption{Type-A Bayesian fit with $(K_{T_1}, K_{T_2}, K_{T_{23}}) = (3, 3, 3)$ to tensor form factor data calculated by HPQCD~\cite{PhysRevD.109.094515}.}
    \label{fig:TFF_BI_fit}
\end{figure}

The \BGL parameterisations of the tensor form factors proceed exactly in the same way as for the \SM form factors, and are outlined in \cref{Appendix:FFbasis}. A unitarity-constrained $K = 3$ Bayesian inference \BGL fit to the tensor form factors computed by HPQCD~\cite{PhysRevD.109.094515} is shown in Figure~\ref{fig:TFF_BI_fit}, with the corresponding coefficients for different orders at the bottom of \tref{tab:TFF_freq_BI}. As discussed in \cref{Appendix:datasets}, the data point at $w = w_{\textrm{max}}$ was removed for each tensor form factor to reduce the condition number of the correlation matrix. With 12 data points, a frequentist fit is possible provided that $K_{T_1} + K_{T_2} + K_{T_{23}} \leq 12$ since the coefficient $a_{T_1,0}$ is eliminated by the kinematic constraint in \cref{eq:T12endpoint,eq:a_T1_0_relation}. Results for Type-A frequentist fits for various truncation orders are shown at the top of \tref{tab:TFF_freq_BI}. Comparing \cref{tab:TypeA_fit_3_3_3_3,tab:TFF_freq_BI} shows that the leading-order \BGL coefficients converge much quicker for the tensor form factors $\{T_1, T_2, T_{23}\}$ than for $\FFs$ in the \SM, with almost no variation in the leading coefficient from the lowest order fit. These same tensor form factors have been analysed in Ref.~\cite{Bordone:2025jur} within the framework of the heavy quark expansion with additional constraints.

\begin{table}[t]
    \centering
    \resizebox{.75\linewidth}{!}{
\begin{tabular}{cS[table-format = 1.6(2)]S[table-format = -1.4(2)]S[table-format = -1.2(2)]S[table-format = -1.2(2)]cc}
    \bottomrule
    \multicolumn{7}{c}{\cellcolor{gray!25}\textbf{Frequentist Fits}} \\
    \toprule
    $(K_{T_1}, K_{T_2}, K_{T_{23}})$ & \multicolumn{1}{c}{$a_{T_1,0}$} & \multicolumn{1}{c}{$a_{T_1,1}$} & \multicolumn{1}{c}{$a_{T_1,2}$} & $a_{T_1,3}$ & $p$ & $\chi^2 / N_\textrm{dof}$ \\
    \midrule
    (2, 1, 1) & 0.01346(35) & -0.0230(37) & {--} & {--} & 0.89 & 0.49/9 \\
    (2, 2, 2) & 0.01346(35) & -0.026(17) & {--} & {--} & 0.74 & 0.62/7 \\
    (3, 3, 3) & 0.01354(36) & -0.032(23) & -0.21(65) & {--} & 0.51 & 0.82/4 \\
    (4, 4, 4) & 0.01342(37) & -0.007(46) & -0.9(3.0) & \multicolumn{1}{c}{$-12.4(51.2)$} & 0.31 & 1.01/1 \\
    \toprule
    $(K_{T_1}, K_{T_2}, K_{T_{23}})$ & \multicolumn{1}{c}{$a_{T_2,0}$} & \multicolumn{1}{c}{$a_{T_2,1}$} & \multicolumn{1}{c}{$a_{T_2,2}$} & $a_{T_2,3}$ & $p$ & $\chi^2 / N_\textrm{dof}$ \\
    \midrule
    (2, 1, 1) & 0.003534(81) & {--} & {--} & {--} & 0.89 & 0.49/9 \\
    (2, 2, 2) & 0.003537(82) & -0.0010(48) & {--} & {--} & 0.74 & 0.62/7 \\
    (3, 3, 3) & 0.003536(82) & 0.0008(57) & -0.12(19) & {--} & 0.51 & 0.82/4 \\
    (4, 4, 4) & 0.003536(82) & -0.0035(97) & 0.39(77) & \multicolumn{1}{c}{$-12.9(15.1)$} & 0.31 & 1.01/1 \\
    \toprule
    $(K_{T_1}, K_{T_2}, K_{T_{23}})$ & \multicolumn{1}{c}{$a_{T_{23},0}$} & \multicolumn{1}{c}{$a_{T_{23},1}$} & \multicolumn{1}{c}{$a_{T_{23},2}$} & $a_{T_{23},3}$ & $p$ & $\chi^2 / N_\textrm{dof}$ \\
    \midrule
    (2, 1, 1) & 0.01054(24) & {--} & {--} & {--} & 0.89 & 0.49/9 \\
    (2, 2, 2) & 0.01055(25) & -0.002(18) & {--} & {--} & 0.74 & 0.62/7 \\
    (3, 3, 3) & 0.01055(25) & 0.004(23) & -0.40(91) & {--} & 0.51 & 0.82/4 \\
    (4, 4, 4) & 0.01055(25) & 0.009(46) & -1.1(4.0) & \multicolumn{1}{c}{$15.1(75.4)$} & 0.31 & 1.01/1 \\
    \bottomrule
    \multicolumn{7}{c}{\cellcolor{gray!25}\textbf{Bayesian Fits}} \\
    \toprule
    $(K_{T_1}, K_{T_2}, K_{T_{23}})$ & \multicolumn{1}{c}{$a_{T_1,0}$} & \multicolumn{1}{c}{$a_{T_1,1}$} & \multicolumn{1}{c}{$a_{T_1,2}$} & \multicolumn{1}{c}{$a_{T_1,3}$} \\
        \midrule
        (2, 1, 1) & 0.01346(35) & -0.0230(38) & \multicolumn{1}{c}{--} & \multicolumn{1}{c}{--} \\
        (2, 2, 2) & 0.01347(34) & -0.026(16) & \multicolumn{1}{c}{--} & \multicolumn{1}{c}{--} \\
        (3, 3, 3) & 0.01355(36) & -0.034(22) & -0.13(48) & \multicolumn{1}{c}{--} \\
        (4, 4, 4) & 0.01355(36) & -0.034(22) & -0.09(43) & 0.01(52) \\
        \toprule
        $(K_{T_1}, K_{T_2}, K_{T_{23}})$ & \multicolumn{1}{c}{$a_{T_2,0}$} & \multicolumn{1}{c}{$a_{T_2,1}$} & \multicolumn{1}{c}{$a_{T_2,2}$} & \multicolumn{1}{c}{$a_{T_2,3}$} \\
        \midrule
        (2, 1, 1) & 0.003535(81) & \multicolumn{1}{c}{--} & \multicolumn{1}{c}{--} & \multicolumn{1}{c}{--} \\
        (2, 2, 2) & 0.003537(80) & -0.0010(47) & \multicolumn{1}{c}{--} & \multicolumn{1}{c}{--} \\
        (3, 3, 3) & 0.003538(81) & 0.0004(54) & -0.09(15) & \multicolumn{1}{c}{--} \\
        (4, 4, 4) & 0.003537(81) & 0.0004(53) & -0.08(14) & -0.04(44) \\
        \toprule
        $(K_{T_1}, K_{T_2}, K_{T_{23}})$ & \multicolumn{1}{c}{$a_{T_{23},0}$} & \multicolumn{1}{c}{$a_{T_{23},1}$} & \multicolumn{1}{c}{$a_{T_{23},2}$} & \multicolumn{1}{c}{$a_{T_{23},3}$} \\
        \midrule
        (2, 1, 1) & 0.01055(24) & \multicolumn{1}{c}{--} & \multicolumn{1}{c}{--} & \multicolumn{1}{c}{--} \\
        (2, 2, 2) & 0.01055(24) & -0.002(18) & \multicolumn{1}{c}{--} & \multicolumn{1}{c}{--} \\
        (3, 3, 3) & 0.01055(24) & -0.000(20) & -0.13(52) & \multicolumn{1}{c}{--} \\
        (4, 4, 4) & 0.01055(24) & -0.001(19) & -0.07(41) & 0.00(45) \\
        \bottomrule
\end{tabular}}  
    \caption{Results for the Type-A frequentist and Bayesian fits to HPQCD tensor form-factor data~\cite{PhysRevD.109.094515} for various truncation orders. Note that a (1, 1, 1) fit is not possible because the \BGL coefficient $a_{T_1, 0}$ has been eliminated by the kinematic constraint in \eqref{eq:T12endpoint}.}
    \label{tab:TFF_freq_BI}
\end{table}

As for the \SM form-factor fits, the frequentist approach in this work does not implement unitarity, often causing higher-order coefficients to exceed unity in magnitude -- see \tref{tab:TFF_freq_BI}. Conversely, unitarity acts as a regulator in the Bayesian framework, constraining higher-order coefficients to be essentially compatible with zero. The results in \cref{tab:TFF_freq_BI} suggest that, to a very good approximation, the tensor form factors $T_{1}, T_{2}$, and $T_{23}$ are constant in the variable $z$. Allowing higher-order terms in the $z$ expansion improves the quality of the fit in the frequentist framework. Taking the results in \tref{tab:TFF_freq_BI} into account, we decide to parameterise each tensor form factor at the order $K = 3$ in subsequent analyses. 

With form-factor data for $\FFs$ and $\{T_1, T_2, T_{23}\}$, the angular coefficients in the presence of scalar, vector, and tensor currents may be constructed and compared with the experimental data. The domain of parameter space compatible with Belle's measurements has implications for \NP models. Therefore, we perform two types of analysis: We first extract the low-energy Wilson coefficients defined in \eqref{eq:SM_BSM_currents}, and then connect them to simplified models at high energies. To this end, in the next section, we introduce the formalism required to match the low-energy description onto simplified models, with a single new heavy bosonic particle that could mediate \NP in $b\to c\ell\bar{\nu}_\ell$ decays.

\subsection[New Physics in \texorpdfstring{$b \to c\ell\bar{\nu}_\ell$}{b -> clnu} Decays]{New Physics in \texorpdfstring{\boldmath$b \to c\ell\bar{\nu}_\ell$}{b -> clnu} Decays}

In order to study possible simplified models that can affect $b\to c\ell\bar{\nu}_\ell$ data, we relate the Wilson coefficients in the \WET to dimension-six operators in the \SMEFT~\cite{Brivio:2017vri,Buchmuller:1985jz}. In the Warsaw basis~\cite{Grzadkowski:2010es}, and considering dimension-six operators only, semileptonic $b\to c\ell\bar{\nu}_\ell$ decays are described by the following effective Lagrangian
\begin{equation}
\begin{aligned}
    \mathcal{L}_\mathrm{eff}(b\to c\ell\bar{\nu}_\ell) = -\frac{1}{\Lambda^2}\bigg\{& [C_{\ell q}^{(3)}]^{\alpha\beta ij}[Q_{\ell q}^{(3)}]_{\alpha\beta ij} + [C_{\ell edq}]^{\alpha\beta ij}[Q_{\ell edq}]_{\alpha\beta ij} \\
    &+ [C_{\ell equ}^{(1)}]^{\alpha\beta ij}[Q_{\ell equ}^{(1)}]_{\alpha\beta ij} + [C_{\ell equ}^{(3)}]^{\alpha\beta ij}[Q_{\ell equ}^{(3)}]_{\alpha\beta ij}\bigg\}\,,
\end{aligned}
\label{eq:Leff_SMEFT}
\end{equation}
where the \SMEFT operators are
\begin{equation}\label{eq:SMEFT_currents}
\begin{aligned}\relax
    [Q_{\ell q}^{(3)}]_{\alpha\beta ij} &= (\bar\ell_\alpha\gamma_\mu\tau^a \ell_\beta)(\bar q_i\gamma^\mu\tau^a q_j), & [Q_{\ell edq}]_{\alpha\beta ij} &= (\bar\ell_\alpha^k e_\beta)(\bar d_i q_j^k), \\
    [Q_{\ell equ}^{(1)}]_{\alpha\beta ij} &= (\bar\ell_\alpha^k e_\beta)\epsilon_{kl}(\bar q_i^l u_j), & [Q_{\ell equ}^{(3)}]_{\alpha\beta ij} &= (\bar\ell_\alpha^k\sigma_{\mu\nu} e_\beta)\epsilon_{kl}(\bar q_i^l \sigma^{\mu\nu} u_j),
\end{aligned}
\end{equation}
which do not modify $v$ nor $G_\textrm{F}$, i.e. we do not consider operators which can modify the W boson couplings to quarks and leptons. Here, $\tau^a$ are the Pauli matrices and $\epsilon_{kl} = i\tau^2_{kl}$ is totally anti-symmetric, with $\epsilon_{12} = +1$. In this notation, $q$ and $\ell$ are the left-handed quark and lepton SU$(2)_L$ doublets, and $u,\,d,\,e$ are the right-handed singlets. 

\subsubsection{Leptoquark Scenarios}

We assess the sensitivity to \NP based on various simplified scenarios. In particular, we restrict our study to scenarios in which we add either a scalar or a vector boson that generates only tree-level contributions to the effective Lagrangian in \eqref{eq:Leff_SMEFT}. The possible mediators, along with their $\textrm{SU}(3)_C \otimes \textrm{SU}(2)_L \otimes \textrm{U}(1)_Y$ quantum numbers, are:
\begin{itemize}
    \item Colourless scalar boson $H^\prime \sim 
    \left(1,2,-1/2\right)$,
    \item Colourless vector boson $W^\prime\sim (1,3,0)$,
    \item Scalar leptoquarks $S_1\sim(\bar{3},1,1/3)$, $S_3\sim(\bar{3},3,1/3)$, and $R_2\sim (3,2,7/6)$,
    \item Vector leptoquarks $U_1\sim(3,1,2/3)$, and $U_3\sim(3,3,2/3)$.
\end{itemize}
In \tref{tab:simp_models}, we list candidate mediators, along with their quantum numbers in the \SM gauge group, which can participate in $b \to c$ transitions. At the low energy $\mu_b = m_b$, these heavy states contribute to different sets of Wilson coefficients in the \WET effective Lagrangian in \eqref{eq:eff_hamiltonian}. In the last four rows of \tref{tab:simp_models}, we list which Wilson coefficients receive contributions from each of the new degrees of freedom. Some comments are in order: None of these states generates a non-zero contribution to $C_{V_R}$ because this low-energy operator can only be generated by higher-order corrections or by non-standard mechanisms for the electroweak symmetry breaking~\cite{Cata:2015lta}. Hence, studying the $C_{V_R} \neq 0$ scenario allows us to test the very intrinsic structure of the Higgs sector of the \SM. This is why $C_{V_R}$ does not appear in any of the scenarios we consider in \tref{tab:simp_models}. Nevertheless, we study the impact of considering $C_{V_R} \neq 0$, as we will see in the following section. On top of this, we notice that the vector boson $W^\prime$, the scalar leptoquark $S_3$ and the vector leptoquark $U_3$ generate only the \SM-like operator $\mathcal{O}_{V_L}$. Therefore, with normalised observables as we are using in this analysis, we cannot probe these states. With this, we are left with the following four possibilities: $H^\prime$, $S_1$, $R_2$, and $U_1$. For convenience, these are highlighted in \tref{tab:simp_models}.

\begin{table}[t]
    \centering
    \setlength{\aboverulesep}{0pt}
\setlength{\belowrulesep}{0pt}
\begin{tabular}{c>{\columncolor{gray!25}} c c >{\columncolor{gray!25}}c c >{\columncolor{gray!25}}c >{\columncolor{gray!25}}c c}
\toprule
Mediator & $H^\prime$ &  $W^\prime$ & $S_1$ & $S_3$ & $R_2$ & $U_1$ & $U_3$ \\
Spin & 0 & 1 & 0 & 0 & 0 & 1 & 1 \\
Q.N. & $\left(1,2,-\frac{1}{2}\right)$ & $(1,3,0)$ & $\left(\bar{3},1,\frac{1}{3}\right)$ & $\left(\bar{3},3,\frac{1}{3}\right)$ & $\left(3,2,\frac{7}{6}\right)$ & $\left(3,1,\frac{2}{3}\right)$ & $\left(3,3,\frac{2}{3}\right)$ \\[2pt]
 \midrule
$\mathcal{O}_{V_L}$ & & $\checkmark$ & $\checkmark$ & $\checkmark$ & & $\checkmark$ & $\checkmark$  \\ 
$\mathcal{O}_{S_L}$ & $\checkmark$ & & $\checkmark$ &  & $\checkmark$ & & \\
$\mathcal{O}_{S_R}$ & $\checkmark$ &  & &   & & $\checkmark$ &   \\
$\mathcal{O}_{T}$ & $\checkmark^\dagger$ &  & $\checkmark$ &   & $\checkmark$ &  &  \\
\bottomrule
\end{tabular}
    \caption{\WET operator contributions for each of the possible $b \to c$ mediators. Quantum numbers (Q.N.) shown are for the \SM gauge group $\textrm{SU}(3)_C \otimes \textrm{SU}(2)_L \otimes \textrm{U}(1)_Y$. The mediators considered in this work are highlighted. $^\dagger$Note that the tensor contribution for the $H^\prime$ is generated at low energies by mixing with $\mathcal{O}_{S_L}$.}
    \label{tab:simp_models}
\end{table}

To connect these scenarios in the \SMEFT (\eqref{eq:Leff_SMEFT}) to the low-energy \WET (\eqref{eq:eff_hamiltonian}), we must account for running effects of the Wilson coefficients with the energy scale. As the description in \eqref{eq:Leff_SMEFT} is valid above the electroweak scale, $\mu_\mathrm{EW} \approx 160$ GeV, we employ the package \texttt{Wilson}~\cite{Aebischer:2018bkb} to run from a high scale $\mu_\mathrm{high} = 2$ TeV down to the pole mass of the $b$ quark, $\mu_b = m_b^\textrm{pole} = 4.8$ GeV. The reason for this specific choice is that the tensor current implemented by the HPQCD collaboration is renormalised at this scale~\cite{PhysRevD.109.094515}. At the electroweak scale $\mu_\mathrm{EW} = 160$ GeV, the \SMEFT description is matched onto the \WET within \texttt{Wilson}~\cite{Aebischer:2018bkb,Aebischer:2017ugx}, following the matching conditions given in Ref.~\cite{Jenkins:2017jig}. For the bottom and charm quark masses, we employ the $\overline{\textrm{MS}}$ value $\overline{m}_b(\mu = \overline{m}_b) = 4.183$ GeV~\cite{ParticleDataGroup:2024cfk} and compute $\overline{m}_c(\mu = \overline{m}_b) = 0.92$ GeV using \texttt{RunDec}~\cite{Chetyrkin:2000yt,Schmidt:2012az,Herren:2017osy}. 

At tree level, ignoring any modifications to the W couplings, matching the \SMEFT operators in \eqref{eq:Leff_SMEFT} onto the \WET operators in \eqref{eq:SM_BSM_currents} yields~\cite{Jenkins:2017jig,Hu:2018veh,Aebischer:2015fzz,Greljo:2023bab,Mohapatra:2024knf,Panda:2024oam}:
\begin{align}
    -\frac{4G_\mathrm{F}}{\sqrt{2}} V_{cb}C_{V_L} &= -\frac{2}{\Lambda^2}\sum_n\left[C_{\ell q}^{(3)} \right]_{ll2n}V_{nb} \,,\\
    -\frac{4G_\mathrm{F}}{\sqrt{2}} V_{cb}C_{S_L} &= -\frac{1}{\Lambda^2}\sum_n\left[C_{\ell equ}^{(1)} \right]_{lln2}^{*}V_{nb}\,, \\
    -\frac{4G_\mathrm{F}}{\sqrt{2}} V_{cb}C_{S_R} &= -\frac{1}{\Lambda^2}\left[C_{\ell edq} \right]_{ll32}^{*}\,, \\
    -\frac{4G_\mathrm{F}}{\sqrt{2}} V_{cb}C_T &= -\frac{1}{\Lambda^2}\sum_n\left[C_{\ell equ}^{(3)} \right]_{lln2}^{*}V_{nb}\,,
\end{align}
where the index $l = 1,2$ denotes the lepton generation. Neglecting the small contributions proportional to $V_{ub}$ ($n = 1$) and $V_{cb}$ ($n = 2$), the relevant relations for $b \to c\ell\bar{\nu}_\ell$ are obtained by setting $n = 3$ and taking $|V_{tb}| \approx 1$. As an example to illustrate the running from $\mu_\textrm{high} = 2$ TeV, setting $l = 1$ yields
\begin{align}
    C_{V_L}(\mu_b) &= \frac{v^2}{\Lambda^2 V_{cb}} 1.050 \, \left[C_{\ell q}^{(3)}\right]_{1123}(\mu_\mathrm{high})\label{aligneq:CVL_low}\,, \\
    C_{S_L}(\mu_b) &= \frac{v^2}{2\Lambda^2 V_{cb}} \left( 1.721 \left[C_{\ell equ}^{(1)}\right]_{1132}(\mu_\mathrm{high}) - 0.306 \left[C_{\ell equ}^{(3)}\right]_{1132}(\mu_\mathrm{high})\right)\label{aligneq:CSL_low} \,,\\
    C_{S_R}(\mu_b) &= \frac{v^2}{2\Lambda^2 V_{cb}} 1.682 \left[C_{\ell edq}\right]_{1132}(\mu_\mathrm{high})\label{aligneq:CSR_low} \,,\\
    C_{T}(\mu_b) &= \frac{v^2}{2\Lambda^2 V_{cb}}\left(0.890 \left[C_{\ell equ}^{(3)}\right]_{1132}(\mu_\mathrm{high}) -0.004 \left[C_{\ell equ}^{(1)}\right]_{1132}(\mu_\mathrm{high}) \right), \label{aligneq:CT_low}
\end{align}
where $v^2 = \frac{1}{\sqrt{2}G_\textrm{F}}$. Although the experimental data we consider involves both a lepton average as well as $\ell = e,\mu$ individually, there is no significant deviation in the \SMEFT-\WET running results between the $b \to ce\bar{\nu}_e$ ($l = 1$) and $b \to c\mu\bar{\nu}_\mu$ ($l = 2$) channels obtained from \texttt{Wilson}~\cite{Aebischer:2018bkb}. We further impose the following matching conditions at the high scale for the two scalar leptoquarks~\cite{deBlas:2017xtg}:
\begin{equation}\label{eq:matching_S1R2}
    C_{\ell e q u}^{(1)}(\mu_\mathrm{high}) = \left\{\begin{array}{cc}
        - 4 \, C_{\ell e q u}^{(3)}(\mu_\mathrm{high}) & \quad \textrm{for} \quad S_1 \\[5pt]
        + 4 \, C_{\ell e q u}^{(3)}(\mu_\mathrm{high}) & \quad \textrm{for} \quad R_2
    \end{array}\right.
\end{equation}
Note that there is no matching condition for $U_1$, as there is no scalar-tensor mixing for $C_{S_R}$. The \BSM fits to the angular coefficients provide values of the low-energy \WET Wilson coefficients, i.e. the left-hand side of \crefrange{aligneq:CVL_low}{aligneq:CT_low}. By imposing \eqref{eq:matching_S1R2} onto \crefrange{aligneq:CVL_low}{aligneq:CT_low}, the relations may be inverted to obtain the \NP scale $\Lambda$ for a given \BSM scenario, assuming that the relevant \SMEFT Wilson coefficient adopts a particular value in magnitude, which we will take as unity. 

\subsubsection{New Physics via the Angular Coefficients}\label{Subsubsection:NP_sensitivity}

\begin{table}[t]
    \centering
    \resizebox{\linewidth}{!}{
\begin{tabular}{cccccccccccc}
\toprule
\textbf{Observable} & $|C_P|^2$ & $|C_{V_L}|^2$ & $|C_{V_R}|^2$ & $|C_T|^2$ & $\mathfrak{Re}(C_P C_{V_L}^*)$ & $\mathfrak{Re}(C_P C_{V_R}^*)$ & $\mathfrak{Re}(C_{V_L} C_{V_R}^*)$ & $\mathfrak{Re}(C_P C_T^*)$ & $\mathfrak{Re}(C_{V_L} C_T^*)$ & $\mathfrak{Re}(C_{V_R} C_T^*)$ \\
\midrule
$J_{1s}$ & -- & $\checkmark$ & $\checkmark$ & $\checkmark$ & -- & -- & $\checkmark$ & -- & $m$ & $m$ \\
$J_{1c}$ & $\checkmark$ & $\checkmark$ & $\checkmark$ & $\checkmark$ & $m$ & $m$ & $\checkmark$ & -- & $m$ & $m$ \\
$J_{2s}$ & -- & $\checkmark$ & $\checkmark$ & $\checkmark$ & -- & -- & $\checkmark$ & -- & -- & -- \\
$J_{2c}$ & -- & $\checkmark$ & $\checkmark$ & $\checkmark$ & -- & -- & $\checkmark$ & -- & -- & -- \\
$J_{3}$ & -- & $\checkmark$ & $\checkmark$ & $\checkmark$ & -- & -- & $\checkmark$ & -- & -- & -- \\
$J_{4}$ & -- & $\checkmark$ & $\checkmark$ & $\checkmark$ & -- & -- & $\checkmark$ & -- & -- & -- \\
$J_{5}$ & -- & $\checkmark$ & $\checkmark$ & $m^2$ & $m$ & $m$ & -- & $\checkmark$ & $m$ & $m$ \\
$J_{6s}$ & -- & $\checkmark$ & $\checkmark$ & $m^2$ & -- & -- & -- & -- & $m$ & $m$ \\
$J_{6c}$ & -- & $m^2$ & $m^2$ & -- & $m$ & $m$ & $m^2$ & $\checkmark$ & $m$ & $m$ \\
\midrule
\textbf{Observable} & -- & -- & -- & -- & $\mathfrak{Im}(C_P C_{V_L}^*)$ & $\mathfrak{Im}(C_P C_{V_R}^*)$ & $\mathfrak{Im}(C_{V_L} C_{V_R}^*)$ & $\mathfrak{Im}(C_P C_T^*)$ & $\mathfrak{Im}(C_{V_L} C_T^*)$ & $\mathfrak{Im}(C_{V_R} C_T^*)$ \\
\midrule
$J_7$ &  &  &  &  & $m$ & $m$ & $m^2$ & $\checkmark$ & $m$ & $m$ \\
$J_8$ &  &  &  &  & --  & -- & $\checkmark$ & -- & -- & -- \\
$J_9$ &  &  &  &  & --  & -- & $\checkmark$ & -- & -- & -- \\
\bottomrule
\end{tabular}}
    \caption{Outline of the angular coefficients' dependence on the \WET Wilson coefficients, where $C_P = C_{S_R} - C_{S_L}$. An entry of $\checkmark$ (--) denotes the presence (absence) of this combination. An entry of $m^n$ denotes the presence of this term, but with kinematic lepton-mass suppression $\propto (m_\ell/\sqrt{q^2})^n$.}
    \label{tab:J_BSM}
\end{table} 

In $\BtoDstar$ decays, the matrix element of the scalar current $\langle D^* | \bar{c}b | \bar{B} \rangle = 0$, leaving a non-zero scalar contribution only from the pseudoscalar current $\bar{c}\gamma_5b$. Thus, the angular coefficients that parameterise the differential decay rate can be written in terms of the pseudoscalar combination:
\begin{equation}\label{eq:CP}
    C_P = C_{S_R} - C_{S_L},
\end{equation}
along with the vector $C_{V_L}, C_{V_R}$ and tensor $C_T$ \WET Wilson coefficients. Outlined in \tref{tab:J_BSM} is the dependence of each angular coefficient on these combinations of Wilson coefficients multiplying the current operators of the various Lorentz structures in \eqref{eq:SM_BSM_currents}. Evident from \tref{tab:J_BSM} is the varying sensitivity to \NP among these Wilson coefficients. For example, many scalar contributions are helicity suppressed by the lepton mass $m_\ell$, making the $\tau$ channel a much better probe for scalar \NP~\cite{Fajfer:2012vx,Celis:2012dk,Sakaki:2013bfa,Bhattacharya:2024zog}. For this reason, we also provide predictions of the angular observables for the $\tau$ channel in \tref{tab:coeffs_mtau} in \cref{Appendix:tau}, for future comparison with prospective experimental analysis. Not all scalar contributions are helicity suppressed; for example, in $J_{1c}$ through $|C_P|^2$, and in the scalar-tensor interference terms $\propto\mathfrak{Re}(C_PC_T^*)$ in $J_5$ and $J_{6c}$. While $|C_{V_L}| = 1$ in the \SM and any non-zero \NP contributions are expected to be small, the most sensitive terms are therefore those proportional to the \SM-\BSM interference terms, such as $\mathfrak{Re}(C_{V_L}C_{V_R}^*)$, $\mathfrak{Re}(C_P C_{V_{L(R)}}^*)$, and $\mathfrak{Re}(C_T C_{V_{L(R)}}^*)$. With the latter two being helicity suppressed, the most sensitive \BSM Wilson coefficient in the fourfold differential decay rate of $\BtoDstar$ is thus $C_{V_R}$. Conversely, the pseudoscalar is expected to be the least constrained by the experimental data, contributing to only four of the twelve angular coefficients with helicity suppression. Note, however, that because the scalar Wilson coefficients enter into $J_{1c}$, they can contribute to the normalisation $\mathcal{N}$ in \eqref{eq:JwithN}, and hence all of the partially integrated, normalised coefficients $\hat{J}_i(w_n) / \mathcal{N}$. Another consequence of working with normalised quantities is the following. Setting $C_{S_L} = C_{S_R} = C_{V_R} = C_T = 0$ recovers the \SM angular coefficients in \crefrange{aligneq:J1s}{aligneq:J789} all multiplied by $|C_{V_L}|^2$. Due to this proportionality, the normalisation $\mathcal{N} \propto |C_{V_L}|^2$, meaning that the partially-integrated, normalised angular coefficients $\hat{J}_i(w_n) / \mathcal{N}$ lose the factor $|C_{V_L}|^2$. Consequently, the \SM signal in the set of normalised observables considered in this work is no longer the dominant contribution to the Wilson coefficient $C_{V_L}$. We therefore determine Wilson coefficients modulo $|C_{V_L}|$ and define $\widetilde{C}_i = C_i / |C_{V_L}|$. While a precise determination of the Wilson coefficients from experimental and \LQCD data is not feasible here, what is of interest is determining the parametrically allowed space for classes of \NP models. 

\subsection{Results}

In the \NP-constraint fits, we consider only the combination of \LQCD datasets (\texttt{Combined}), and individually analyse each non-lepton-averaged $B^0$ and $B^+$ experimental datasets. For the vector and tensor form factors, the truncation order $K = 3$ was selected for the parameterisations, based on the previously discussed results. To aid convergence of the sampling, the Type-B Bayesian fit results in \tref{tab:TypeB_fit_3_3_3_3} are employed as the Bayesian priors for the \BGL coefficients. These priors are uncorrelated uniform distributions spanning the 99.7\% confidence interval for each parameter. For the real/imaginary parts of the Wilson coefficients, the uniform prior in $[-1, 1]$ was used. The same log-likelihood kernel as in the \SM case is employed, with the omission of the normalisation term that determines $|V_{cb}|$. The reason for this is as follows. The total rate on the lattice is calculated modulo $|V_{cb}|^2$ through \eqref{eq:Gamma}. When fitting $|V_{cb}|^2\widetilde{\Gamma}^{\textrm{lat}}$, there are therefore terms of the form $|V_{cb}|^2C_iC_j$, with no way of disentangling the Wilson coefficients $C_{i,j}$ from the \CKM matrix element. Thus, the \BSM analysis presented in this work does not attempt to determine $|V_{cb}|$. When a value of $|V_{cb}|$ is required for the \NP scale in \crefrange{aligneq:CVL_low}{aligneq:CT_low}, the result from the relevant Type-B \SM fit in \tref{tab:TypeB_Vcb_ml} is employed.

\begin{table}[t]
    \centering
    \resizebox{\linewidth}{!}{
\begin{tabular}{cS[table-format = -1.3(2)]S[table-format = -1.3(2)]S[table-format =  -1.3(2)]S[table-format = -1.3(2)]S[table-format = -1.3(2)]S[table-format = -1.3(2)]}
    \toprule
    \textbf{Dataset} & \multicolumn{1}{c}{$\mathfrak{Re} \ \widetilde{C}_{P}$} & \multicolumn{1}{c}{$\mathfrak{Im} \ \widetilde{C}_{P}$} & \multicolumn{1}{c}{$\mathfrak{Re} \ \widetilde{C}_{V_R}$} & \multicolumn{1}{c}{$\mathfrak{Im} \ \widetilde{C}_{V_R}$} & \multicolumn{1}{c}{$\mathfrak{Re} \ \widetilde{C}_T$} & \multicolumn{1}{c}{$\mathfrak{Im} \ \widetilde{C}_T$} \\
    \midrule
    $\langle B^0, B^+, e \rangle$ & 0.00(35) & -0.00(35) & -0.005(44) & -0.052(91) & -0.000(39) & 0.001(38) \\
    $\langle B^0, B^+, \mu \rangle$ & -0.11(34) & -0.02(34) & 0.025(42) & -0.080(77) & 0.012(28) & 0.001(28) \\
    $\langle B^0, e \rangle$ & 0.00(57) & 0.00(57) & 0.033(66) & -0.10(11) & 0.000(35) & 0.000(35) \\
    $\langle B^0, \mu \rangle$ & -0.03(36) & 0.02(36) & 0.068(54) & -0.03(10) & 0.008(31) & 0.001(32) \\
    $\langle B^+, e \rangle$ & -0.00(44) & -0.01(44) & -0.019(59) & -0.01(13) & 0.000(54) & -0.001(54) \\
    $\langle B^+, \mu \rangle$ & -0.15(51) & -0.11(53) & -0.001(59) & -0.11(10) & 0.018(35) & 0.000(35) \\
    \bottomrule
\end{tabular}}
    \caption{Type-B \BSM results for the \WET Wilson coefficients from fits at the truncation order $K = 3$ to \texttt{HPQCD 23}~\cite{PhysRevD.109.094515}, \texttt{JLQCD 23}~\cite{PhysRevD.109.074503}, \texttt{FNAL/MILC 21}~\cite{Bazavov_2022}, and \texttt{HPQCD 23 tensor}~\cite{PhysRevD.109.094515} \LQCD datasets and the \texttt{Belle 23 angular}~\cite{Belle:2023xgj,hepdata.153299} experimental datasets.}
    \label{tab:BSM_WC}
\end{table}

\begin{figure}[t]
    \centering
    \includegraphics[width=\linewidth]{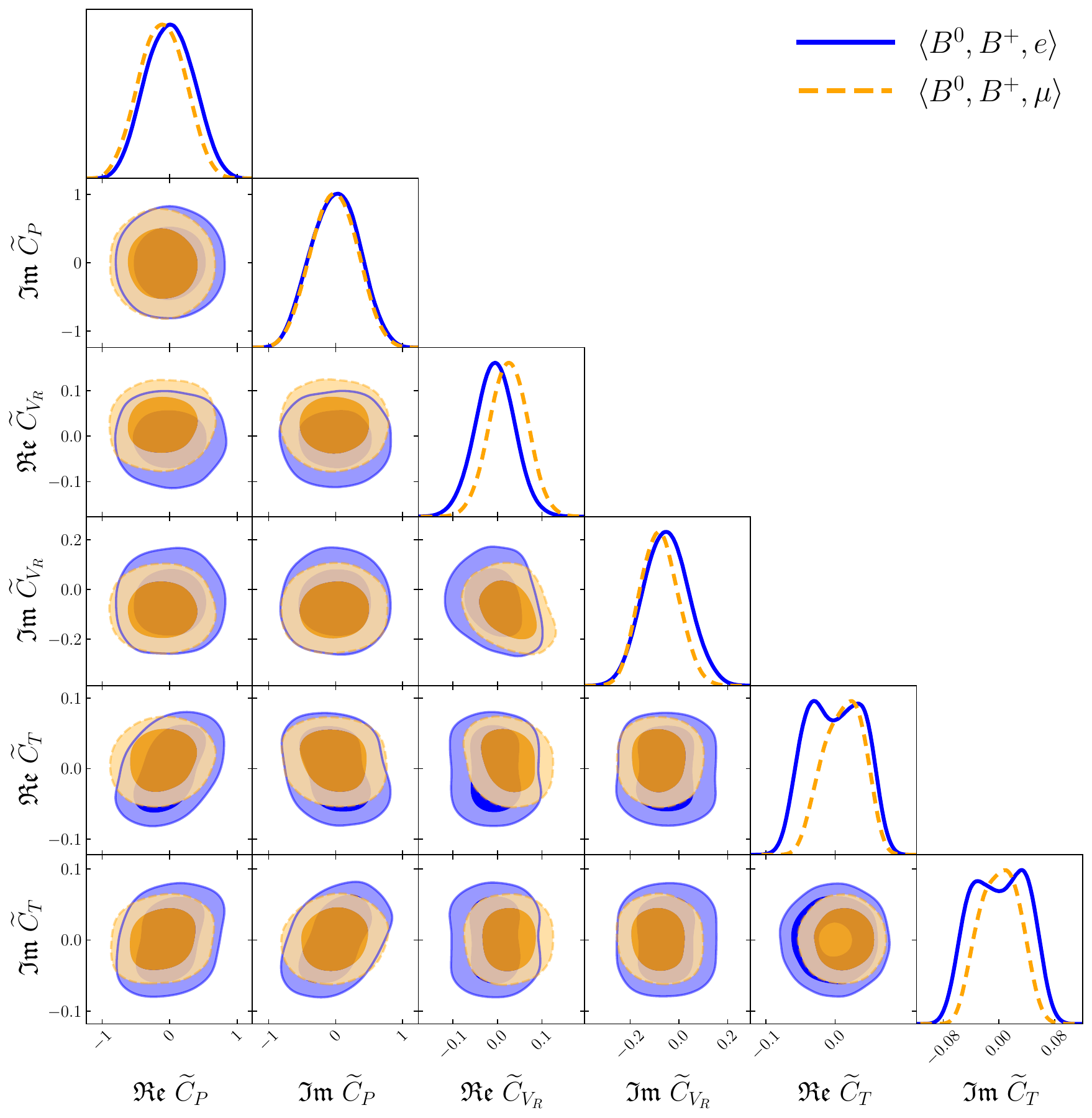}
    \caption{Triangle correlation plot of the posterior samples from Type-B \BSM fits to vector and axial form-factor data (the combination of \texttt{HPQCD 23}~\cite{PhysRevD.109.094515}, \texttt{JLQCD 23}~\cite{PhysRevD.109.074503}, and \texttt{FNAL/MILC 21}~\cite{Bazavov_2022} datasets), tensor form-factor data (\texttt{HPQCD 23 tensor}~\cite{PhysRevD.109.094515}), and the \texttt{Belle 23 angular}~\cite{Belle:2023xgj,hepdata.153299} experimental dataset. Dark (light) shading denotes the 68\% (95\%) confidence region. In solid blue we represent the results associated with the $\langle B^0,B^+,e\rangle$ data and dashed in orange with the $\langle B^0,B^+,\mu\rangle$ ones.}
    \label{fig:BSM_TypeB_triangle_plot}
\end{figure}

To assess the sensitivity of the angular-coefficient datasets to each Wilson coefficient, a global fit with all Wilson coefficients active was performed first. These results are shown in \tref{tab:BSM_WC}, with the corresponding `triangle' correlation plot for the Wilson coefficient subset of the parameter space shown in \fref{fig:BSM_TypeB_triangle_plot}. One-dimensional marginalised probability distributions (kernel density estimators) are shown along the diagonal, and two-dimensional contour plots are shown on the off-diagonals. Enabling these extra degrees of freedom through the Wilson coefficients has little effect on the posterior distributions of the \BGL coefficients parameterising the form factors. As expected, the posterior of $\widetilde{C}_P$ is the broadest, though it did not vary when extending the prior beyond $[-1, 1]$. In addition, it is clear that this fit is only able to discriminate real and imaginary parts of $\widetilde{C}_{V_R}$ irrespective of lepton mass; for $\widetilde{C}_T$, this is true only for $\ell = \mu$. Furthermore, the larger mass of the muon better resolves the posterior of $\widetilde{C}_T$ relative to the electron, refining the bimodal structure into a single maximum. In general, all distributions for the real and imaginary parts of the Wilson coefficients are zero within one standard deviation, except for $\mathfrak{Re} \ \widetilde{C}_{V_R}$ for the $\langle B^0, \mu \rangle$ dataset, and $\mathfrak{Im} \ \widetilde{C}_{V_R}$ for the datasets $\langle B^0,B^+,\mu\rangle$ and $\langle B^+,\mu\rangle$. Although the significance of this finding is low, it highlights the importance of continuing to study these modes to better understand the possible nature of \BSM effects.

Our results can be directly compared with the findings of Ref.~\cite{Fedele:2023ewe}, where fits to the fully averaged experimental datasets are performed, assuming real \NP Wilson coefficients. We therefore compare our results in \tref{tab:BSM_WC} for the real parts of the Wilson coefficients only in the $\langle B^0, B^+ ,e \rangle$ and $\langle B^0, B^+,\mu \rangle$ cases. Our findings are rather similar to those of Ref.~\cite{Fedele:2023ewe}, in that the $1\sigma$ regions for the real parts of the Wilson coefficients are centred around zero. Nevertheless, we find that our $1\sigma$ intervals are consistently larger than those in Ref.~\cite{Fedele:2023ewe}. This is likely due to the fact that our fit contains more parameters, since we allow the Wilson coefficients to be complex, and to our use of datasets that are not averaged over lepton species. A direct comparison with Ref.~\cite{Colangelo:2024mxe} is instead difficult, since the authors do not consider experimental correlations in their analysis.

\subsubsection{Model Comparison}

\begin{table}[t]
    \centering
    \resizebox{\linewidth}{!}{
\begin{tabular}{cS[table-format = -1.2(2)]S[table-format = -1.2(2)]S[table-format = -1.3(2)]S[table-format = -1.3(2)]S[table-format = -1.3(2)]S[table-format = -1.3(2)]}
    \toprule
    \multirow{2}{*}{\textbf{Dataset}} & \multicolumn{2}{c}{\textbf{Scalar Only}} & \multicolumn{2}{c}{\textbf{Tensor Only}} & \multicolumn{2}{c}{\boldmath$C_{V_R} \neq 0$ \textbf{Only}} \\
    & \multicolumn{1}{c}{$\mathfrak{Re} \ \widetilde{C}_{P}$} & \multicolumn{1}{c}{$\mathfrak{Im} \ \widetilde{C}_{P}$} & \multicolumn{1}{c}{$\mathfrak{Re} \ \widetilde{C}_T$} & \multicolumn{1}{c}{$\mathfrak{Im} \ \widetilde{C}_T$} & \multicolumn{1}{c}{$\mathfrak{Im} \ \widetilde{C}_{V_R}$} & \multicolumn{1}{c}{$\mathfrak{Im} \ \widetilde{C}_{V_R}$} \\
    \midrule
    $\langle B^0, B^+, e \rangle$ & -0.00(36) & -0.00(36) & -0.000(39) & 0.000(39) & -0.013(43) & -0.054(94) \\
    $\langle B^0, B^+, \mu \rangle$ & -0.14(34) & 0.01(36) & 0.016(29) & 0.001(31) & 0.026(41) & -0.087(76) \\
    $\langle B^0, e \rangle$ & 0.00(58) & -0.01(58) & 0.001(39) & 0.000(38) & 0.016(62) & -0.11(11) \\
    $\langle B^0, \mu \rangle$ & -0.08(34) & -0.02(36) & 0.013(32) & -0.000(34) & 0.065(51) & -0.04(10) \\
    $\langle B^+, e \rangle$ & -0.01(34) & 0.00(34) & -0.000(49) & 0.000(48) & -0.025(56) & -0.00(13) \\
    $\langle B^+, \mu \rangle$ & -0.09(53) & 0.03(54) & 0.010(36) & 0.001(38) & -0.006(55) & -0.12(10) \\
    \bottomrule
\end{tabular}}
    \caption{Type-B \BSM results for the pseudoscalar,  tensor, and right-handed vector \WET Wilson coefficients from fits at the truncation order $K = 3$ to \texttt{HPQCD 23}~\cite{PhysRevD.109.094515}, \texttt{JLQCD 23}~\cite{PhysRevD.109.074503}, \texttt{FNAL/MILC 21}~\cite{Bazavov_2022} (and \texttt{HPQCD 23 tensor}~\cite{PhysRevD.109.094515} for $C_T$) \LQCD datasets and the \texttt{Belle 23 angular}~\cite{Belle:2023xgj,hepdata.153299} experimental datasets.}
    \label{tab:BSM_WC_scenarios}
\end{table}

To probe the \NP scenarios introduced in the previous section, the relevant Wilson coefficients are activated individually and determined modulo $|C_{V_L}|$, with the rest fixed to zero. Results for the cases of individual Wilson coefficients are reported in \tref{tab:BSM_WC_scenarios}, which remain predominantly unchanged with respect to the global fits in \tref{tab:BSM_WC}. In \tref{tab:Jeffreys_scale}, the natural logarithm of the \INS evidence, $\ln\mathcal{Z}_{\textrm{INS}}$, is calculated by \texttt{PyMultiNest}~\cite{Buchner_2014,Feroz_2008,Feroz_2009,Feroz_2019} during parameter inference and is reported for each of the Type B fits. The ratio of this evidence between two different models, the Bayes factor in \eqref{eq:Bayes_factor_12}, can be interpreted with Jeffreys' scale~\cite{JeffreysScale} as follows. Recalling that strong (decisive) evidence of model 1 over 2 is signalled if $\mathcal{K} > 2.3 (4.6)$, some immediate conclusions may be drawn. With the experimental and \LQCD datasets on the $\BtoDstar$ angular coefficients considered in this work, from a model comparison perspective, there is
\begin{itemize}
    \item Support in favour of \NP contributions from either $C_{V_R}$, $C_P$, or $C_T$.
    \item Support of scalar \NP contributions over tensor. 
\end{itemize}
It is important to reiterate that the fits performed here inherently suffer from the restrictions discussed in \cref{Subsubsection:NP_sensitivity}; consequently, the sensitivity levels differ across the various Wilson coefficients. Moreover, the primary goal of these fits is to establish the regions of parameter space compatible with experimental measurements, rather than precise central values. According to the results in \tref{tab:Jeffreys_scale}, allowing extra degrees of freedom through $C_{S_{L,R}}$ or $C_{T}$ \BSM contributions significantly improves the fit relative to the \SM case. The improvement is less for the tensor current; setting $C_T \neq 0$ requires three additional form factors, which adds eight more parameters to the fit for $K = 3$. Since the Bayesian inference framework naturally integrates Occam's razor, between competing models that describe data, the simpler of the two that makes fewer assumptions is generally preferred. That is, the significantly smaller Bayesian evidence of the tensor current model relative to that of the scalar could very well be due to over-fitting to the data.
\begin{table}[t]
    \centering
    \resizebox{\linewidth}{!}{
\begin{tabular}{cS[table-format = -2.4(2)]S[table-format = -2.4(2)]S[table-format = -2.4(2)]S[table-format = -2.4(2)]S[table-format = -2.4(2)]}
    \toprule
    \multirow{2}{*}{\textbf{Dataset}} & \multicolumn{5}{c}{\textbf{Importance Nested Sampling (INS) Evidence} \boldmath$\ln\mathcal{Z}_{\textrm{\textbf{INS}}}$} \\
     & \multicolumn{1}{c}{\SM with $|V_{cb}|$} & \multicolumn{1}{c}{\BSM Global} & \multicolumn{1}{c}{Only $C_P \neq 0$} & \multicolumn{1}{c}{Only $C_T \neq 0$} & \multicolumn{1}{c}{Only $C_{V_R} \neq 0$} \\
    \midrule
    $\langle B^0, B^+, e \rangle$ & -76.4069(49) & -70.7380(44) & -43.67(19) & -64.3837(46) & -47.320(12) \\
    $\langle B^0, B^+, \mu \rangle$ & -77.6217(53) & -72.1829(36) & -44.201(46) & -65.8576(43) & -47.547(35) \\
    $\langle B^0, e \rangle$ & -77.2337(54) & -67.3428(40) & -40.299(60) & -63.0327(62) & -44.686(41) \\
    $\langle B^0, \mu \rangle$ & -77.9080(58) & -70.7914(33) & -43.697(32) & -65.2044(47) & -46.229(65) \\
    $\langle B^+, e \rangle$ & -76.5462(61) & -70.7837(43) & -45.918(33) & -65.8150(44) & -48.497(40) \\
    $\langle B^+, \mu \rangle$ & -77.3339(63) & -67.9000(40) & -40.943(28) & -63.5570(41) & -44.14(49) \\
    \bottomrule
\end{tabular}}
    \caption{\INS evidence $\ln\mathcal{Z}_{\textrm{INS}}$ outputted by \texttt{PyMultiNest}~\cite{Buchner_2014,Skilling:2004pqw,Feroz_2008,Feroz_2009,Feroz_2019} for each of the Type-B \BSM fits considered in this work (see Tables \ref{tab:BSM_WC} and \ref{tab:BSM_WC_scenarios} for the corresponding fit results), as well as the Type B \SM fits that determined $|V_{cb}|$.}
    \label{tab:Jeffreys_scale}
\end{table}

\subsubsection{Bounds on Effective Scales of New Physics}

\begin{figure}[t]
    \centering
    \includegraphics[width=\linewidth]{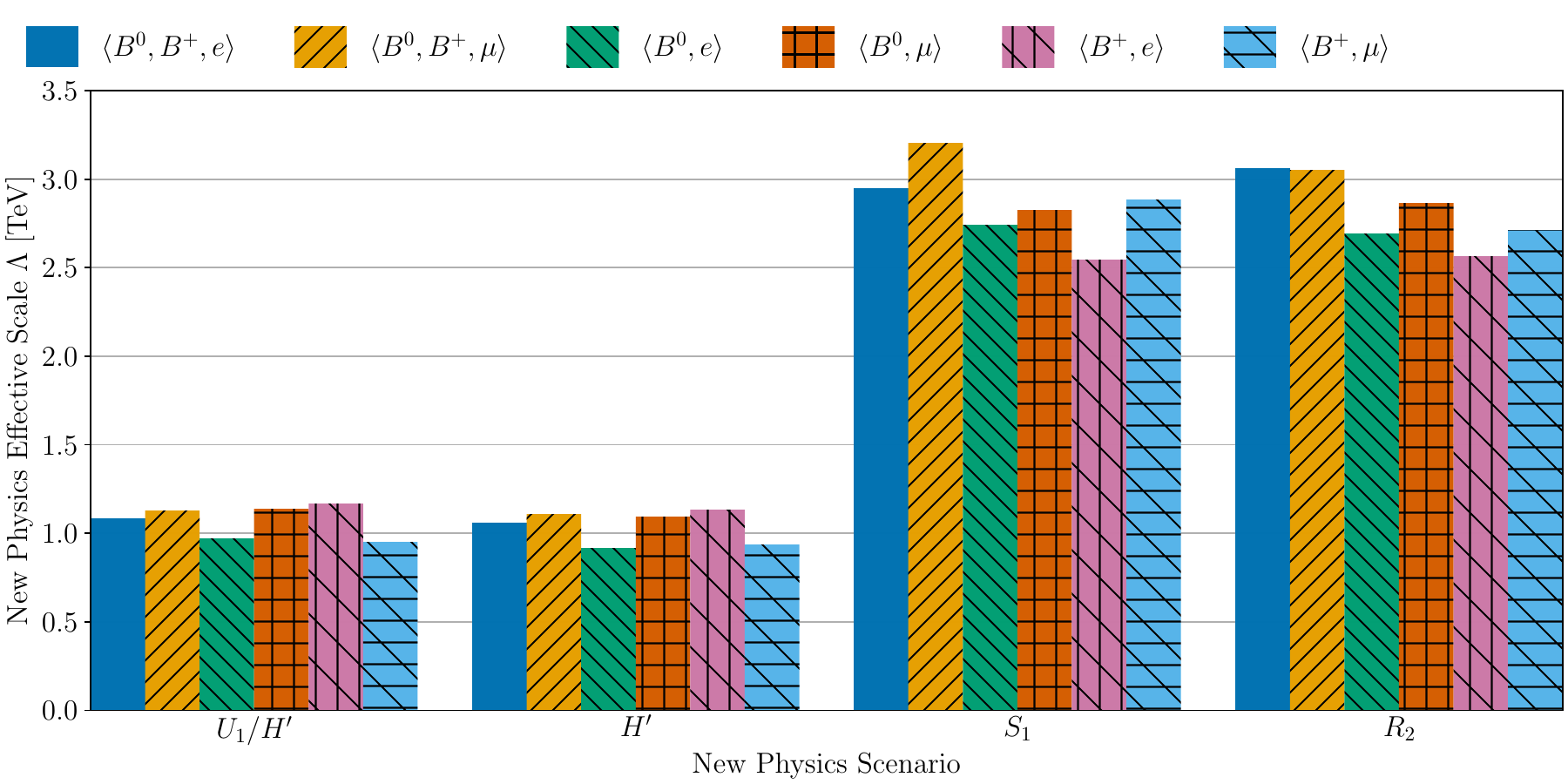}
    \caption{Plot of predicted effective energy scales, where the \SMEFT operator assumes unit magnitude, for various \BSM scenarios. The height of each bar represents the effective scale where \NP is excluded at the 68\% confidence level.}
    \label{fig:NP_exclusion}
\end{figure}

Recall that the results of the \BSM fits in \tref{tab:BSM_WC_scenarios} represent a range of values of the \WET Wilson coefficients at the scale $\mu_b = m_b^{\textrm{pole}}$ compatible with the data. These are related to the \SMEFT Wilson coefficients, after performing \RGE and matching at the electroweak scale, through the relations in \crefrange{aligneq:CVL_low}{aligneq:CT_low}. By applying the matching conditions in \eqref{eq:matching_S1R2} to isolate one \SMEFT Wilson coefficient, \crefrange{aligneq:CVL_low}{aligneq:CT_low} may be inverted for the \NP scale $\Lambda$. Since in all cases shown in \tref{tab:BSM_WC_scenarios} the \NP Wilson coefficients are compatible with zero within one standard deviation, we interpret our results as lower bounds on the possible value of $\Lambda$. Taking, for example, the Wilson coefficient associated with the left-handed scalar current, \eqref{aligneq:CSL_low} reads
\begin{equation}\label{eq:Lambda_scalar_current}
    \Lambda \geq \sqrt{\frac{v^2}{2|C_{S_L}(\mu_b)| |V_{cb}|}|(\mp 4)1.721 - 0.306|}
\end{equation}
for $S_1$ and $R_2$, respectively. \eqref{eq:Lambda_scalar_current} is applied to each posterior sample of $C_{S_L}(\mu_b) = \mathfrak{Re} \ C_{S_L}(\mu_b) + i\mathfrak{Im} \ C_{S_L}(\mu_b)$, which can then yield a central value estimate and error on the \NP scale $\Lambda$ via statistical bootstrapping. The bound in \eqref{eq:Lambda_scalar_current} is the effective energy scale below which we can rule out \NP. The results for these scale predictions for the $S_1, R_2$, and $U_1$ \BSM scenarios are shown in \fref{fig:NP_exclusion} for the \texttt{Combined} \LQCD dataset. These results assume the \SMEFT Wilson coefficients take unit magnitude; for a different value, the \NP scale is proportional to the square root of the magnitude of \SMEFT coefficient. 

The bars in \fref{fig:NP_exclusion} represent the scales at which \NP mediated by the relevant Wilson coefficient is excluded at the $68\%$ confidence level. The height of each bar is given by the central value of $\Lambda$ minus one standard deviation, yielding a lower bound. In the case of the $U_1$ vector leptoquark, only $C_{S_R}$ is non-zero at low energies. Therefore, we can easily derive bounds from each experimental dataset. These bounds are shown in the first set of limits in \fref{fig:NP_exclusion}, where all datasets consistently constrain the \NP scale to be above roughly 1 TeV. Concerning the $S_1$ and $R_2$ states, even though they generate contributions to both $C_{S_L}$ and $C_T$ at low energies, the matching conditions at high energies in \eqref{eq:matching_S1R2} impose a proportionality between the two of them. Using these relations, we find the lower bounds of 2.5--3.2 TeV and 2.6--3.1 TeV for the $S_1$ and $R_2$, respectively. Finally, in the case of the $H^\prime$, there are three non-zero operators at low energies: $C_{S_R}$, $C_{S_L}$, and $C_T$, while, at high energies, only the operators $C_{\ell equ}^{(1)}$ and $C_{\ell edq}$ appear. We also notice from \eqref{aligneq:CT_low} that $C_T\approx 0$ for this scenario. Imposing a precise proportionality between $C_{S_R}$ and $C_{S_L}$ requires specifying the UV properties of a model with an $H^\prime$ state. We refrain from doing so, and we recall that the $\BtoDstar$ decay mode only allows one to extract the difference  $C_{S_R}-C_{S_L}$. This implies that, without adding other decay modes, it is not feasible to perform a two-dimensional fit that can disentangle $C_{S_R}$ and $C_{S_L}$. Therefore, we provide bounds when only one coefficient is non-zero at a time. Given that numerically the proportionality factors of $C_{S_R}$ and $C_{S_L}$ to $C_{\ell edq}$ and $C_{\ell equ}^{(1)}$, respectively, are very close, we find no significant difference in the lower bounds extracted when considering one or the other. Therefore, we obtain the same picture as in the $U_1$ case, and the corresponding bounds are shown together in \fref{fig:NP_exclusion}. As a final general remark, we stress that these results can be easily reinterpreted in specific \BSM scenarios, for example, by modifying Equations~(\ref{aligneq:CVL_low})-(\ref{aligneq:CT_low}) with higher-order effects and appropriately rescaling the results according to the flavour structure for the Wilson coefficients predicted in the specific benchmark. Furthermore, the range of lower bounds on the effective \NP scale that we find is in line with those obtained from other flavour observables, as well as direct searches~\cite{Allwicher:2023shc,deBlas:2025xhe}. 

\section{Summary and Conclusions}

In this work, we have extended our previous analysis of Ref.~\cite{Bordone:2024weh} to the full angular distribution of exclusive semileptonic $\BtoDstar$ decays, utilising recent experimental data from the Belle collaboration~\cite{Belle:2023xgj,hepdata.153299} and \LQCD form-factor data~\cite{PhysRevD.109.094515,PhysRevD.109.074503,Bazavov_2022}. Lepton masses are explicitly taken into account and compared with the electron-muon average that is seen in the literature. 

A central element of this study is the determination of the CKM matrix element~$|V_{cb}|$. Compared to its determination based on the differential decay rates using the same analysis strategy in Ref.~\cite{Bordone:2024weh} and the same lattice input, the angular data set turned out to be less problematic, as far as inconsistencies during the data-analysis chain are concerned. While the analysis in Ref.~\cite{Bordone:2024weh} discarded three quarters of the available experimental bins to obtain a consistent and artefact-free analysis, excellent fits including all experimental bins were possible here. We nevertheless raise a concern for consideration by the experimental and theory communities concerning data representation. For instance, results for angular coefficients in the experimental dataset~\cite{Belle:2023xgj,hepdata.153299} that in the \SM are manifestly positive (e.g. $J_{2s}$) have a negative central value, or central values with  Gaussian errors that allow fluctuations to negative values with high probability. A more fine-grained presentation of central values (mean, median) and the error beyond the Gaussian approximation, e.g. in terms of histograms or resampling bins, would substantially improve the quality of analysis chains, and allow others to probe the data more accurately and in more detail. 

With our central result $|V_{cb}|=0.03997(71)$ we confirm our previous determination $|V_{cb}|=0.04042(71)$~\cite{Bordone:2024weh}, which was based on the same lattice input but experimental results for the differential decay rate~\cite{Belle:2023bwv,Belle-II:2023okj,hepdata.137767,hepdata.145129}, and also other previous $|V_{cb}|$ determinations~\cite{Martinelli:2021onb,Martinelli:2021myh,Martinelli:2022xir,Martinelli:2023fwm,Fedele:2023ewe,Fang:2026hru}. A slight tension between results from the two different $B$-meson isospin channels is observed and warrants monitoring in the analysis of future datasets. All in all, our result confirms the $|V_{cb}|$ puzzle. Our full error budget suggests that further reductions in the error on $|V_{cb}|$ will primarily be obtained from more precise \LQCD calculations, and to a similar extent from more precise determinations of the $\BtoDstar$ total branching fraction. Hence, our analysis shows that improved determinations of the normalised angular coefficients will only further reduce the error on $|V_{cb}|$ once these two contributions to the error budget are better controlled. 

The second central element in this paper is the first determination of \BSM constraints for a small set of \NP scenarios derived from the angular coefficients. We present the results of a global fit to the \WET Wilson coefficients, accessible with normalised angular observables, including their imaginary parts. We find no statistical indication for any of the Wilson coefficients being non-zero, apart from $\mathfrak{Im} \ C_{V_R}$, which deviates from zero by one standard deviation. We then reinterpret these results in terms of \NP particles, focusing on the $S_1$, $R_2$, and $U_1$ leptoquarks, and on a scalar $H^\prime$ with the same quantum numbers as the \SM Higgs boson. This required incorporating the \RGE of the \WET Wilson coefficients in the \SMEFT to obtain bounds on the effective scales of each scenario for heavy \NP mediated by a scalar and tensor current based on a Bayesian analysis of \LQCD and experimental data. We find that for the $U_1$ leptoquark and the $H^\prime$, the lower bound on the \NP scale is $\sim 1$ TeV, while for the $S_1$ and $R_2$ they reach $2.5-3.2$ TeV. We note that working with normalised experimental measurements of the angular coefficients as provided by Belle in Refs.~\cite{Belle:2023xgj,hepdata.153299} restricts the analysis to Wilson coefficients modulo $C_{V_L}$. By employing Jeffreys' scale to interpret the Bayesian evidence computed during parameter inference, the quality of simultaneous fits to \LQCD and experimental data sets is compared. This analysis indicates a preference for \NP of scalar nature over tensor. While the scalar current is not constrained precisely, allowing contributions from $C_{P}$ significantly improves the fits.  

We are looking forward to new experimental data to become available not only for $\BtoDstar$, but also for other channels such as $B\to D\ell\bar\nu_\ell$. It would be natural to incorporate additional channels into our analysis, in this way improving the constraining power of potential scalar \NP contributions.

\section{Acknowledgements}
We thank Claudia Cornella, Christopher Sachrajda and Peter Stangl for useful suggestions. The work of M.B. is supported by the Cluster of Excellence \textit{PRISMA}$^{++}$ (EXC 2118/2) funded by the German Research Foundation (DFG) under Germany’s Excellence Strategy (Project ID 390831469). A.J. is supported by the Eric \& Wendy Schmidt Fund for Strategic Innovation (grant agreement SIF-2023-004).

\appendix

\section{Definitions of the Angular Coefficients}\label{App:ang_coeff_defs}

The helicity amplitudes for $\BtoDstar$~\cite{Hagiwara:1989gza} may be written in terms of the \BGL form factors $f, \mathcal{F}_1, \mathcal{F}_2$, and $g$ as:
\begin{align}
    H_\pm(w) &= f(w) \mp m_B^2 r\sqrt{w^2 - 1}g(w), \label{aligneq:Hpm} \\
    H_0(w) &= \frac{\mathcal{F}_1(w)}{m_B\sqrt{r^2 - 2rw + 1}}, \label{aligneq:H0} \\
    H_t(w) &= \frac{m_B r\sqrt{w^2 - 1}}{\sqrt{r^2 - 2rw + 1}}\mathcal{F}_2(w) \label{aligneq:Ht},
\end{align}
where $r = m_{D^*} / m_B$. They correspond to definite polarisations of the virtual W boson and are obtained by projecting the hadronic matrix element of the weak $V - A$ current onto the different polarisation states of the W boson. For massless leptons, the three helicity amplitudes $H_{\pm}, H_0$ contribute, with the scalar/time-like amplitude $H_t$ contributing only for massive leptons. The form factors are related at the kinematic endpoints $w = 1$ and $w = w_\textrm{max}^{(\ell)}$:
\begin{align}\label{aligneq:fF1F2F1_constraints}
    \mathcal{F}_1(1) = m_B(1 - r)f(1), && \mathcal{F}_2(w_\textrm{max}^{(\ell)}) = \frac{1 + r}{m_B^2 r(1 - r)(1 + w_\textrm{max})}\mathcal{F}_1(w_\textrm{max}^{(\ell)}),
\end{align}
where, referring back to \eqref{eq:w(q^2)}, $w_{\textrm{max}}^{(\ell)}$ occurs at the minimum of $q^2$:
\begin{equation}\label{eq:w_max^ell}
    w_\textrm{max}^{(\ell)} = \frac{m_B^2 + m_{D^*}^2 - m_\ell^2}{2m_Bm_{D^*}} = \frac{1 + r^2 - m_\ell^2 / m_B^2}{2r}.
\end{equation}
In the \SM, the angular coefficients $J_i(w)$ that parameterise the $w$-dependence of the fourfold differential decay rate in \eqref{eq:ang_distribution} are given in terms of these helicity amplitudes~\cite{Duraisamy:2014sna,Ivanov:2016qtw,Becirevic:2019tpx}:
\begin{align}
    J_{1s}(w) &= \frac{3}{2}F(w)\big[H_+^2(w) + H_-^2(w)\big]\left(1 + \frac{m_\ell^2}{3q^2}\right) \label{aligneq:J1s} \\
    J_{1c}(w) &= 2F(w)\left[ H_0^2(w)\left(1 + \frac{m_\ell^2}{q^2}\right) + 2\frac{m_\ell^2}{q^2}H_t^2(w) \right] \label{aligneq:J1c} \\
    J_{2s}(w) &= \frac{1}{2}F(w)\big[ H_+^2(w) + H_-^2(w) \big]\left(1 - \frac{m_\ell^2}{q^2}\right) \label{aligneq:J2s} \\
    J_{2c}(w) &= -2F(w)H_0^2(w)\left(1 - \frac{m_\ell^2}{q^2}\right) \label{aligneq:J2c} \\
    J_{3}(w) &= -2F(w)H_+(w)H_-(w)\left(1 - \frac{m_\ell^2}{q^2}\right) \label{aligneq:J3} \\
    J_{4}(w) &= F(w)H_0(w)\big[ H_+(w) + H_-(w) \big]\left(1 - \frac{m_\ell^2}{q^2}\right) \label{aligneq:J4} \\
    J_{5}(w) &= 2F(w)\left( H_0(w)\big[H_-(w) - H_+(w) \big] + \frac{m_\ell^2}{q^2}H_t(w)\big[H_+(w) + H_-(w) \big] \right) \label{aligneq:J5} \\
    J_{6s}(w) &= -2F(w)\big[ H_+^2(w) - H_-^2(w) \big] \label{aligneq:J6s} \\
    J_{6c}(w) &= -8\frac{m_\ell^2}{q^2}F(w)H_0(w)H_t(w) \label{aligneq:J6c} \\
    J_{7}(w) &= J_8(w) = J_9(w) = 0\,, \label{aligneq:J789}
\end{align}
where the kinematical factor $F(w)$ is given by
\begin{equation}
    F(w) \equiv \sqrt{w^2 - 1}(1 - 2rw + r^2)\left(1 - \frac{m_\ell^2}{q^2}\right)^2\,.
\end{equation}

\section{Experimental and Lattice Datasets}\label{Appendix:datasets}

\begin{figure}[t]
    \centering
    \includegraphics[width=\linewidth]{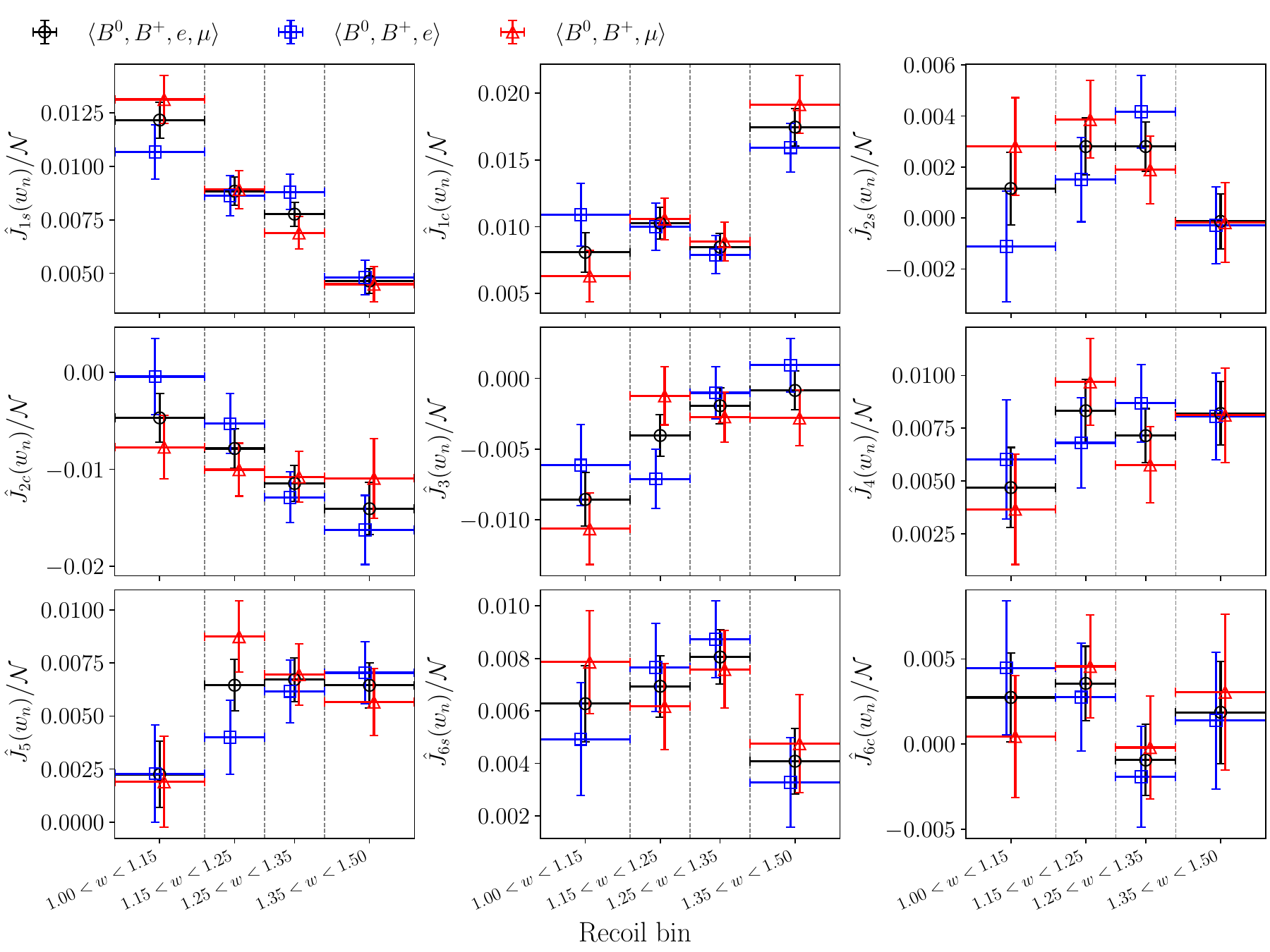}
    \caption{Partially-integrated, normalised angular coefficients from the \texttt{Belle 23 angular} datasets for the $B$ meson lepton averages $\langle B^0, B^+, e, \mu \rangle$, $\langle B^0, B^+, e \rangle$, and $\langle B^0, B^+, \mu \rangle$. The points for $\ell = e$ and $\ell = \mu$ have been horizontally shifted to improve visibility and reduce overlap.}
    \label{fig:exp_ang_coeffs_ml}
\end{figure}

\begin{figure}[t]
    \centering
    \includegraphics[width=\linewidth]{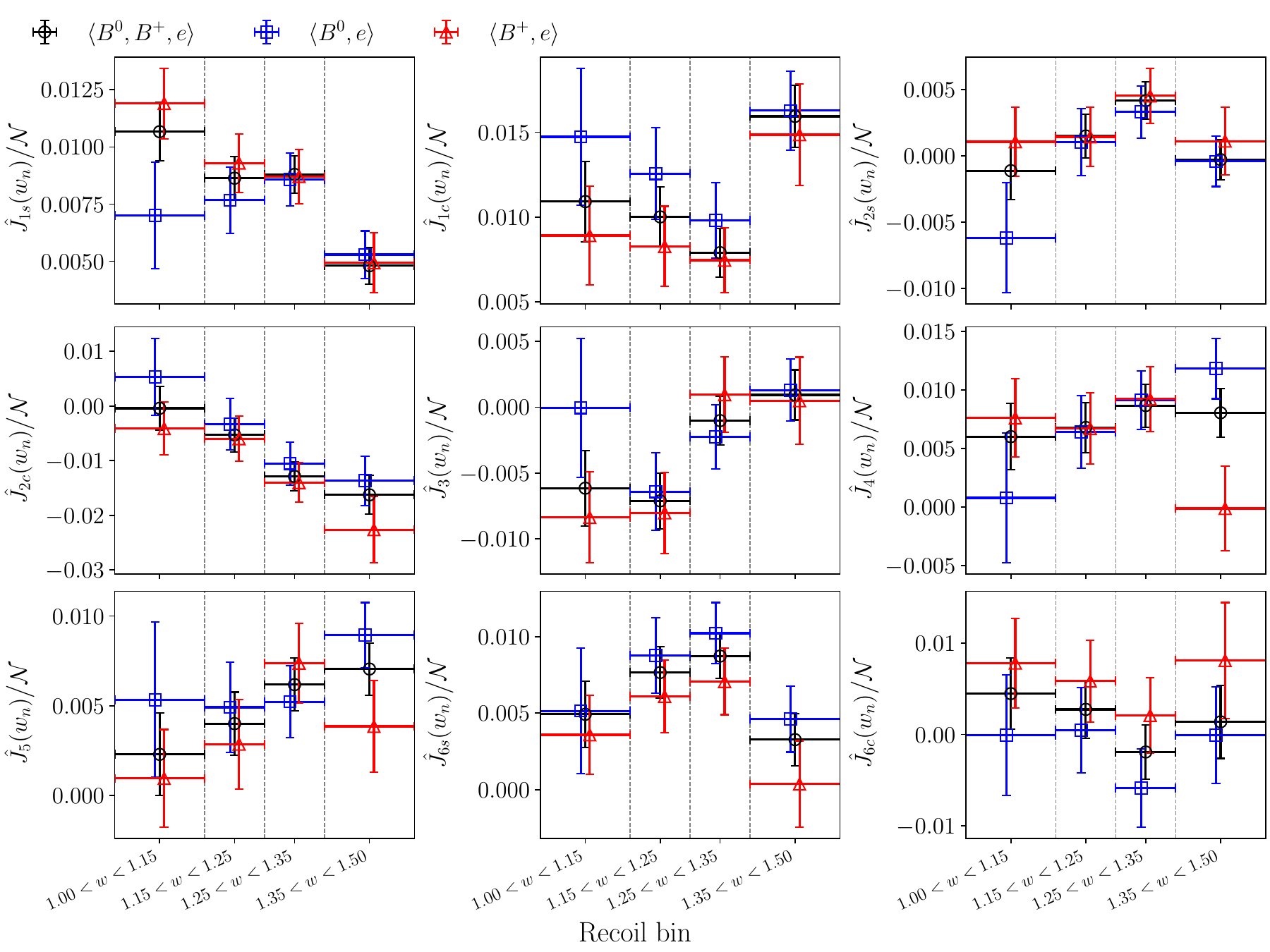}
    \caption{Partially-integrated, normalised angular coefficients from the \texttt{Belle 23 angular} datasets for $\ell = e$. The points for $\ell = e$ and $\ell = \mu$ have been horizontally shifted to improve visibility and reduce overlap.}
    \label{fig:exp_ang_coeffs_B0B+_e}
\end{figure}

\begin{figure}[t]
    \centering
    \includegraphics[width=\linewidth]{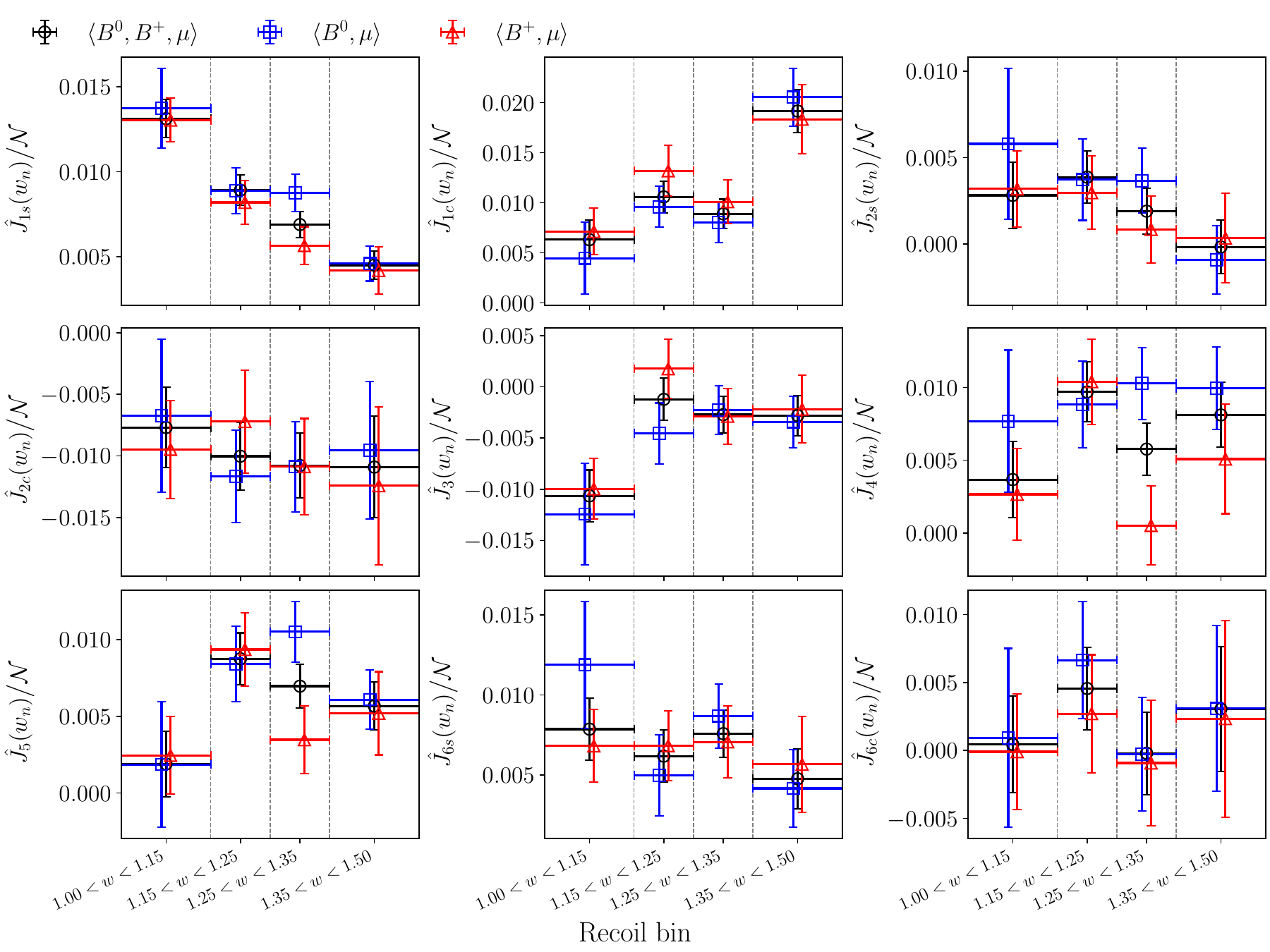}
    \caption{Partially-integrated, normalised angular coefficients from the \texttt{Belle 23 angular} datasets for $\ell = \mu$. The points for $\ell = e$ and $\ell = \mu$ have been horizontally shifted to improve visibility and reduce overlap.}
    \label{fig:exp_ang_coeffs_B0B+_mu}
\end{figure}

In this work, we consider three \SM \LQCD datasets (\texttt{HPQCD 23}~\cite{PhysRevD.109.094515}, \texttt{JLQCD 23}~\cite{PhysRevD.109.074503} and \texttt{FNAL/MILC 21}~\cite{Bazavov_2022}) for the $\BtoDstar$ form factors, one experimental dataset (\texttt{Belle 23 angular}~\cite{Belle:2023xgj,hepdata.153299}) of the angular coefficients, and one \LQCD dataset for the $\BtoDstar$ tensor form factors provided in \texttt{HPQCD 23 tensor}~\cite{PhysRevD.109.094515}. 
\begin{description}
    \item[\texttt{HPQCD 23}\normalfont~\cite{PhysRevD.109.094515}] -- \SM form factors for $\BtoDstar$ and $\BstoDstar$ computed in the \HQET basis $\{h_V, h_{A_1}, h_{A_2}, h_{A_3}\}$ for the vector and axial currents. These are converted to the $\FFs$ basis via the following relations:
    \begin{align}
        f &= m_B\sqrt{r}(w + 1)h_{A_1} , \\
        \mathcal{F}_1 &= m_B^2 \sqrt{r}(w + 1)[(w - r)h_{A_1} - (w - 1)(rh_{A_2} + h_{A_3})], \\
        \mathcal{F}_2 &= \frac{1}{\sqrt{r}}[(w + 1)h_{A_1} + (rw - 1)h_{A_2} + (r - w)h_{A_3}], \\
        g &= \frac{h_V}{m_B\sqrt{r}}.
    \end{align}
    Form factor data and the full covariance matrix are provided at the kinematic points 
    \begin{equation*}
        w \in \{1.00, 1.13, 1.25, 1.38, 1.504\}.
    \end{equation*}
    Due to the high correlation between the data for $f, \mathcal{F}_1, \mathcal{F}_2$, and $g$, the points at the kinematic endpoints, $w = 1.00, 1.504$, were removed from the analysis, reducing the condition number of the correlation matrix from $10^{15}$ to $10^5$. 
    
    \item[\texttt{JLQCD 23}\normalfont~\cite{PhysRevD.109.074503}] -- \SM form factors for $\BtoDstar$ are computed in the $\FFs$ basis at the following kinematic points: 
    \begin{equation*}
        w \in \{1.025, 1.060, 1.100\}.
    \end{equation*}
    No points have been omitted as the condition number of the correlation matrix is $10^4$. 
    
    \item[\texttt{FNAL/MILC 21}\normalfont~\cite{Bazavov_2022}] -- \SM form factors for $\BtoDstar$ are computed in the $\FFs$ basis at the following kinematic points:
    \begin{equation*}
        w \in \{1.03, 1.10, 1.17\}.
    \end{equation*}
    All data points are retained, with a correlation matrix condition number of $3 \times 10^4$. 
    \item[\texttt{Combined}\normalfont~\cite{PhysRevD.109.094515,PhysRevD.109.074503,Bazavov_2022}] -- a combination of the above three lattice datasets. These are treated as statistically independent because each employs a different gauge configuration and QCD discretisation. As such, the covariance matrix is constructed as the block diagonal of the three individual covariance matrices.

    \item[\texttt{HPQCD 23 tensor}\normalfont~\cite{PhysRevD.109.094515}] -- tensor form factors for $\BtoDstar$ and $\BstoDstar$ computed in the \HQET basis $\{h_{T_1}, h_{T_2}, h_{T_3}\}$, converted to the $\{T_1, T_2, T_3\}$ basis by
    \begin{align}
        T_1 &= -\frac{1}{2\sqrt{r}}\left[(1 - r) h_{T_2} - (1 + r) h_{T_1}\right],\label{aligneq:T1} \\
        T_2 &= \frac{1}{2\sqrt{r}}\left[\frac{2 r (w + 1)}{1 + r} h_{T_1} - \frac{2r(w - 1)}{1 - r}h_{T_2}\right],\label{aligneq:T2} \\
        T_3 &= \frac{1}{2\sqrt{r}}\left[(1 - r) h_{T_1} - (1 + r) h_{T_2} + \left(1 - r^2\right) h_{T_3} \right]. \label{aligneq:T3}
    \end{align}
    For each tensor form factor, the data point at $w = w_{\textrm{max}}$ was removed, reducing the condition number of the correlation matrix from $1 \times 10^{17}$ to $9 \times 10^5$. 
    
    \item[\texttt{Belle 23 angular}\normalfont~\cite{Belle:2023xgj,hepdata.153299}] -- central values and full covariance matrices of the twelve bin-averaged, partially-integrated, normalised angular coefficients for $\BtoDstar$, with $\ell = e, \mu$, measured over the following decay chains:

    \begin{itemize}
        \item $\BotoDstar$, with $D^{*+} \to D^0 \pi^+$, $D^{*+} \to D^+ \pi^0$, and
        \item $\BptoDstar$, with $\bar{D}^{*0} \to \bar{D}^0 \pi^0$,
    \end{itemize}

    with charge conjugation implied throughout. Bin-averaged measurements of the angular coefficients are provided in the following intervals:
    \begin{equation}\label{eq:Belle_w_bins}
        w \in \{[1.00, 1.15], [1.15, 1.25], [1.25, 1.35], [1.35, 1.50]\},
    \end{equation}
    provided for each mode, or averaged over $B^+/B^0$ and $\ell = e, \mu$. The provided uncertainties on the angular coefficients are the combination of statistical and systematic uncertainties. These were determined using the full Belle dataset (an integrated luminosity of 711 fb$^{-1}$ at the $\Upsilon(4S)$ resonance), which was obtained by the Belle detector at the KEKB $e^+e^-$ collider~\cite{Kurokawa:2001nw,Abe:2013kxa}. The condition numbers of the full (all twelve angular coefficients) correlation matrices are:
    \begin{itemize}
        \item $\langle B^0, B^+, e, \mu \rangle$: $1.6 \times 10^6$.
        \item $\langle B^0, B^+, e \rangle$: $1.8 \times 10^6$, $\langle B^0, B^+, \mu \rangle$: $1.7 \times 10^6$.
        \item $\langle B^0, e \rangle$: $2.2 \times 10^6$, $\langle B^0, \mu \rangle$: $6.0 \times 10^7$.
        \item $\langle B^+, e \rangle$: $7.9 \times 10^6$, $\langle B^+, \mu \rangle$: $1.9 \times 10^8$.
    \end{itemize}
\end{description}
Plots of the partially-integrated, normalised angular coefficients $\hat{J}_i / \mathcal{N}$ for the aforementioned datasets are shown in \fref{fig:angcoeffs}. For completeness, we also include in \cref{fig:exp_ang_coeffs_ml,fig:exp_ang_coeffs_B0B+_e,fig:exp_ang_coeffs_B0B+_mu} plots of the \texttt{Belle 23 angular} datasets for the different lepton mass cases considered in the main text. 

\section{Form Factors and the BGL Parameterisation}\label{Appendix:FFbasis}

The pseudoscalar to vector transition in $\BtoDstar$ admits the following Lorentz decompositions~\cite{Ball:2004rg}. We denote by $k$ ($p$) the four momentum of the $D^*$ ($B$) meson, and the $D^*$ polarisation by $\eta$. For the vector current,
\begin{equation}\label{eq:BVVFF}
    \langle D^*(k, \eta)| \bar{c} \gamma^\mu b | \bar{B}(p) \rangle = i\epsilon^{\mu\nu\rho\sigma} \eta^*_{\nu}(k)\, p_\rho\, k_\sigma \frac{2\, V(q^2)}{m_B + m_{D^*}}.
\end{equation}
For the axial vector current,
\begin{align}
    \begin{split} \label{eq:BVAFF}
    \langle D^*(k, \eta)| \bar{c} \gamma^\mu \gamma_5 b |\bar{B}(p)\rangle &= \eta^*_{\nu} \bigg\lbrace 2 m_{D^*} A_0(q^2) \frac{q^\mu q^\nu}{q^2} \\
    &\quad + 16\frac{m_B  m_{D^*}^2}{\lambda} A_{12} \left[2 p^\mu q^\nu - \frac{m_B^2 - m_{D^*}^2 + q^2}{q^2} q^\mu q^\nu\right] \\
    &\phantom{=} + (m_B + m_{D^*})\, A_1(q^2) \left[g^{\mu\nu} + \frac{2(m_B^2 + m_{D^*}^2 - q^2)}{\lambda} q^\mu q^\nu \right. \\
    &\quad \left. - \frac{2(m_B^2 - m_{D^*}^2 - q^2)}{\lambda} p^\mu q^\nu\right]\bigg\rbrace\,,
\end{split}
\end{align}
where $\lambda \equiv \lambda(q^2)$ is the K\"{a}ll\'{e}n function
\begin{equation}\label{eq:Kallen}
    \lambda(q^2) = (m_B^2 - m_{D^*}^2 - q^2)^2 - 4m_{D^*}^2 q^2.
\end{equation}
Finally, for the tensor current:
\begin{align}
\begin{split}\label{eq:BVTFF}
    \langle D^*(k, \eta) | \bar{c} \sigma^{\mu\nu} b |\bar{B}(p)\rangle
        & = i
        \eta^*_{\alpha}\epsilon^{\mu\nu}\!{}_{\rho\sigma}\left\{-\left[\left((p+k)^\rho-\frac{m_B^2-m_{D^*}^2}{q^2}q^\rho\right)g^{\alpha\sigma} \right.\right. \\
        & \left.\left. \qquad\qquad\quad\,\,\, +
        \frac{2}{q^2} p^\alpha p^\rho k^\sigma\right]T_1(q^2)\right.\\
        &  \qquad\qquad\quad\,\,\,-\left.\left(\frac{2}{q^2}p^\alpha p^\rho k^\sigma-\frac{m_B^2-m_{D^*}^2}{q^2}q^\rho
        g^{\alpha\sigma}\right) T_2(q^2) \right. \\
        & \qquad\qquad\quad\,\,\, \left. + \frac{2}{m_B^2 - m_{D^*}^2}p^\alpha p^\rho k^\sigma T_3(q^2)\right\}\,.
\end{split}
\end{align}
The HPQCD dataset~\cite{PhysRevD.109.094515} extracts \HQET form factors $h_i(w)$. Omitting the form factor dependence on $w$, they are related to the QCD and $\FFs$ bases by the following relations
\begin{align}
    V &= \frac{1 + r}{2 \sqrt{r}} h_V = \frac{m_B + m_{D^*}}{2}g \\
    A_0 &= \frac{1}{2 \sqrt{r}} \left[ (w + 1) h_{A_1} + (r w - 1) h_{A_2} + (r - w) h_{A_3}\right] = \frac{1}{2}\mathcal{F}_2 \\
    A_1 &= \frac{\sqrt{r}\,(1 + w)}{1 + r} h_{A_1} = \frac{1}{m_B + m_{D^*}}f \\
    A_{12} &= \frac{1 + w}{8 \sqrt{r}}\left[(w - r) h_{A_1} - (w - 1) r h_{A_2} - (w - 1) h_{A_3}\right] = \frac{1}{8 m_Bm_{D^*}}\mathcal{F}_1 \\
    T_1 &= -\frac{1}{2\sqrt{r}}\left[(1 - r) h_{T_2} - (1 + r) h_{T_1}\right] \label{eq:T1}\\
    T_2 &= \frac{1}{2\sqrt{r}}\left[\frac{2 r (w + 1)}{1 + r} h_{T_1} - \frac{2r(w - 1)}{1 - r} h_{T_2}\right] \label{eq:T2}\\
    T_3 &= \frac{1}{2\sqrt{r}}\left[(1 - r) h_{T_1} - (1 + r) h_{T_2} + \left(1 - r^2\right) h_{T_3} \right],
\end{align}
where $r \equiv m_{D^*} / m_B$. The tensor form factors are related by the endpoint relation $T_1(0) = T_2(0)$. Furthermore, in the helicity basis, it is more convenient to use
\begin{equation}\label{eq:T23}
    T_{23} = \frac{(m_B^2-m_{D^*}^2)(m_B^2+3m_{D^*}^2-q^2)T_2-\lambda T_3}{8 m_B m_{D^*}^2(m_B-m_{D^*})}.
\end{equation}
\subsection{Form Factor Parameterisations}\label{Subsection:BGL}

In this work, we parameterise the hadronic transition form factors using the \BGL ansatz~\cite{Boyd:1994tt}. At its core are the two fundamental properties of unitarity and analyticity. To begin, it is useful to define
\begin{equation}
    t_\pm = (m_B \pm m_{D^*})^2,
\end{equation}
where $t_+$ is the squared invariant mass of the $BD^*$ system, and $t_-$ is the upper limit of the squared momentum transfer in $\BtoDstar$. The \BGL parameterisation relies on the conformal transformation
\begin{equation}\label{eq:z_transformation}
    z(q^2; t_0) := \frac{\sqrt{t_+ - q^2} - \sqrt{t_+ - t_0}}{\sqrt{t_+ - q^2} + \sqrt{t_+ - t_0}},
\end{equation}
which maps the complex $q^2$ plane (with a branch cut along the real axis above $q^2 = t_+$) to the unit disk in the variable $z$. The choice of the parameter $t_0 \in (-\infty, t_+)$ amounts to defining which $q^2$ value is mapped to the point $z = 0$. Taking $t_0 = t_-$ and translating $q^2 \to w$ with \eqref{eq:w(q^2)} gives \eqref{eq:z_transformation} in terms of the recoil variable:
\begin{equation}\label{eq:z_transformation_w}
    z(w) = \frac{\sqrt{w + 1} - \sqrt{2}}{\sqrt{w + 1} + \sqrt{2}}.
\end{equation}
Typical values of the variable $z$ within the physical semileptonic range for $\BtoDstar$ are $z \in [0, 0.056]$, making it well-suited as a series expansion parameter with fast convergence. To translate the unitarity constraints in the pair production region into the physical semileptonic region $m_\ell^2 \leq q^2 \leq t_-$ requires the form factors to be analytic below the branch cut on the $q^2$ axis. Sub-threshold poles located at the invariant squared mass of the intermediate states are cancelled by Blaschke products, defined as
\begin{align}\label{eq:Blaschke_product}
    B(z; \{z_p\}) = \prod_p \frac{z - z_p}{1 - zz_p^*}, && z_p = \frac{\sqrt{t_+ - m_p^2} - \sqrt{t_+ - t_0}}{\sqrt{t_+ - m_p^2} + \sqrt{t_+ - t_0}},
\end{align}
which have unit modulus on the unit circle: $|B(e^{i\theta}; \{z_p\})| = 1$, and vanish at the poles. Note that $B(z; \varnothing) = 1$ in the case of no poles. A form factor with spin-parity $J^P$ that has poles below the $t_+$ threshold may have its analyticity restored by multiplying by a Blaschke factor:
\begin{equation}
    B_{J^P}(w) = \prod_j \frac{z(w) - z(m_{J^P,\, j}^2)}{1 - z(w)z^*(m_{J^P,\, j}^2)},
\end{equation}
where $m_{J^P,\, j}$ are the masses of the $\bar{c}b$ QCD bound states considered in this work, and are given in \tref{tab:cb_masses}. They are provided in Ref.~\cite{Bordone:2025jur}, which were taken from Table III of Ref.~\cite{Bigi:2017jbd}, updated by the comments in Footnote 3 of Ref.~\cite{Gambino:2019sif}. Since these masses, and hence the Blaschke factors, depend on the spin-parity values of the resonance, the $\FFs$ form factor basis is convenient because they also assume definite $J^P$ values. 

With any sub-threshold poles removed, the product
$B(z;\{z_p\})F(z)$ is thus analytic in $q^2 < t_+$. Outer functions are introduced to essentially mediate the change of variables from $w \to z(w)$, and simplify the resulting constraints on the parameterisation of the form factors. They are analytic with no zeroes in $|z| < 1$, and have unit modulus on the boundary $|z| = 1$. Their introduction modifies the unitarity constraint to the form
\begin{equation}\label{eq:unitarity_unit_circle}
    \frac{1}{2\pi i}\oint\limits_{|z| = 1}|\phi(z)B(z)F(z)|^2 \frac{\textrm{d}z}{z} \leq 1.
\end{equation}
The outer functions for the vector form factors are given by~\cite{Bordone:2024weh}:
\begin{table}[t]
    \centering
    \resizebox{1.0\textwidth}{!}{%
    \begin{tabular}{c ccc ccc cccc}
    \toprule
    $J^P$ & \multicolumn{3}{c}{$0^-$} & \multicolumn{3}{c}{$1^-$} & \multicolumn{4}{c}{$1^+$} \\
    $j$ & 1 & 2 & 3 & 1 & 2 & 3 & 1 & 2 & 3 & 4 \\
    \midrule
    Mass [GeV] & 6.275 & 6.871 & 7.250 & 6.329 & 6.910 & 7.020 & 6.739 & 6.750 & 7.145 & 7.150 \\
    \bottomrule
    \end{tabular}}
    \caption{Pole masses of the $\bar{c}b$ QCD bound states in the $B_c$ mass spectrum required for the construction of the Blaschke factors, organised according to their spin-parity quantum numbers~\cite{Bordone:2025jur,Bigi:2017jbd,Gambino:2019sif}.}
    \label{tab:cb_masses}
\end{table}
\begin{equation}
    \phi_X^\textrm{V}(z) = N\sqrt{\frac{n_I}{K\pi\chi}}\frac{r^a(1 + z)^a(1 - z)^{b/2}}{[(1 + r)(1 - z) + 2\sqrt{r}(1 + z)]^c},
\end{equation}
where $\chi$ are susceptibilities~\cite{Bigi:2016mdz,Bigi_2017,Bigi:2017jbd,Martinelli_2021,harrison2024bc_chi}, and the other parameters are given in \tref{tab:outer_func_pars_ff}. The factor $n_I$ arises due to the flavour SU(3) symmetry that counts the number of massless spectator quarks, with a conservative subtraction to account for the symmetry breaking of the $s$ quark. For the tensor form factors, the outer functions are parameterised as~\cite{Boyd_1997}:
\begin{table}[t]
    \centering
    \begin{tabular}{c ccccccc}
    \toprule
    \multirow{2}{*}{Form Factor} & \multicolumn{7}{c}{Parameters} \\
     & $N$ & $n_I$ & $K$ & $\chi$ & $a$ & $b$ & $c$ \\
    \midrule
    $f$ & $4/m_B^2$ & 2.6 & $3$ & $3.894 \times 10^{-4}$ GeV$^{-2}$ & 1 & 3 & 4 \\
    $\mathcal{F}_1$ & $4/m_B^3$ & 2.6 & $6$ & $3.894 \times 10^{-4}$ GeV$^{-2}$ & 1 & 5 & 5 \\
    $\mathcal{F}_2$ & $8\sqrt{2}$ & 2.6 & $1$ & $1.9421 \times 10^{-2}$ & 2 & $-1$ & 4 \\
    $g$ & 16 & 2.6 & $3$ & $5.131 \times 10^{-4}$ GeV$^{-2}$ & 2 & $-1$ & 4 \\
    \bottomrule
    \end{tabular}
    \caption{Parameters of the outer functions for the vector form factors~\cite{Bordone:2024weh}.}
    \label{tab:outer_func_pars_ff}
\end{table}
\begin{align}
\begin{split}
    \phi_X^\textrm{T}(t) &= \sqrt{\frac{n_I}{K\pi\chi}}\left(\frac{t_+ - t}{t_+ - t_0}\right)^\frac{1}{4}\big( \sqrt{t_+ - t} + \sqrt{t_+ - t_0} \big)(t_+ - t)^\frac{a}{4} \\
    &\quad\times\big(\sqrt{t_+ - t} + \sqrt{t_+ - t_-} \big)^\frac{b}{2}\big(\sqrt{t_+ - t} + \sqrt{t_+} \big)^{-(c+3)},
\end{split}
\end{align}
where the $n_f$ factor in Ref.~\cite{Bordone:2025jur} has been set to $n_I$, since the strange quark channel is not analysed separately in this work. The parameters for the outer functions for the tensor form factors are in \tref{tab:outer_func_pars_tensor}. The tensor and axial tensor susceptibilities are calculated in a local operator product expansion, and have been numerically evaluated at the scale $\mu = \sqrt{m_c m_b}$~\cite{Bharucha:2010im}. 

\begin{table}[t]
    \centering
    \begin{tabular}{c cccccc}
    \toprule
    \multirow{2}{*}{Form Factor} & \multicolumn{6}{c}{Parameters} \\
     & $n_I$ & $K$ & $\chi$ & $a$ & $b$ & $c$ \\
    \midrule
    $T_1$ & 2.6 & 24 & $4.98(40) \times 10^{-4}$ GeV$^{-2}$ & 3 & 3 & 2 \\
    $T_2$ & 2.6 & $24 / (t_+t_-)$ & $2.77(22) \times 10^{-4}$ GeV$^{-2}$ & 1 & 1 & 2 \\
    $T_{23}$ & 2.6 & $48t_+/(t_+ - t_-)^2$ & $2.77(22) \times 10^{-4}$ GeV$^{-2}$ & 1 & 1 & 1 \\
    \bottomrule
    \end{tabular}
    \caption{Parameters of the outer functions for the tensor form factors~\cite{Bordone:2025jur}.}
    \label{tab:outer_func_pars_tensor}
\end{table}

With the inclusion of Blaschke factors and outer functions, the product $F(z)B_{J^P}(z)\phi(z)$ is, by construction, analytic in the open unit disk; the form factors may thus be written as
\begin{equation}\label{eq:FF_B_phi_series}
    F(z) = \frac{1}{B_{J^P}(z)\phi_F(z)}\sum_{n=0}^{\infty}a_n z^n,
\end{equation}
for a form factor $F$ with spin-parity $J^P$. The \BGL coefficients are restricted by a unitarity constraint for each spin-parity channel; on substituting \eqref{eq:FF_B_phi_series} into \eqref{eq:unitarity_unit_circle} yields \eqref{eq:BGL_unitarity_vector}
for the form factors $\FFs$, and
\begin{align}\label{eq:BGL_unitarity_tensor}
    \sum_{n = 0}^\infty |a_{T_1,n}|^2 \leq 1, && \sum_{n = 0}^\infty |a_{T_2,n}|^2 + |a_{T_{23},n}|^2 \leq 1
\end{align}
for the tensor form factors.

\subsection{Endpoint Kinematic Constraints}\label{App:Z_constraint}

In the following, we outline the implementation of the kinematic constraints in \cref{aligneq:fF1F2F1_constraints} as done in Ref.~\cite{Bordone:2024weh}, and the analogous case for the tensor form factors. The kinematic constraints are imposed exactly by eliminating the leading-order coefficients in the $z$-expansion of the form factors $f$ and $\mathcal{F}_2$ (recalling that $z(w = 1) = 0$):
\begin{align}
    a_{f,0} &= \frac{\phi_f(z = 0)}{\phi_{\mathcal{F}_1}(z = 0)}a_{\mathcal{F}_1, 0} \\
    a_{\mathcal{F}_2,0} &= \frac{B_{\mathcal{F}_2}(z_{\textrm{max}})\phi_{\mathcal{F}_2}(z_{\textrm{max}})}{B_{\mathcal{F}_1}(z_{\textrm{max}})\phi_{\mathcal{F}_1}(z_{\textrm{max}})}\sum_{k = 0}^{K_{\mathcal{F}_1} - 1}a_{\mathcal{F}_1, k}z_{\textrm{max}}^k - \sum_{k = 1}^{K_{\mathcal{F}_2} - 1}a_{\mathcal{F}_2, k}z_{\textrm{max}}^k,
\end{align}
where $z_{\textrm{max}} := z(w_{\textrm{max}}) = z(q^2 = 0)$. That the form factors are linear in the \BGL coefficients $\mathbf{a}$ allows one to write the form factors in the form $\mathbf{f}^{\textrm{BGL}}(\mathbf{a}) = \mathsf{Z}\mathbf{a}$. Denoting by $X \in \FFs$ a form factor, the design matrix has block-diagonal entries given by
\begin{equation}
    (\mathsf{Z}_{XX})_{ij} = \frac{1}{\phi_X(z_i)B_X(z_i)}(z_i)^j,
\end{equation}
where the index $i$ runs over the available discrete $z$ values, dictated by the form factor data in question. The aforementioned kinematic constraints introduce the following off-diagonal block entries:
\begin{align}
    (\mathsf{Z}_{f\mathcal{F}_1})_{i0} &= \frac{1}{m_B(1 - r)}\frac{\phi_f(z = 0)}{\phi_{\mathcal{F}_1}(z = 0)}\frac{1}{\phi_f(z_i)B_f(z_i)} \\
    (\mathsf{Z}_{\mathcal{F}_2\mathcal{F}_1})_{ij} &= \frac{1 + r}{m_B^2 r(1 - r)(1 + w_{\textrm{max}})}\frac{\phi_{\mathcal{F}_2}(z_{\textrm{max}})B_{\mathcal{F}_2}(z_{\textrm{max}})}{\phi_{\mathcal{F}_1}(z_{\textrm{max}})B_{\mathcal{F}_1}(z_{\textrm{max}})}\frac{1}{\phi_{\mathcal{F}_2}(z_i)B_{\mathcal{F}_2}(z_i)}(z_i)^j.
\end{align}
The tensor form factors $T_1(q^2)$ and $T_2(q^2)$ are related at zero recoil by
\begin{equation}\label{eq:T12endpoint}
    T_1(q^2 = 0) = T_2(q^2 = 0).
\end{equation}
The \BGL parameterisation (truncated at order $K_F$) of the form factors can be written as
\begin{equation}
    F(w) = \frac{1}{B_{F}(z)\phi_F(z)}\sum_{k=0}^{K_F - 1}a_{F,k}[z(w)]^k,
\end{equation}
where $F \in \{T_1, T_2, T_{23}\}$, and the parameter $z$ is given in terms of the recoil variable $w$ in \eqref{eq:z_transformation_w}. Vanishing momentum transfer $q^2 = 0$ corresponds to the maximum value of the recoil parameter, $w_\textrm{max}^{(\ell)}$. The endpoint relation in \eqref{eq:T12endpoint} reads
\begin{equation}
    \frac{1}{B_{T_1}(z_\textrm{max})\phi_{T_1}(z_\textrm{max})}\sum_{k=0}^{K_{T_1} - 1}a_{T_1,k}z_\textrm{max}^k = \frac{1}{B_{T_2}(z_\textrm{max})\phi_{T_2}(z_\textrm{max})}\sum_{k=0}^{K_{T_2} - 1}a_{T_2,k}z_\textrm{max}^k
\end{equation}
This constraint effectively eliminates one coefficient. Choosing this to be $a_{T_1,0}$:
\begin{equation}\label{eq:a_T1_0_relation}
    a_{T_1,0} = \frac{B_{T_1}(z_\textrm{max})\phi_{T_1}(z_\textrm{max})}{B_{T_2}(z_\textrm{max})\phi_{T_2}(z_\textrm{max})}\sum_{k=0}^{K_{T_2} - 1}a_{T_2,k}z_\textrm{max}^k - \sum_{k=1}^{K_{T_1} - 1}a_{T_1,k}z_\textrm{max}^k.
\end{equation}
The form factor $T_1(w)$ thus reads
\begin{align}
\begin{split}
    T_1(w) &= \frac{1}{B_{T_1}(z)\phi_{T_1}(z)}\left[ \frac{B_{T_1}(z_\textrm{max})\phi_{T_1}(z_\textrm{max})}{B_{T_2}(z_\textrm{max})\phi_{T_2}(z_\textrm{max})}\sum_{k=0}^{K_{T_2} - 1}a_{T_2,k}z_\textrm{max}^k \right. \\
    &\qquad\qquad\qquad\qquad \left. + \sum_{k=1}^{K_{T_1} - 1}a_{T_1,k}\{[z(w)]^k\ - z_\textrm{max}^k\} \right]
\end{split}
\end{align}
For a linear fit, the design matrix $\mathsf{Z}_{XX}$ contracts with the fit parameters such that $F_X(w) = \sum_{j}(\mathsf{Z}_{XX})_{ij}a_{j}$, where $a_j$ are components of the \BGL coefficient vector $\mathbf{a}$. With no constraint, the design matrix $\mathsf{Z}_{XX}$ has block-diagonal entries
\begin{equation}
    (\mathsf{Z}_{XX})_{ij} = \frac{1}{B_X(z_i)\phi_X(z_i)}(z_i)^j.
\end{equation}
The endpoint constraint introduces a non-zero off-diagonal block originating from the constraint
\begin{equation}
    (\mathsf{Z}_{T_1T_2})_{ij} = \frac{B_{T_1}(z_\textrm{max})\phi_{T_1}(z_\textrm{max})}{B_{T_2}(z_\textrm{max})\phi_{T_2}(z_\textrm{max})}\frac{1}{B_{T_1}(z_i)\phi_{T_1}(z_i)}(z_\textrm{max})^j
\end{equation}
and alters the $Z_{T_1T_1}$ diagonal element as a result of eliminating $a_{T_1,0}$:
\begin{equation}
    (\mathsf{Z}_{T_1T_1})_{in} = \frac{1}{B_{T_1}(z_i)\phi_{T_1}(z_i)}\left[(z_i)^n - (z_\textrm{max})^n\right]
\end{equation}
for $n \geq 1$.

\section{Predictions of \texorpdfstring{\boldmath$B \to D^*\tau\bar{\nu}_\tau$}{B->D*taunutau} Angular Coefficients from LQCD}\label{Appendix:tau}

While no experimental data for $B \to D^*\tau\bar{\nu}_\tau$ from Belle is available at this time, \LQCD predictions may also be computed for $m_\ell = m_\tau$. These have been tabulated for each \LQCD dataset in \tref{tab:coeffs_mtau}, and follow from the same Type-A $K = 3$ fits to form-factor data as in \cref{Subsec:Type_A_results}. Due to the large mass of the $\tau$, the maximum value of the recoil variable is $w_{\textrm{max}}^{(\tau)} \approx 1.35$, reducing the number of bins compared with the case of light leptons. 

\begin{table}[t]
    \centering
    \resizebox{\linewidth}{!}{
\begin{tabular}{ccS[table-format=-1.6(2)]S[table-format=-1.6(2)]S[table-format=-1.6(2)]S[table-format=-1.7(2)]}
    \toprule
    \textbf{Angular} & \multirow{2}{*}{$\bm{w}$} & \multicolumn{4}{c}{\textbf{\LQCD Dataset}} \\
    \textbf{Coefficient} & & \texttt{HPQCD 23} & \texttt{JLQCD 23} & \texttt{FNAL/MILC 21} & \texttt{Combined} \\
    \midrule
    \multirow{3}{*}{$\hat{J}_{1s}(w_n) / \mathcal{N}$} & $1.00 < w < 1.15$ & 0.02276(88) & 0.02079(63) & 0.02225(45) & 0.02171(30) \\
     & $1.15 < w < 1.25$ & 0.01213(57) & 0.01122(43) & 0.01147(26) & 0.01163(19) \\
     & $1.25 < w < w_{\textrm{max}}^{(\tau)}$ & 0.00321(23) & 0.00302(18) & 0.00295(14) & 0.003100(96) \\
    \midrule
    \multirow{3}{*}{$\hat{J}_{1c}(w_n) / \mathcal{N}$} & $1.00 < w < 1.15$ & 0.02102(66) & 0.02151(38) & 0.02192(37) & 0.02140(22) \\
     & $1.15 < w < 1.25$ & 0.0156(14) & 0.0185(10) & 0.01677(58) & 0.01718(45) \\
     & $1.25 < w < w_{\textrm{max}}^{(\tau)}$ & 0.00620(99) & 0.00828(76) & 0.00669(46) & 0.00723(35) \\
    \midrule
    \multirow{3}{*}{$\hat{J}_{2s}(w_n) / \mathcal{N}$} & $1.00 < w < 1.15$ & 0.00437(17) & 0.00399(12) & 0.004276(88) & 0.004171(59) \\
     & $1.15 < w < 1.25$ & 0.001799(83) & 0.001663(63) & 0.001703(38) & 0.001724(28) \\
     & $1.25 < w < w_{\textrm{max}}^{(\tau)}$ & 0.000292(20) & 0.000274(16) & 0.000268(13) & 0.0002816(84) \\
    \midrule
    \multirow{3}{*}{$\hat{J}_{2c}(w_n) / \mathcal{N}$} & $1.00 < w < 1.15$ & -0.00901(25) & -0.00901(16) & -0.00922(16) & -0.009005(94) \\
     & $1.15 < w < 1.25$ & -0.00418(29) & -0.00479(25) & -0.00438(13) & -0.00447(10) \\
     & $1.25 < w < w_{\textrm{max}}^{(\tau)}$ & -0.00082(10) & -0.001044(89) & -0.000860(52) & -0.000921(39) \\
    \midrule
    \multirow{3}{*}{$\hat{J}_{3}(w_n) / \mathcal{N}$} & $1.00 < w < 1.15$ & -0.00772(34) & -0.00713(24) & -0.00749(18) & -0.00734(12) \\
     & $1.15 < w < 1.25$ & -0.00275(18) & -0.00259(12) & -0.002554(84) & -0.002590(63) \\
     & $1.25 < w < w_{\textrm{max}}^{(\tau)}$ & -0.000409(43) & -0.000393(32) & -0.000368(23) & -0.000383(17) \\
    \midrule
    \multirow{3}{*}{$\hat{J}_{4}(w_n) / \mathcal{N}$} & $1.00 < w < 1.15$ & 0.00860(22) & 0.00824(15) & 0.00859(15) & 0.008396(92) \\
     & $1.15 < w < 1.25$ & 0.003634(80) & 0.003759(56) & 0.003611(48) & 0.003672(31) \\
     & $1.25 < w < w_{\textrm{max}}^{(\tau)}$ & 0.000634(37) & 0.000699(25) & 0.000623(24) & 0.000659(15) \\
    \midrule
    \multirow{3}{*}{$\hat{J}_{5}(w_n) / \mathcal{N}$} & $1.00 < w < 1.15$ & 0.01066(50) & 0.01033(37) & 0.01126(28) & 0.01085(19) \\
     & $1.15 < w < 1.25$ & 0.00968(54) & 0.01018(37) & 0.01006(30) & 0.01026(20) \\
     & $1.25 < w < w_{\textrm{max}}^{(\tau)}$ & 0.00359(31) & 0.00406(19) & 0.00366(19) & 0.00393(11) \\
    \midrule
    \multirow{3}{*}{$\hat{J}_{6s}(w_n) / \mathcal{N}$} & $1.00 < w < 1.15$ & 0.01242(99) & 0.01086(76) & 0.01250(49) & 0.01199(36) \\
     & $1.15 < w < 1.25$ & 0.00897(84) & 0.00807(66) & 0.00872(43) & 0.00883(31) \\
     & $1.25 < w < w_{\textrm{max}}^{(\tau)}$ & 0.00249(30) & 0.00230(22) & 0.00235(17) & 0.00249(11) \\
    \midrule
    \multirow{3}{*}{$\hat{J}_{6c}(w_n) / \mathcal{N}$} & $1.00 < w < 1.15$ & -0.00872(64) & -0.00963(31) & -0.00978(23) & -0.00949(16) \\
     & $1.15 < w < 1.25$ & -0.0105(12) & -0.01296(81) & -0.01164(47) & -0.01195(36) \\
     & $1.25 < w < w_{\textrm{max}}^{(\tau)}$ & -0.00522(92) & -0.00709(67) & -0.00570(41) & -0.00618(31) \\
    \midrule\multicolumn{2}{c}{\textbf{Decay rate} $\widetilde{\Gamma}^{\textrm{lat}}$ $(\times 10^{-12})$} & 3.19(17) & 3.36(18) & 3.24(15) & 3.279(93) \\
    \bottomrule
\end{tabular}}
    \caption{\LQCD predictions of the angular coefficients and decay rate $\widetilde{\Gamma}^{\textrm{lat}}$ for $B \to D^*\tau\bar{\nu}_\tau$ from $K = 3$ Type-A fits to the \texttt{HPQCD 23}~\cite{PhysRevD.109.094515}, \texttt{JLQCD 23}~\cite{PhysRevD.109.074503}, \texttt{FNAL/MILC 21}~\cite{Bazavov_2022} datasets, and their combination.}
    \label{tab:coeffs_mtau}
\end{table}

\bibliographystyle{JHEP}
\bibliography{biblio.bib}

@article{Cabibbo:1963yz,
    author = "Cabibbo, Nicola",
    title = "{Unitary Symmetry and Leptonic Decays}",
    doi = "10.1103/PhysRevLett.10.531",
    journal = "Phys. Rev. Lett.",
    volume = "10",
    pages = "531--533",
    year = "1963"
}

@article{Kobayashi:1973fv,
    author = "Kobayashi, Makoto and Maskawa, Toshihide",
    title = "{CP Violation in the Renormalizable Theory of Weak Interaction}",
    reportNumber = "KUNS-242",
    doi = "10.1143/PTP.49.652",
    journal = "Prog. Theor. Phys.",
    volume = "49",
    pages = "652--657",
    year = "1973"
}

@article{Bazavov_2022,
   title="{Semileptonic form factors for $B\rightarrow D^*\ell \nu $ at nonzero recoil from $2+1$-flavor lattice QCD: Fermilab Lattice and MILC Collaborations}",
   eprint = "2105.14019",
   archivePrefix = "arXiv",
   primaryClass = "hep-lat",
   volume={82},
   ISSN={1434-6052},
   url={http://dx.doi.org/10.1140/epjc/s10052-022-10984-9},
   DOI={10.1140/epjc/s10052-022-10984-9},
   number={12},
   journal={The European Physical Journal C},
   publisher={Springer Science and Business Media LLC},
   author={Bazavov, A. and DeTar, C. E. and Du, D. and El-Khadra, A. X. and Gámiz, E. and Gelzer, Z. and Gottlieb, S. and Heller, U. M. and Kronfeld, A. S. and Laiho, J. and Mackenzie, P. B. and Simone, J. N. and Sugar, R. and Toussaint, D. and Van de Water, R. S. and Vaquero, A.},
   year={2022},
   month=dec }

@article{PhysRevD.109.094515,
  title = "{$B\ensuremath{\rightarrow}{D}^{*}$ and ${B}_{s}\ensuremath{\rightarrow}{D}_{s}^{*}$ vector, axial-vector and tensor form factors for the full ${q}^{2}$ range from lattice QCD}",
  author = {Harrison, Judd and Davies, Christine T. H.},
  collaboration = {HPQCD},
  eprint = "2304.03137",
  archivePrefix = "arXiv",
  primaryClass = "hep-lat",
  journal = {Phys. Rev. D},
  volume = {109},
  issue = {9},
  pages = {094515},
  numpages = {46},
  year = {2024},
  month = {05},
  publisher = {American Physical Society},
  doi = {10.1103/PhysRevD.109.094515},
  url = {https://link.aps.org/doi/10.1103/PhysRevD.109.094515}
}

@article{PhysRevD.109.074503,
  title = "{$B\ensuremath{\rightarrow}{D}^{*}\ensuremath{\ell}{\ensuremath{\nu}}_{\ensuremath{\ell}}$ semileptonic form factors from lattice QCD with M\"obius domain-wall quarks}",
  author = {Aoki, Y. and Colquhoun, B. and Fukaya, H. and Hashimoto, S. and Kaneko, T. and Kellermann, R. and Koponen, J. and Kou, E.},
  collaboration = {JLQCD},
  eprint = "2306.05657",
  archivePrefix = "arXiv",
  primaryClass = "hep-lat",
  journal = {Phys. Rev. D},
  volume = {109},
  issue = {7},
  pages = {074503},
  numpages = {24},
  year = {2024},
  month = {04},
  publisher = {American Physical Society},
  doi = {10.1103/PhysRevD.109.074503},
  url = {https://link.aps.org/doi/10.1103/PhysRevD.109.074503}
}

@article{Bordone:2024weh,
    author = {Bordone, Marzia and J\"{u}ttner, Andreas},
    title = "{New strategies for probing $B\rightarrow D^*\ell \bar{\nu }_\ell $ lattice and experimental data}",
    eprint = "2406.10074",
    archivePrefix = "arXiv",
    primaryClass = "hep-ph",
    reportNumber = "CERN-TH-2024-083",
    doi = "10.1140/epjc/s10052-025-13773-2",
    journal = "Eur. Phys. J. C",
    volume = "85",
    number = "2",
    pages = "129",
    year = "2025"
}

@article{Bordone:2025jur,
    author = "Bordone, Marzia and Gubernari, Nico and Jung, Martin and van Dyk, Danny",
    title = "{Challenging $ {\overline{B}}_{(s)}\to {D}_{(s)}^{\left(\ast \right)} $ form factors with the heavy quark expansion}",
    eprint = "2507.03569",
    archivePrefix = "arXiv",
    primaryClass = "hep-ph",
    reportNumber = "CERN-TH-2025-092, EOS-2025-03, IPPP/25/25, P3H-25-032, SI-HEP-2025-10, ZU-TH 35/25",
    doi = "10.1007/JHEP11(2025)051",
    journal = "JHEP",
    volume = "11",
    pages = "051",
    year = "2025"
}

@article{Feroz_2008,
   title="{Multimodal nested sampling: an efficient and robust alternative to Markov Chain Monte Carlo methods for astronomical data analyses: Multimodal nested sampling}",
   volume={384},
   ISSN={1365-2966},
   url={http://dx.doi.org/10.1111/j.1365-2966.2007.12353.x},
   DOI={10.1111/j.1365-2966.2007.12353.x},
   number={2},
   journal={Monthly Notices of the Royal Astronomical Society},
   publisher={Oxford University Press (OUP)},
   author={Feroz, F. and Hobson, M. P.},
   year={2008},
   month=jan, pages={449–463} 
}

@article{Feroz_2009,
   title="{MultiNest: an efficient and robust Bayesian inference tool for cosmology and particle physics}",
   volume={398},
   ISSN={1365-2966},
   url={http://dx.doi.org/10.1111/j.1365-2966.2009.14548.x},
   DOI={10.1111/j.1365-2966.2009.14548.x},
   number={4},
   journal={Monthly Notices of the Royal Astronomical Society},
   publisher={Oxford University Press (OUP)},
   author={Feroz, F. and Hobson, M. P. and Bridges, M.},
   year={2009},
   month=oct, pages={1601–1614} 
}

@article{Feroz_2019,
   title="{Importance Nested Sampling and the MultiNest Algorithm}",
   volume={2},
   ISSN={2565-6120},
   url={http://dx.doi.org/10.21105/astro.1306.2144},
   DOI={10.21105/astro.1306.2144},
   number={1},
   journal={The Open Journal of Astrophysics},
   publisher={Maynooth University},
   author={Feroz, Farhan and Hobson, Michael P. and Cameron, Ewan and Pettitt, Anthony N.},
   year={2019},
   month=nov 
}

@article{Buchner_2014,
   title="{X-ray spectral modelling of the AGN obscuring region in the CDFS: Bayesian model selection and catalogue}",
   volume={564},
   ISSN={1432-0746},
   url={http://dx.doi.org/10.1051/0004-6361/201322971},
   DOI={10.1051/0004-6361/201322971},
   journal={Astronomy \& Astrophysics},
   publisher={EDP Sciences},
   author={Buchner, J. and Georgakakis, A. and Nandra, K. and Hsu, L. and Rangel, C. and Brightman, M. and Merloni, A. and Salvato, M. and Donley, J. and Kocevski, D.},
   year={2014},
   month=apr, pages={A125} 
}

@article{Belle:2023xgj,
    author = "Prim, M. T. and others",
    collaboration = "Belle",
    title = "{Measurement of Angular Coefficients of $\bar{B} \to D^*\ell\bar{\nu}_\ell$: Implications for $|V_{cb}|$ and Tests of Lepton Flavor Universality}",
    eprint = "2310.20286",
    archivePrefix = "arXiv",
    primaryClass = "hep-ex",
    reportNumber = "Belle Preprint 2023-18; KEK Preprint 2023-32",
    doi = "10.1103/PhysRevLett.133.131801",
    journal = "Phys. Rev. Lett.",
    volume = "133",
    number = "13",
    pages = "131801",
    year = "2024"
}

@misc{hepdata.153299,
    author = "{\textsc{Belle} collaboration}",
    title = "{Measurement of Angular Coefficients of $\bar{B} \to D^* \ell \bar{\nu}_\ell$: Implications for $|V_{cb}|$ and Tests of Lepton Flavor Universality}",
    howpublished = "{\href{https://doi.org/10.17182/hepdata.153299}{HEPData (collection)}}",
    year = 2024,
}

@article{Grzadkowski:2010es,
    author = "Grzadkowski, B. and Iskrzynski, M. and Misiak, M. and Rosiek, J.",
    title = "{Dimension-Six Terms in the Standard Model Lagrangian}",
    eprint = "1008.4884",
    archivePrefix = "arXiv",
    primaryClass = "hep-ph",
    reportNumber = "IFT-9-2010, TTP10-35",
    doi = "10.1007/JHEP10(2010)085",
    journal = "JHEP",
    volume = "10",
    pages = "085",
    year = "2010"
}

@article{Jenkins:2017jig,
    author = "Jenkins, Elizabeth E. and Manohar, Aneesh V. and Stoffer, Peter",
    title = "{Low-Energy Effective Field Theory below the Electroweak Scale: Operators and Matching}",
    eprint = "1709.04486",
    archivePrefix = "arXiv",
    primaryClass = "hep-ph",
    doi = "10.1007/JHEP03(2018)016",
    journal = "JHEP",
    volume = "03",
    pages = "016",
    year = "2018",
    note = "[Erratum: JHEP 12, 043 (2023)]"
}

@article{Aebischer:2018bkb,
    author = "Aebischer, Jason and Kumar, Jacky and Straub, David M.",
    title = "{Wilson: a Python package for the running and matching of Wilson coefficients above and below the electroweak scale}",
    eprint = "1804.05033",
    archivePrefix = "arXiv",
    primaryClass = "hep-ph",
    doi = "10.1140/epjc/s10052-018-6492-7",
    journal = "Eur. Phys. J. C",
    volume = "78",
    number = "12",
    pages = "1026",
    year = "2018"
}

@article{Cata:2015lta,
    author = "Cat{\`a}, Oscar and Jung, Martin",
    title = "{Signatures of a nonstandard Higgs boson from flavor physics}",
    eprint = "1505.05804",
    archivePrefix = "arXiv",
    primaryClass = "hep-ph",
    reportNumber = "LMU-ASC-34-15, FLAVOUR(267104)-ERC-99",
    doi = "10.1103/PhysRevD.92.055018",
    journal = "Phys. Rev. D",
    volume = "92",
    number = "5",
    pages = "055018",
    year = "2015"
}

@article{Duraisamy:2014sna,
    author = "Duraisamy, Murugeswaran and Sharma, Preet and Datta, Alakabha",
    title = "{Azimuthal $B \to D^{*} \tau^{-} \bar{\nu}_\tau$ angular distribution with tensor operators}",
    eprint = "1405.3719",
    archivePrefix = "arXiv",
    primaryClass = "hep-ph",
    doi = "10.1103/PhysRevD.90.074013",
    journal = "Phys. Rev. D",
    volume = "90",
    number = "7",
    pages = "074013",
    year = "2014"
}

@article{Martinelli:2024vde,
    author = "Martinelli, G. and Simula, S. and Vittorio, L.",
    title = "{What we can learn from the angular differential rates from semileptonic $B \to D^*\ell\nu_\ell$ decays}",
    eprint = "2409.10492",
    archivePrefix = "arXiv",
    primaryClass = "hep-ph",
    doi = "10.1103/PhysRevD.111.013005",
    journal = "Phys. Rev. D",
    volume = "111",
    number = "1",
    pages = "013005",
    year = "2025"
}

@article{LHCb:2023uiv,
    author = "Aaij, Roel and others",
    collaboration = "LHCb",
    title = "{Test of lepton flavor universality using $B^0\to D^{\ast-}\tau^+\nu_\tau$ decays with hadronic {\ensuremath{\tau}} channels}",
    eprint = "2305.01463",
    archivePrefix = "arXiv",
    primaryClass = "hep-ex",
    reportNumber = "LHCb-PAPER-2022-052, CERN-EP-2023-062",
    doi = "10.1103/PhysRevD.108.012018",
    journal = "Phys. Rev. D",
    volume = "108",
    number = "1",
    pages = "012018",
    year = "2023",
    note = "[Erratum: Phys.Rev.D 109, 119902 (2024)]"
}

@article{Fedele:2022iib,
    author = "Fedele, Marco and Blanke, Monika and Crivellin, Andreas and Iguro, Syuhei and Kitahara, Teppei and Nierste, Ulrich and Watanabe, Ryoutaro",
    title = "{Impact of $\Lambda_b \to \Lambda_c \tau\nu$ measurement on New Physics in $b \to cl\nu$ transitions}",
    eprint = "2211.14172",
    archivePrefix = "arXiv",
    primaryClass = "hep-ph",
    reportNumber = "PSI-PR-22-34, ZU-TH 56/22, TTP22-069, P3H-22-113, KEK-TH-2474",
    doi = "10.1103/PhysRevD.107.055005",
    journal = "Phys. Rev. D",
    volume = "107",
    number = "5",
    pages = "055005",
    year = "2023"
}

@article{Blanke:2019qrx,
    author = "Blanke, Monika and Crivellin, Andreas and Kitahara, Teppei and Moscati, Marta and Nierste, Ulrich and Ni{\v{s}}and{\v{z}}i{\'c}, Ivan",
    title = "{Addendum to {\textquotedblleft}Impact of polarization observables and $B_c\to \tau \nu$ on new physics explanations of the $b\to c \tau \nu$ anomaly''}",
    eprint = "1905.08253",
    archivePrefix = "arXiv",
    primaryClass = "hep-ph",
    reportNumber = "PSI-PR-19-09; ZU-TH 26/19; TTP-19-012; P3H-19-011",
    doi = "10.1103/PhysRevD.100.035035",
    month = "5",
    year = "2019",
    note = "[Addendum: Phys.Rev.D 100, 035035 (2019)]"
}

@article{Blanke:2018yud,
    author = "Blanke, Monika and Crivellin, Andreas and de Boer, Stefan and Kitahara, Teppei and Moscati, Marta and Nierste, Ulrich and Ni{\v{s}}and{\v{z}}i{\'c}, Ivan",
    title = "{Impact of polarization observables and $ B_c\to \tau \nu$ on new physics explanations of the $b\to c \tau \nu$ anomaly}",
    eprint = "1811.09603",
    archivePrefix = "arXiv",
    primaryClass = "hep-ph",
    reportNumber = "PSI-PR-18-16; TTP-18-42, PSI-PR--18--16, TTP--18--42",
    doi = "10.1103/PhysRevD.99.075006",
    journal = "Phys. Rev. D",
    volume = "99",
    number = "7",
    pages = "075006",
    year = "2019"
}

@article{LHCb:2023zxo,
    author = "Aaij, Roel and others",
    collaboration = "LHCb",
    title = "{Measurement of the ratios of branching fractions $\mathcal{R}(D^{*})$ and $\mathcal{R}(D^{0})$}",
    eprint = "2302.02886",
    archivePrefix = "arXiv",
    primaryClass = "hep-ex",
    reportNumber = "LHCb-PAPER-2022-039, CERN-EP-2022-284",
    doi = "10.1103/PhysRevLett.131.111802",
    journal = "Phys. Rev. Lett.",
    volume = "131",
    pages = "111802",
    year = "2023"
}

@article{Belle:2019rba,
    author = "Caria, G. and others",
    collaboration = "Belle",
    title = "{Measurement of $\mathcal{R}(D)$ and $\mathcal{R}(D^*)$ with a semileptonic tagging method}",
    eprint = "1910.05864",
    archivePrefix = "arXiv",
    primaryClass = "hep-ex",
    reportNumber = "Belle-2019-18, KEK-2019-40",
    doi = "10.1103/PhysRevLett.124.161803",
    journal = "Phys. Rev. Lett.",
    volume = "124",
    number = "16",
    pages = "161803",
    year = "2020"
}

@misc{Tsaklidis:2025BelleIILFU,
  author        = {Tsaklidis, Ilias},
  title         = {{\href{https://indico.cern.ch/event/1440982/contributions/6576608/attachments/3136874/5566279/2025_CKM_LFU_belleII_itsaklid.pdf}{Status and prospects of LFU and angular measurements at Belle II}}},
  collaboration = {Belle II},
  year          = {2025},
  month         = sep
}

@article{Belle-II:2025yjp,
    author = "Adachi, I. and others",
    collaboration = "Belle-II",
    title = "{Test of lepton flavor universality with measurements of $R(D^+)$ and $R(D^{*+})$ using semileptonic $B$ tagging at the Belle II experiment}",
    eprint = "2504.11220",
    archivePrefix = "arXiv",
    primaryClass = "hep-ex",
    reportNumber = "Belle II Preprint 2025-011, KEK Preprint 2025-9",
    doi = "10.1103/fmn3-h8fy",
    journal = "Phys. Rev. D",
    volume = "112",
    number = "3",
    pages = "032010",
    year = "2025"
}

@article{LHCb:2024jll,
    author = "Aaij, Roel and others",
    collaboration = "LHCb",
    title = "{Measurement of the Branching Fraction Ratios $R(D^+)$ and $R(D^{*+})$ Using Muonic {\ensuremath{\tau}} Decays}",
    eprint = "2406.03387",
    archivePrefix = "arXiv",
    primaryClass = "hep-ex",
    reportNumber = "LHCb-PAPER-2024-007, CERN-EP-2024-125",
    doi = "10.1103/PhysRevLett.134.061801",
    journal = "Phys. Rev. Lett.",
    volume = "134",
    number = "6",
    pages = "061801",
    year = "2025"
}

@article{LHCb:2023ssl,
    author = "Aaij, R. and others",
    collaboration = "LHCb",
    title = "{Measurement of the D* longitudinal polarization in $B^0\to D^{\ast-}\tau^+\nu_\tau$ decays}",
    eprint = "2311.05224",
    archivePrefix = "arXiv",
    primaryClass = "hep-ex",
    reportNumber = "LHCb-PAPER-2023-020, CERN-EP-2023-225",
    doi = "10.1103/PhysRevD.110.092007",
    journal = "Phys. Rev. D",
    volume = "110",
    number = "9",
    pages = "092007",
    year = "2024"
}

@article{Mandal:2020htr,
    author = "Mandal, Rusa and Murgui, Clara and Pe{\~n}uelas, Ana and Pich, Antonio",
    title = "{The role of right-handed neutrinos in $b \to c \tau \bar{\nu}$ anomalies}",
    eprint = "2004.06726",
    archivePrefix = "arXiv",
    primaryClass = "hep-ph",
    reportNumber = "IFIC/20-14; FTUV/20-0414; SI-HEP-2020-XX, IFIC/20-14; FTUV/20-0414; SI-HEP-2020-10",
    doi = "10.1007/JHEP08(2020)022",
    journal = "JHEP",
    volume = "08",
    number = "08",
    pages = "022",
    year = "2020"
}

@misc{Hitlin:2025babarRD,
  author        = {Hitlin, David G.},
  title         = {{\href{https://indico.fnal.gov/event/71119/contributions/325774/attachments/194109/269256/Hitlin_Fermilab_2512215_final.pdf}{Measurement of $R(D)$ and $R(D^*)$ using semileptonic $B$ tags and leptonic $\tau$ decays}}},
  collaboration = {BABAR},
  year          = {2025},
  month         = dec
}

@article{Bernlochner:2024xiz,
    author = "Bernlochner, Florian U. and Fedele, Marco and Kretz, Tim and Nierste, Ulrich and Prim, Markus T.",
    title = "{Model independent bounds on heavy sterile neutrinos from the angular distribution of B {\textrightarrow} D$^{*}${\ensuremath{\ell}}{\ensuremath{\nu}} decays}",
    eprint = "2410.11945",
    archivePrefix = "arXiv",
    primaryClass = "hep-ph",
    reportNumber = "TTP24-041",
    doi = "10.1007/JHEP01(2025)040",
    journal = "JHEP",
    volume = "01",
    pages = "040",
    year = "2025"
}

@article{Boyd:1995cf,
    author = "Boyd, C. Glenn and Grinstein, Benjamin and Lebed, Richard F.",
    title = "{Model independent extraction of $|V_{cb}|$ using dispersion relations}",
    eprint = "hep-ph/9504235",
    archivePrefix = "arXiv",
    reportNumber = "UCSD-PTH-95-03",
    doi = "10.1016/0370-2693(95)00480-9",
    journal = "Phys. Lett. B",
    volume = "353",
    pages = "306--312",
    year = "1995"
}

@article{Boyd:1995sq,
    author = "Boyd, C. Glenn and Grinstein, Benjamin and Lebed, Richard F.",
    title = "{Model independent determinations of $\overline{B} \to D\ell\,\overline{\nu}, D^{*}\ell\,\overline{\nu}$ form-factors}",
    eprint = "hep-ph/9508211",
    archivePrefix = "arXiv",
    reportNumber = "UCSD-PTH-95-11",
    doi = "10.1016/0550-3213(95)00653-2",
    journal = "Nucl. Phys. B",
    volume = "461",
    pages = "493--511",
    year = "1996"
}

@article{Boyd_1997,
   title={Precision corrections to dispersive bounds on form factors},
   volume={56},
   ISSN={1089-4918},
   url={http://dx.doi.org/10.1103/PhysRevD.56.6895},
   DOI={10.1103/physrevd.56.6895},
   number={11},
   journal={Physical Review D},
   publisher={American Physical Society (APS)},
   author={Boyd, C. Glenn and Grinstein, Benjamín and Lebed, Richard F.},
   year={1997},
   month=dec, 
   pages={6895–6911} 
}

@article{ParticleDataGroup:2024cfk,
    author = "Navas, S. and others",
    collaboration = "Particle Data Group",
    title = "{Review of particle physics}",
    doi = "10.1103/PhysRevD.110.030001",
    journal = "Phys. Rev. D",
    volume = "110",
    number = "3",
    pages = "030001",
    year = "2024"
}

@article{Flynn:2023qmi,
    author = {Flynn, J. and J{\"u}ttner, A. and Tsang, J. T.},
    title = "{Bayesian inference for form-factor fits regulated by unitarity and analyticity}",
    eprint = "2303.11285",
    archivePrefix = "arXiv",
    primaryClass = "hep-ph",
    reportNumber = "CERN-TH-2023-047",
    doi = "10.1007/JHEP12(2023)175",
    journal = "JHEP",
    volume = "12",
    pages = "175",
    year = "2023"
}

@article{Flynn:2023eok,
    author = "Flynn, Jonathan and J{\"u}ttner, Andreas and Tsang, Tobias",
    title = "{Extrapolating semileptonic form factors using Bayesian-inference fits regulated by unitarity and analyticity}",
    eprint = "2312.14631",
    archivePrefix = "arXiv",
    primaryClass = "hep-lat",
    reportNumber = "CERN-TH-2023-248",
    doi = "10.22323/1.453.0239",
    journal = "PoS",
    volume = "LATTICE2023",
    pages = "239",
    year = "2024"
}

@misc{BFF_code,
    author       = "Andreas Jüttner",
    title        = "{Bayesian Form factor Fit (BFF) library}",
    howpublished         = "{Available at: \url{https://github.com/andreasjuettner/BFF}, \url{https://zenodo.org/records/7799543}}"
}

@article{Sirlin:1981ie,
    author = "Sirlin, A.",
    title = "{Large $m_W$, $m_Z$ Behavior of the $\mathcal{O}(\alpha)$ Corrections to Semileptonic Processes Mediated by W}",
    reportNumber = "NYU/TR8/81",
    doi = "10.1016/0550-3213(82)90303-0",
    journal = "Nucl. Phys. B",
    volume = "196",
    pages = "83--92",
    year = "1982"
}

@article{Bigi_2017,
    author = "Bigi, Dante and Gambino, Paolo and Schacht, Stefan",
    title = "{A fresh look at the determination of $|V_{cb}|$ from $B\to D^{*} \ell \nu$}",
    eprint = "1703.06124",
    archivePrefix = "arXiv",
    primaryClass = "hep-ph",
    doi = "10.1016/j.physletb.2017.04.022",
    journal = "Phys. Lett. B",
    volume = "769",
    pages = "441--445",
    year = "2017"
}

@article{Bigi:2017jbd,
    author = "Bigi, Dante and Gambino, Paolo and Schacht, Stefan",
    title = "{$R(D^*)$, $|V_{cb}|$, and the Heavy Quark Symmetry relations between form factors}",
    eprint = "1707.09509",
    archivePrefix = "arXiv",
    primaryClass = "hep-ph",
    doi = "10.1007/JHEP11(2017)061",
    journal = "JHEP",
    volume = "11",
    pages = "061",
    year = "2017"
}

@article{Bigi:2016mdz,
    author = "Bigi, Dante and Gambino, Paolo",
    title = "{Revisiting $B\to D \ell \nu$}",
    eprint = "1606.08030",
    archivePrefix = "arXiv",
    primaryClass = "hep-ph",
    doi = "10.1103/PhysRevD.94.094008",
    journal = "Phys. Rev. D",
    volume = "94",
    number = "9",
    pages = "094008",
    year = "2016"
}

@article{Martinelli_2021,
    author = "Martinelli, G. and Simula, S. and Vittorio, L.",
    title = "{Constraints for the semileptonic $B \to D^{(*)}$ form factors from lattice QCD simulations of two-point correlation functions}",
    eprint = "2105.07851",
    archivePrefix = "arXiv",
    primaryClass = "hep-lat",
    doi = "10.1103/PhysRevD.104.094512",
    journal = "Phys. Rev. D",
    volume = "104",
    number = "9",
    pages = "094512",
    year = "2021"
}

@article{harrison2024bc_chi,
    author = "Harrison, Judd",
    title = "{$\bar{b}c$ susceptibilities from fully relativistic lattice QCD}",
    eprint = "2405.01390",
    archivePrefix = "arXiv",
    primaryClass = "hep-lat",
    doi = "10.1103/PhysRevD.110.054506",
    journal = "Phys. Rev. D",
    volume = "110",
    number = "5",
    pages = "054506",
    year = "2024"
}

@article{Bharucha:2010im,
    author = "Bharucha, Aoife and Feldmann, Thorsten and Wick, Michael",
    title = "{Theoretical and Phenomenological Constraints on Form Factors for Radiative and Semi-Leptonic B-Meson Decays}",
    eprint = "1004.3249",
    archivePrefix = "arXiv",
    primaryClass = "hep-ph",
    reportNumber = "IPPP-10-31, DCPT-10-62, TUM-HEP-756-10",
    doi = "10.1007/JHEP09(2010)090",
    journal = "JHEP",
    volume = "09",
    pages = "090",
    year = "2010"
}

@article{Belle:2023bwv,
    author = "Prim, M. T. and others",
    collaboration = "Belle",
    title = "{Measurement of differential distributions of $B \to D^*\ell\bar{\nu}_\ell$ and implications on $|V_{cb}|$}",
    eprint = "2301.07529",
    archivePrefix = "arXiv",
    primaryClass = "hep-ex",
    reportNumber = "Belle Preprint 2022-34; KEK Preprint 2022-47",
    doi = "10.1103/PhysRevD.108.012002",
    journal = "Phys. Rev. D",
    volume = "108",
    number = "1",
    pages = "012002",
    year = "2023"
}

@misc{hepdata.137767,
    author = "{Belle Collaboration}",
    title = "{Measurement of Differential Distributions of $B \to D^* \ell \bar \nu_\ell$ and Implications on $|V_{cb}|$}",
    howpublished = "{\href{https://doi.org/10.17182/hepdata.137767}{HEPData (collection)}}",
    year = 2023,
}

@article{Belle-II:2023okj,
    author = "Adachi, I. and others",
    collaboration = "Belle-II",
    title = "{Determination of $|V_{cb}|$ using $\overline{B}^0 \to D^{*+}\ell^-\bar{\nu}_\ell$ decays with Belle II}",
    eprint = "2310.01170",
    archivePrefix = "arXiv",
    primaryClass = "hep-ex",
    reportNumber = "Belle II Preprint 2023-014, KEK Preprint 2023-28",
    doi = "10.1103/PhysRevD.108.092013",
    journal = "Phys. Rev. D",
    volume = "108",
    number = "9",
    pages = "092013",
    year = "2023"
}

@misc{hepdata.145129,
    author = "{Belle-II Collaboration}",
    title = "{Determination of $|V_{cb}|$ using $\overline{B}^0\to D^{*+}\ell^-\bar\nu_\ell$ decays with Belle II}",
    howpublished = "{\href{https://doi.org/10.17182/hepdata.145129}{HEPData (collection)}}",
    year = 2023,
}

@article{Kurokawa:2001nw,
    author = "Kurokawa, S. and Kikutani, Eiji",
    title = "{Overview of the KEKB accelerators}",
    reportNumber = "KEK-PREPRINT-2001-157A",
    doi = "10.1016/S0168-9002(02)01771-0",
    journal = "Nucl. Instrum. Meth. A",
    volume = "499",
    pages = "1--7",
    year = "2003"
}

@article{Abe:2013kxa,
    author = "Abe, Tetsuo and others",
    title = "{Achievements of KEKB}",
    doi = "10.1093/ptep/pts102",
    journal = "PTEP",
    volume = "2013",
    pages = "03A001",
    year = "2013"
}

@misc{HFLAV_spectrum,
  collaboration= {Heavy Flavor Averaging Group (HFLAV)},
  title        = "{Average of exclusive $B \to D^* \ell \nu$}",
  month        = jul,
  year         = 2024,
  publisher    = {Zenodo},
  version      = {1.1.0},
  doi          = {\href{https://doi.org/10.5281/zenodo.12696548}{{\texttt{doi:10.5281/zenodo.12696548}}}},
}

@article{Jung:2018lfu,
    author = "Jung, Martin and Straub, David M.",
    title = "{Constraining new physics in $b\to c\ell\nu$ transitions}",
    eprint = "1801.01112",
    archivePrefix = "arXiv",
    primaryClass = "hep-ph",
    doi = "10.1007/JHEP01(2019)009",
    journal = "JHEP",
    volume = "01",
    pages = "009",
    year = "2019"
}

@article{Buchmuller:1985jz,
    author = "Buchmuller, W. and Wyler, D.",
    title = "{Effective Lagrangian Analysis of New Interactions and Flavor Conservation}",
    reportNumber = "CERN-TH-4254/85",
    doi = "10.1016/0550-3213(86)90262-2",
    journal = "Nucl. Phys. B",
    volume = "268",
    pages = "621--653",
    year = "1986"
}

@article{Brivio:2017vri,
    author = "Brivio, Ilaria and Trott, Michael",
    title = "{The Standard Model as an Effective Field Theory}",
    eprint = "1706.08945",
    archivePrefix = "arXiv",
    primaryClass = "hep-ph",
    doi = "10.1016/j.physrep.2018.11.002",
    journal = "Phys. Rept.",
    volume = "793",
    pages = "1--98",
    year = "2019"
}

@article{Aebischer:2017ugx,
    author = "Aebischer, Jason and others",
    title = "{\texttt{WCxf}: an exchange format for Wilson coefficients beyond the Standard Model}",
    eprint = "1712.05298",
    archivePrefix = "arXiv",
    primaryClass = "hep-ph",
    reportNumber = "IFIC-17-61, TUM-HEP-1117-17, LMU-ASC-74-17, IFIC/17-61, KA-TP-38-2017, TUM-HEP-1117/17, LMU-ASC 74/17",
    doi = "10.1016/j.cpc.2018.05.022",
    journal = "Comput. Phys. Commun.",
    volume = "232",
    pages = "71--83",
    year = "2018"
}

@article{Korner:1989qb,
    author = "{K\"{o}rner, J. G. and Schuler, G. A.}",
    title = "{Exclusive semileptonic heavy meson decays including lepton mass effects}",
    reportNumber = "DESY-89-122, MZ-TH-88-14",
    doi = "10.1007/BF02440838",
    journal = "Z. Phys. C",
    volume = "46",
    pages = "93",
    year = "1990"
}

@article{Korner:1989ve,
    author = "{K\"{o}rner, J. G. and Schuler, G. A.}",
    title = "{Lepton Mass Effects in Semileptonic $B$ Meson Decays}",
    reportNumber = "DESY-89-062-REV, MZ-TH-89-16-REV",
    doi = "10.1016/0370-2693(89)90220-7",
    journal = "Phys. Lett. B",
    volume = "231",
    pages = "306--311",
    year = "1989"
}

@article{Gilman:1989uy,
    author = "Gilman, Frederick J. and Singleton, Robert L.",
    title = "{Analysis of Semileptonic Decays of Mesons Containing Heavy Quarks}",
    reportNumber = "SLAC-PUB-5065",
    doi = "10.1103/PhysRevD.41.142",
    journal = "Phys. Rev. D",
    volume = "41",
    pages = "142",
    year = "1990"
}

@article{BaBar:2006taf,
    author = "Aubert, Bernard and others",
    collaboration = "BaBar",
    title = "{Measurements of the $B \to D^{*}$ form-factors using the decay $\bar{B}^0 \to D^{*+} e^{-}\overline{\nu}_e$}",
    eprint = "hep-ex/0602023",
    archivePrefix = "arXiv",
    reportNumber = "SLAC-PUB-11672, BABAR-PUB-05-052",
    doi = "10.1103/PhysRevD.74.092004",
    journal = "Phys. Rev. D",
    volume = "74",
    pages = "092004",
    year = "2006"
}

@article{Duraisamy:2013pia,
    author = "Duraisamy, Murugeswaran and Datta, Alakabha",
    title = "{The Full $B \to D^{*} \tau^{-} \bar{\nu}_\tau$ Angular Distribution and CP violating Triple Products}",
    eprint = "1302.7031",
    archivePrefix = "arXiv",
    primaryClass = "hep-ph",
    doi = "10.1007/JHEP09(2013)059",
    journal = "JHEP",
    volume = "09",
    pages = "059",
    year = "2013"
}

@article{Becirevic:2019tpx,
    author = "Be{\v{c}}irevi{\'c}, Damir and Fedele, Marco and Ni{\v{s}}and{\v{z}}i{\'c}, Ivan and Tayduganov, Andrey",
    title = "{Lepton Flavor Universality tests through angular observables of $\overline{B}\to D^{(\ast)}\ell\overline{\nu}$ decay modes}",
    eprint = "1907.02257",
    archivePrefix = "arXiv",
    primaryClass = "hep-ph",
    reportNumber = "DO-TH 19/10, LPT-Orsay-19-26, QFET-2019-06, TTP19-022",
    month = "7",
    year = "2019"
}

@article{Huang:2021fuc,
    author = {Huang, Zhuo-Ran and Kou, Emi and L{\"u}, Cai-Dian and Tang, Ru-Ying},
    title = "{Un-binned Angular Analysis of $B\to D^*\ell \nu_\ell$ and the Right-handed Current}",
    eprint = "2106.13855",
    archivePrefix = "arXiv",
    primaryClass = "hep-ph",
    doi = "10.1103/PhysRevD.105.013010",
    journal = "Phys. Rev. D",
    volume = "105",
    number = "1",
    pages = "013010",
    year = "2022"
}

@article{Altmannshofer:2008dz,
    author = "Altmannshofer, Wolfgang and Ball, Patricia and Bharucha, Aoife and Buras, Andrzej J. and Straub, David M. and Wick, Michael",
    title = "{Symmetries and Asymmetries of $B \to K^{*} \mu^{+} \mu^{-}$ Decays in the Standard Model and Beyond}",
    eprint = "0811.1214",
    archivePrefix = "arXiv",
    primaryClass = "hep-ph",
    reportNumber = "IPPP-08-58, DCPT-08-116, TUM-HEP-696-08",
    doi = "10.1088/1126-6708/2009/01/019",
    journal = "JHEP",
    volume = "01",
    pages = "019",
    year = "2009"
}

@article{Colangelo:2018cnj,
    author = "Colangelo, Pietro and De Fazio, Fulvia",
    title = "{Scrutinizing $ \overline{B}\to {D}^{\ast}\left(D\pi \right){\ell}^{-}{\overline{\nu}}_{\ell } $ and $ \overline{B}\to {D}^{\ast}\left(D\gamma \right){\ell}^{-}{\overline{\nu}}_{\ell } $ in search of new physics footprints}",
    eprint = "1801.10468",
    archivePrefix = "arXiv",
    primaryClass = "hep-ph",
    reportNumber = "BARI-TH-18-715",
    doi = "10.1007/JHEP06(2018)082",
    journal = "JHEP",
    volume = "06",
    pages = "082",
    year = "2018"
}

@article{Bhattacharya:2019olg,
    author = "Bhattacharya, Bhubanjyoti and Datta, Alakabha and Kamali, Saeed and London, David",
    title = "{CP Violation in ${\bar B}^0\to D^{*+}\mu^-{\bar\nu}_\mu$}",
    eprint = "1903.02567",
    archivePrefix = "arXiv",
    primaryClass = "hep-ph",
    reportNumber = "UdeM-GPP-TH-19-269, MITP/19-016",
    doi = "10.1007/JHEP05(2019)191",
    journal = "JHEP",
    volume = "05",
    pages = "191",
    year = "2019"
}

@article{Colangelo:2024mxe,
    author = "Colangelo, Pietro and De Fazio, Fulvia and Loparco, Francesco and Losacco, Nicola",
    title = "{New physics couplings from angular coefficient functions of $\bar{B} \to D^*(D\pi)\ell\bar{\nu}_\ell$}",
    eprint = "2401.12304",
    archivePrefix = "arXiv",
    primaryClass = "hep-ph",
    reportNumber = "BARI-TH/754-24",
    doi = "10.1103/PhysRevD.109.075047",
    journal = "Phys. Rev. D",
    volume = "109",
    number = "7",
    pages = "075047",
    year = "2024"
}

@article{Kapoor:2024ufg,
    author = "Kapoor, Tejhas and Huang, Zhuo-Ran and Kou, Emi",
    title = "{New physics search via angular distribution of $B \to D^* \ell\nu_\ell$ decay in the light of the new lattice data}",
    eprint = "2401.11636",
    archivePrefix = "arXiv",
    primaryClass = "hep-ph",
    doi = "10.1007/JHEP02(2025)053",
    journal = "JHEP",
    volume = "02",
    pages = "053",
    year = "2025"
}

@article{HeavyFlavorAveragingGroupHFLAV:2024ctg,
    author = "Banerjee, Sw. and others",
    collaboration = "Heavy Flavor Averaging Group (HFLAV)",
    title = "{Averages of $b$-hadron, $c$-hadron, and $\tau$-lepton properties as of 2023}",
    eprint = "2411.18639",
    archivePrefix = "arXiv",
    primaryClass = "hep-ex",
    doi = "10.1103/x87q-tld5",
    journal = "Phys. Rev. D",
    volume = "113",
    number = "1",
    pages = "012008",
    year = "2026"
}

@article{Aebischer:2015fzz,
    author = "Aebischer, Jason and Crivellin, Andreas and Fael, Matteo and Greub, Christoph",
    title = "{Matching of gauge invariant dimension-six operators for $b\to s$ and $b\to c$ transitions}",
    eprint = "1512.02830",
    archivePrefix = "arXiv",
    primaryClass = "hep-ph",
    reportNumber = "CERN-PH-TH-2015-278",
    doi = "10.1007/JHEP05(2016)037",
    journal = "JHEP",
    volume = "05",
    pages = "037",
    year = "2016"
}

@article{Hu:2018veh,
    author = "Hu, Quan-Yi and Li, Xin-Qiang and Yang, Ya-Dong",
    title = "{$b\to c\tau\nu$ Transitions in the Standard Model Effective Field Theory}",
    eprint = "1810.04939",
    archivePrefix = "arXiv",
    primaryClass = "hep-ph",
    doi = "10.1140/epjc/s10052-019-6766-8",
    journal = "Eur. Phys. J. C",
    volume = "79",
    number = "3",
    pages = "264",
    year = "2019"
}

@article{Jung:2026ewj,
    author = "Jung, Martin and Schacht, Stefan",
    title = "{$\bar B\to D^{(*)}\ell\bar\nu$ Branching Ratios and Evidence for Isospin Breaking in $\Upsilon(4S)$ Decays}",
    eprint = "2604.08391",
    archivePrefix = "arXiv",
    primaryClass = "hep-ph",
    reportNumber = "IPPP/26/30",
    month = "4",
    year = "2026"
}

@article{deBlas:2025xhe,
    author = "de Blas, Jorge and Goncalves, Angelica and Miralles, V{\'\i}ctor and Reina, Laura and Silvestrini, Luca and Valli, Mauro",
    title = "{Constraining new physics effective interactions via a global fit of electroweak, Drell-Yan, Higgs, top, and flavour observables}",
    eprint = "2507.06191",
    archivePrefix = "arXiv",
    primaryClass = "hep-ph",
    doi = "10.1007/JHEP03(2026)013",
    journal = "JHEP",
    volume = "03",
    pages = "013",
    year = "2026"
}

@article{Allwicher:2023shc,
    author = "Allwicher, Lukas and Cornella, Claudia and Isidori, Gino and Stefanek, Ben A.",
    title = "{New physics in the third generation. A comprehensive SMEFT analysis and future prospects}",
    eprint = "2311.00020",
    archivePrefix = "arXiv",
    primaryClass = "hep-ph",
    reportNumber = "ZU-TH 71/23, MITP-23-060, KCL-PH-TH/2023-59",
    doi = "10.1007/JHEP03(2024)049",
    journal = "JHEP",
    volume = "03",
    pages = "049",
    year = "2024"
}

@article{Sakaki:2013bfa,
    author = "Sakaki, Yasuhito and Tanaka, Minoru and Tayduganov, Andrey and Watanabe, Ryoutaro",
    title = "{Testing leptoquark models in $\bar B \to D^{(*)} \tau \bar\nu$}",
    eprint = "1309.0301",
    archivePrefix = "arXiv",
    primaryClass = "hep-ph",
    reportNumber = "OU-HET-791, KEK-TH-1660, OU-HET 791",
    doi = "10.1103/PhysRevD.88.094012",
    journal = "Phys. Rev. D",
    volume = "88",
    number = "9",
    pages = "094012",
    year = "2013"
}

@article{Celis:2012dk,
    author = "Celis, Alejandro and Jung, Martin and Li, Xin-Qiang and Pich, Antonio",
    title = "{Sensitivity to charged scalars in $B\to D^{(*)}\tau\nu_\tau$ and $B\to\tau\nu_\tau$ decays}",
    eprint = "1210.8443",
    archivePrefix = "arXiv",
    primaryClass = "hep-ph",
    reportNumber = "FTUV-12-1031, IFIC-12-59, DO-TH-12-32",
    doi = "10.1007/JHEP01(2013)054",
    journal = "JHEP",
    volume = "01",
    pages = "054",
    year = "2013"
}

@article{Bhattacharya:2024zog,
    author = "Bhattacharya, Bhubanjyoti and Browder, Thomas E. and Datta, Alakabha and Kapoor, Tejhas and Kou, Emi and Mukherjee, Lopamudra",
    title = "{New physics searches via angular distributions of $ \overline{B}\to {D}^{\ast}\left(\to D\pi \right)\tau \left(\to \ell {\nu}_{\tau }{\overline{\nu}}_{\ell}\right){\overline{\nu}}_{\tau } $ decays}",
    eprint = "2411.09414",
    archivePrefix = "arXiv",
    primaryClass = "hep-ph",
    doi = "10.1007/JHEP04(2025)135",
    journal = "JHEP",
    volume = "04",
    pages = "135",
    year = "2025"
}

@article{Greljo:2023bab,
    author = "Greljo, Admir and Salko, Jakub and Smolkovi{\v{c}}, Aleks and Stangl, Peter",
    title = "{SMEFT restrictions on exclusive b {\textrightarrow} u{\ensuremath{\ell}}{\ensuremath{\nu}} decays}",
    eprint = "2306.09401",
    archivePrefix = "arXiv",
    primaryClass = "hep-ph",
    reportNumber = "CERN-TH-2023-111",
    doi = "10.1007/JHEP11(2023)023",
    journal = "JHEP",
    volume = "11",
    pages = "023",
    year = "2023"
}

@article{Mohapatra:2024knf,
    author = "Mohapatra, Manas Kumar and Panda, Dhiren and Mohanta, Rukmani",
    title = "{Imprints of new physics operators in the semileptonic $B \to a_1(1260)\ell^-\bar{\nu}_\ell$ process in SMEFT approach}",
    eprint = "2402.18410",
    archivePrefix = "arXiv",
    primaryClass = "hep-ph",
    doi = "10.1016/j.physletb.2024.138866",
    journal = "Phys. Lett. B",
    volume = "855",
    pages = "138866",
    year = "2024"
}

@article{Panda:2024oam,
    author = "Panda, Dhiren and Mohapatra, Manas Kumar and Mohanta, Rukmani",
    title = "{Analysis of $b \rightarrow c \ell \nu$ baryonic decay modes in SMEFT approach}",
    eprint = "2411.19044",
    archivePrefix = "arXiv",
    primaryClass = "hep-ph",
    doi = "10.1140/epjp/s13360-025-07024-2",
    journal = "Eur. Phys. J. Plus",
    volume = "140",
    number = "11",
    pages = "1078",
    year = "2025"
}

@article{deBlas:2017xtg,
    author = "de Blas, J. and Criado, J. C. and Perez-Victoria, M. and Santiago, J.",
    title = "{Effective description of general extensions of the Standard Model: the complete tree-level dictionary}",
    eprint = "1711.10391",
    archivePrefix = "arXiv",
    primaryClass = "hep-ph",
    reportNumber = "CERN-TH-2017-251",
    doi = "10.1007/JHEP03(2018)109",
    journal = "JHEP",
    volume = "03",
    pages = "109",
    year = "2018"
}

@book{JeffreysScale,
    author = "Harold Jeffreys",
    title = "{The Theory of Probability}",
    publisher = "{Oxford University Press}",
    isbn = "ISBN: 9780198503682",
    year = "1998",
}

@article{Skilling:2004pqw,
    author = "Skilling, John",
    title = "{Nested Sampling}",
    doi = "10.1063/1.1835238",
    journal = "AIP Conf. Proc.",
    volume = "735",
    number = "1",
    pages = "395",
    year = "2004"
}

@techreport{LHCb-CONF-2026-001,
      author       = "Collaboration, The",
      collaboration = "LHCb",
      title         = "{Measurement of hadronic form-factor parameters with an angular analysis of $B^0 \rightarrow D^{*-} \mu^{+}\nu_{\mu}$ decays}",
      institution   = "CERN",
      reportNumber  = "LHCb-CONF-2026-001, CERN-LHCb-CONF-2026-001",
      address       = "Geneva",
      year          = "2026",
      url           = "https://cds.cern.ch/record/2957157",
}

@article{Akaike:1974vps,
    author = "Akaike, H.",
    title = "{A new look at the statistical model identification}",
    doi = "10.1109/TAC.1974.1100705",
    journal = "IEEE Trans. Automatic Control",
    volume = "19",
    number = "6",
    pages = "716--723",
    year = "1974"
}

@article{Ray:2023xjn,
    author = "Ray, Ipsita and Nandi, Soumitra",
    title = "{Test of new physics effects in $ \overline{B}\to \left({D}^{\left(\ast \right)},\pi \right){\ell}^{-}{\overline{\nu}}_{\ell } $ decays with heavy and light leptons}",
    eprint = "2305.11855",
    archivePrefix = "arXiv",
    primaryClass = "hep-ph",
    doi = "10.1007/JHEP01(2024)022",
    journal = "JHEP",
    volume = "01",
    pages = "022",
    year = "2024"
}

@article{Belle-II:2023svm,
    author = "Adachi, I. and others",
    collaboration = "Belle-II",
    title = "{Tests of Light-Lepton Universality in Angular Asymmetries of $B^0 \to D^{*-}\ell\nu$ Decays}",
    eprint = "2308.02023",
    archivePrefix = "arXiv",
    primaryClass = "hep-ex",
    reportNumber = "Belle II Preprint 2023-013, KEK Preprint 2023-18",
    doi = "10.1103/PhysRevLett.131.181801",
    journal = "Phys. Rev. Lett.",
    volume = "131",
    number = "18",
    pages = "181801",
    year = "2023"
}

@article{Datta:2022czw,
    author = "Datta, Alakabha and Liu, Hongkai and Marfatia, Danny",
    title = "{$B^- \to D^{(\ast)}\ell X^-$ decays in effective field theory with massive right-handed neutrinos}",
    eprint = "2204.01818",
    archivePrefix = "arXiv",
    primaryClass = "hep-ph",
    doi = "10.1103/PhysRevD.106.L011702",
    journal = "Phys. Rev. D",
    volume = "106",
    number = "1",
    pages = "L011702",
    year = "2022"
}

@article{Huang:2025kof,
    author = "Huang, Zhuo-Ran and Bhutta, Faisal Munir and Farooq, Nimra and Paracha, M. Ali and Li, Ying",
    title = "{Reinvestigating the semileptonic $B\to D^{(\ast)}\tau\bar\nu_\tau$ decays in the model independent scenarios and leptoquark models}",
    eprint = "2501.03734",
    archivePrefix = "arXiv",
    primaryClass = "hep-ph",
    doi = "10.1103/PhysRevD.111.115035",
    journal = "Phys. Rev. D",
    volume = "111",
    number = "11",
    pages = "115035",
    year = "2025"
}

@article{DAgostini:1993arp,
    author = "D'Agostini, G.",
    title = "{On the use of the covariance matrix to fit correlated data}",
    reportNumber = "DESY-93-175",
    doi = "10.1016/0168-9002(94)90719-6",
    journal = "Nucl. Instrum. Meth. A",
    volume = "346",
    pages = "306--311",
    year = "1994"
}

@article{Belle:2018ezy,
    author = "Waheed, E. and others",
    collaboration = "Belle",
    title = "{Measurement of the CKM matrix element $|V_{cb}|$ from $B^0\to D^{*-}\ell^ {+} \nu_\ell$ at Belle}",
    eprint = "1809.03290",
    archivePrefix = "arXiv",
    primaryClass = "hep-ex",
    doi = "10.1103/PhysRevD.100.052007",
    journal = "Phys. Rev. D",
    volume = "100",
    number = "5",
    pages = "052007",
    year = "2019",
    note = "[Erratum: Phys.Rev.D 103, 079901 (2021)]"
}

@article{FlavourLatticeAveragingGroupFLAG:2024oxs,
    author = "Aoki, Y. and others",
    collaboration = "Flavour Lattice Averaging Group (FLAG)",
    title = "{FLAG review 2024}",
    eprint = "2411.04268",
    archivePrefix = "arXiv",
    primaryClass = "hep-lat",
    reportNumber = "CERN-TH-2024-192, FERMILAB-PUB-24-0785-T",
    doi = "10.1103/nfzp-p5dn",
    journal = "Phys. Rev. D",
    volume = "113",
    number = "1",
    pages = "014508",
    year = "2026"
}

@article{BaBar:2012obs,
    author = "Lees, J. P. and others",
    collaboration = "BaBar",
    title = "{Evidence for an excess of $\bar{B} \to D^{(*)} \tau^-\bar{\nu}_\tau$ decays}",
    eprint = "1205.5442",
    archivePrefix = "arXiv",
    primaryClass = "hep-ex",
    reportNumber = "BABAR-PUB-12-012, SLAC-PUB-15028",
    doi = "10.1103/PhysRevLett.109.101802",
    journal = "Phys. Rev. Lett.",
    volume = "109",
    pages = "101802",
    year = "2012"
}

@article{BaBar:2013mob,
    author = "Lees, J. P. and others",
    collaboration = "BaBar",
    title = "{Measurement of an Excess of $\bar{B} \to D^{(*)}\tau^- \bar{\nu}_\tau$ Decays and Implications for Charged Higgs Bosons}",
    eprint = "1303.0571",
    archivePrefix = "arXiv",
    primaryClass = "hep-ex",
    reportNumber = "BABAR-PUB-13-001, SLAC-PUB-15381",
    doi = "10.1103/PhysRevD.88.072012",
    journal = "Phys. Rev. D",
    volume = "88",
    number = "7",
    pages = "072012",
    year = "2013"
}

@article{Belle:2016dyj,
    author = "Hirose, S. and others",
    collaboration = "Belle",
    title = "{Measurement of the $\tau$ lepton polarization and $R(D^*)$ in the decay $\bar{B} \to D^* \tau^- \bar{\nu}_\tau$}",
    eprint = "1612.00529",
    archivePrefix = "arXiv",
    primaryClass = "hep-ex",
    reportNumber = "KEK-PREPRINT-2016-53, BELLE-PREPRINT-2016-14",
    doi = "10.1103/PhysRevLett.118.211801",
    journal = "Phys. Rev. Lett.",
    volume = "118",
    number = "21",
    pages = "211801",
    year = "2017"
}

@article{Hagiwara:1989gza,
    author = "Hagiwara, Kaoru and Martin, Alan D. and Wade, M. F.",
    title = "{Helicity Amplitude Analysis of $B \to D^* \ell\nu$ Decays}",
    reportNumber = "DTP/89/12",
    doi = "10.1016/0370-2693(89)90541-8",
    journal = "Phys. Lett. B",
    volume = "228",
    pages = "144--148",
    year = "1989"
}

@article{Gambino:2019sif,
    author = "Gambino, Paolo and Jung, Martin and Schacht, Stefan",
    title = "{The $V_{cb}$ puzzle: An update}",
    eprint = "1905.08209",
    archivePrefix = "arXiv",
    primaryClass = "hep-ph",
    doi = "10.1016/j.physletb.2019.06.039",
    journal = "Phys. Lett. B",
    volume = "795",
    pages = "386--390",
    year = "2019"
}

@article{Ivanov:2016qtw,
    author = {Ivanov, Mikhail A. and K{\"o}rner, J{\"u}rgen G. and Tran, Chien-Thang},
    title = "{Analyzing new physics in the decays $\bar{B}^0 \to D^{(\ast)}\tau^-\bar\nu_{\tau}$ with form factors obtained from the covariant quark model}",
    eprint = "1607.02932",
    archivePrefix = "arXiv",
    primaryClass = "hep-ph",
    reportNumber = "MITP-16-073",
    doi = "10.1103/PhysRevD.94.094028",
    journal = "Phys. Rev. D",
    volume = "94",
    number = "9",
    pages = "094028",
    year = "2016"
}

@article{Tanaka:1994ay,
    author = "Tanaka, M.",
    title = "{Charged Higgs effects on exclusive semitauonic $B$ decays}",
    eprint = "hep-ph/9411405",
    archivePrefix = "arXiv",
    reportNumber = "KEK-TH-422, KEK-PREPRINT-94-155",
    doi = "10.1007/BF01571294",
    journal = "Z. Phys. C",
    volume = "67",
    pages = "321--326",
    year = "1995"
}

@article{Biancofiore:2013ki,
    author = "Biancofiore, Pietro and Colangelo, Pietro and De Fazio, Fulvia",
    title = "{On the anomalous enhancement observed in $B \to D^{(*)}\tau{\bar \nu}_\tau$ decays}",
    eprint = "1302.1042",
    archivePrefix = "arXiv",
    primaryClass = "hep-ph",
    reportNumber = "BARI-TH-670-2013",
    doi = "10.1103/PhysRevD.87.074010",
    journal = "Phys. Rev. D",
    volume = "87",
    number = "7",
    pages = "074010",
    year = "2013"
}

@article{Fajfer:2012vx,
    author = "Fajfer, Svjetlana and Kamenik, Jernej F. and Nisandzic, Ivan",
    title = "{On the $B \to D^* \tau \bar \nu_{\tau}$ Sensitivity to New Physics}",
    eprint = "1203.2654",
    archivePrefix = "arXiv",
    primaryClass = "hep-ph",
    doi = "10.1103/PhysRevD.85.094025",
    journal = "Phys. Rev. D",
    volume = "85",
    pages = "094025",
    year = "2012"
}

@article{Tanaka:2012nw,
    author = "Tanaka, Minoru and Watanabe, Ryoutaro",
    title = "{New physics in the weak interaction of $\bar B\to D^{(*)}\tau\bar\nu$}",
    eprint = "1212.1878",
    archivePrefix = "arXiv",
    primaryClass = "hep-ph",
    reportNumber = "OU-HET-772-2012",
    doi = "10.1103/PhysRevD.87.034028",
    journal = "Phys. Rev. D",
    volume = "87",
    number = "3",
    pages = "034028",
    year = "2013"
}

@article{Bobeth:2021lya,
    author = "Bobeth, Christoph and Bordone, Marzia and Gubernari, Nico and Jung, Martin and van Dyk, Danny",
    title = "{Lepton-flavour non-universality of ${\bar{B}}\rightarrow D^*\ell {{\bar{\nu }}}$ angular distributions in and beyond the Standard Model}",
    eprint = "2104.02094",
    archivePrefix = "arXiv",
    primaryClass = "hep-ph",
    reportNumber = "EOS-2021-03, TUM-HEP 1305/20, P3H-21-021, SI-HEP-2021-12",
    doi = "10.1140/epjc/s10052-021-09724-2",
    journal = "Eur. Phys. J. C",
    volume = "81",
    number = "11",
    pages = "984",
    year = "2021"
}

@article{Ball:2004rg,
    author = "Ball, Patricia and Zwicky, Roman",
    title = "{$B_{d,s} \to  \rho, \omega, K^*, \phi$ Decay Form Factors from Light-Cone Sum Rules Revisited}",
    eprint = "hep-ph/0412079",
    archivePrefix = "arXiv",
    reportNumber = "IPPP-04-74, DCPT-04-48, TPI-MINN-04-39",
    doi = "10.1103/PhysRevD.71.014029",
    journal = "Phys. Rev. D",
    volume = "71",
    pages = "014029",
    year = "2005"
}

@article{Tsang:2023nay,
    author = "Tsang, J. Tobias and Della Morte, Michele",
    title = "{B-physics from lattice gauge theory}",
    eprint = "2310.02705",
    archivePrefix = "arXiv",
    primaryClass = "hep-lat",
    reportNumber = "CERN-TH-2023-175",
    doi = "10.1140/epjs/s11734-023-01011-3",
    journal = "Eur. Phys. J. ST",
    volume = "233",
    number = "2",
    pages = "253--270",
    year = "2024"
}

@article{Martinelli:2024bov,
    author = "Martinelli, G. and Simula, S. and Vittorio, L.",
    title = "{Semileptonic $B \rightarrow D^*$ decays from light to $\tau $ leptons: the extraction of the form factor $F_2$ from data}",
    eprint = "2410.17974",
    archivePrefix = "arXiv",
    primaryClass = "hep-ph",
    doi = "10.1140/epjc/s10052-025-13972-x",
    journal = "Eur. Phys. J. C",
    volume = "85",
    number = "3",
    pages = "242",
    year = "2025"
}

@article{Bigi:2017njr,
    author = "Bigi, Dante and Gambino, Paolo and Schacht, Stefan",
    title = "{A fresh look at the determination of $|V_{cb}|$ from $B\to D^{*} \ell \nu$}",
    eprint = "1703.06124",
    archivePrefix = "arXiv",
    primaryClass = "hep-ph",
    doi = "10.1016/j.physletb.2017.04.022",
    journal = "Phys. Lett. B",
    volume = "769",
    pages = "441--445",
    year = "2017"
}

@article{Jaiswal:2020wer,
    author = "Jaiswal, Sneha and Nandi, Soumitra and Patra, Sunando Kumar",
    title = "{Updates on extraction of $|V_{cb}|$ and SM prediction of $R(D^*)$ in $B\to D^{*}\ell\nu_\ell$ decays}",
    eprint = "2002.05726",
    archivePrefix = "arXiv",
    primaryClass = "hep-ph",
    doi = "10.1007/JHEP06(2020)165",
    journal = "JHEP",
    volume = "06",
    pages = "165",
    year = "2020"
}

@article{Iguro:2020cpg,
    author = "Iguro, Syuhei and Watanabe, Ryoutaro",
    title = "{Bayesian fit analysis to full distribution data of $ \overline{{B}}\to {{D}}^{\left(\ast \right)}\mathrm{\ell}\overline{\nu }:\left|{\mathrm{V}}_{\mathrm{cb}}\right| $ determination and new physics constraints}",
    eprint = "2004.10208",
    archivePrefix = "arXiv",
    primaryClass = "hep-ph",
    doi = "10.1007/JHEP08(2020)006",
    journal = "JHEP",
    volume = "08",
    number = "08",
    pages = "006",
    year = "2020"
}

@article{Fedele:2023ewe,
    author = "Fedele, Marco and Blanke, Monika and Crivellin, Andreas and Iguro, Syuhei and Nierste, Ulrich and Simula, Silvano and Vittorio, Ludovico",
    title = "{Discriminating $B \to D^* \ell\nu$ form factors via polarization observables and asymmetries}",
    eprint = "2305.15457",
    archivePrefix = "arXiv",
    primaryClass = "hep-ph",
    reportNumber = "PSI-PR-23-15, PSI-PR-23-12, ZU-TH 22/23, TTP23-019, P3H-23-033, LAPTH-020/23",
    doi = "10.1103/PhysRevD.108.055037",
    journal = "Phys. Rev. D",
    volume = "108",
    number = "5",
    pages = "055037",
    year = "2023"
}

@article{DiCarlo:2021dzg,
    author = "Di Carlo, M. and Martinelli, G. and Naviglio, M. and Sanfilippo, F. and Simula, S. and Vittorio, L.",
    title = "{Unitarity bounds for semileptonic decays in lattice QCD}",
    eprint = "2105.02497",
    archivePrefix = "arXiv",
    primaryClass = "hep-lat",
    doi = "10.1103/PhysRevD.104.054502",
    journal = "Phys. Rev. D",
    volume = "104",
    number = "5",
    pages = "054502",
    year = "2021"
}

@article{Martinelli:2021onb,
    author = "Martinelli, G. and Simula, S. and Vittorio, L.",
    title = "{$\vert V_{cb} \vert$ and $R(D^{(*)})$ using lattice QCD and unitarity}",
    eprint = "2105.08674",
    archivePrefix = "arXiv",
    primaryClass = "hep-ph",
    doi = "10.1103/PhysRevD.105.034503",
    journal = "Phys. Rev. D",
    volume = "105",
    number = "3",
    pages = "034503",
    year = "2022"
}

@article{Martinelli:2021myh,
    author = "Martinelli, G. and Simula, S. and Vittorio, L.",
    title = "{Exclusive determinations of $\vert V_{cb} \vert $ and $R(D^{*})$ through unitarity}",
    eprint = "2109.15248",
    archivePrefix = "arXiv",
    primaryClass = "hep-ph",
    doi = "10.1140/epjc/s10052-022-11050-0",
    journal = "Eur. Phys. J. C",
    volume = "82",
    number = "12",
    pages = "1083",
    year = "2022"
}

@article{Martinelli:2022xir,
    author = "Martinelli, G. and Naviglio, M. and Simula, S. and Vittorio, L.",
    title = "{$|V_{cb}|$, lepton flavor universality and $SU(3)_F$ symmetry breaking in $B_s\to D_s^{(\ast)}\ell\nu$ decays through unitarity and lattice QCD}",
    eprint = "2204.05925",
    archivePrefix = "arXiv",
    primaryClass = "hep-ph",
    doi = "10.1103/PhysRevD.106.093002",
    journal = "Phys. Rev. D",
    volume = "106",
    number = "9",
    pages = "093002",
    year = "2022"
}

@article{Martinelli:2023fwm,
    author = "Martinelli, G. and Simula, S. and Vittorio, L.",
    title = "{Updates on the determination of $\vert V_{cb} \vert ,R(D^{*})$ and $\vert V_{ub} \vert /\vert V_{cb} \vert $}",
    eprint = "2310.03680",
    archivePrefix = "arXiv",
    primaryClass = "hep-ph",
    doi = "10.1140/epjc/s10052-024-12742-5",
    journal = "Eur. Phys. J. C",
    volume = "84",
    number = "4",
    pages = "400",
    year = "2024"
}

@article{Gopal:2024mgb,
    author = "Gopal, Abinand and Gubernari, Nico",
    title = "{Unitarity bounds with subthreshold and anomalous cuts for b-hadron decays}",
    eprint = "2412.04388",
    archivePrefix = "arXiv",
    primaryClass = "hep-ph",
    doi = "10.1103/PhysRevD.111.L031501",
    journal = "Phys. Rev. D",
    volume = "111",
    number = "3",
    pages = "L031501",
    year = "2025"
}

@article{Simula:2025fft,
    author = "Simula, Silvano and Vittorio, Ludovico",
    title = "{Multiple dispersive bounds. II. Subthreshold branch cuts}",
    eprint = "2509.00412",
    archivePrefix = "arXiv",
    primaryClass = "hep-ph",
    doi = "10.1103/g1q2-3pmq",
    journal = "Phys. Rev. D",
    volume = "113",
    number = "7",
    pages = "074018",
    year = "2026"
}

@article{Gubernari:2020eft,
    author = "Gubernari, Nico and van Dyk, Danny and Virto, Javier",
    title = "{Non-local matrix elements in $B_{(s)}\to \{K^{(*)},\phi\}\ell^+\ell^-$}",
    eprint = "2011.09813",
    archivePrefix = "arXiv",
    primaryClass = "hep-ph",
    reportNumber = "EOS-2020-01, TUM-HEP-1292/20, P3H-20-066, SI-HEP-2020-27",
    doi = "10.1007/JHEP02(2021)088",
    journal = "JHEP",
    volume = "02",
    pages = "088",
    year = "2021"
}

@article{Boyd:1994tt,
    author = "Boyd, C. Glenn and Grinstein, Benjamin and Lebed, Richard F.",
    title = "{Constraints on form-factors for exclusive semileptonic heavy to light meson decays}",
    eprint = "hep-ph/9412324",
    archivePrefix = "arXiv",
    reportNumber = "UCSD-PTH-94-27",
    doi = "10.1103/PhysRevLett.74.4603",
    journal = "Phys. Rev. Lett.",
    volume = "74",
    pages = "4603--4606",
    year = "1995"
}

@article{Chetyrkin:2000yt,
    author = "Chetyrkin, K. G. and Kuhn, Johann H. and Steinhauser, M.",
    title = "{RunDec: A Mathematica package for running and decoupling of the strong coupling and quark masses}",
    eprint = "hep-ph/0004189",
    archivePrefix = "arXiv",
    reportNumber = "DESY-00-034, TTP-00-05",
    doi = "10.1016/S0010-4655(00)00155-7",
    journal = "Comput. Phys. Commun.",
    volume = "133",
    pages = "43--65",
    year = "2000"
}

@article{Schmidt:2012az,
    author = "Schmidt, Barbara and Steinhauser, Matthias",
    title = "{CRunDec: a C++ package for running and decoupling of the strong coupling and quark masses}",
    eprint = "1201.6149",
    archivePrefix = "arXiv",
    primaryClass = "hep-ph",
    reportNumber = "SFB-CPP-12-03, TTP12-02",
    doi = "10.1016/j.cpc.2012.03.023",
    journal = "Comput. Phys. Commun.",
    volume = "183",
    pages = "1845--1848",
    year = "2012"
}

@article{Herren:2017osy,
    author = "Herren, Florian and Steinhauser, Matthias",
    title = "{Version 3 of RunDec and CRunDec}",
    eprint = "1703.03751",
    archivePrefix = "arXiv",
    primaryClass = "hep-ph",
    reportNumber = "TTP17-011",
    doi = "10.1016/j.cpc.2017.11.014",
    journal = "Comput. Phys. Commun.",
    volume = "224",
    pages = "333--345",
    year = "2018"
}

@article{Christ:2024xzj,
    author = "Christ, Norman H. and Feng, Xu and Jin, Luchang and Sachrajda, Christopher T. and Wang, Tianle",
    title = "{Lattice Calculation of Electromagnetic Corrections to Kl3 decay}",
    eprint = "2402.08915",
    archivePrefix = "arXiv",
    primaryClass = "hep-lat",
    doi = "10.22323/1.453.0266",
    journal = "PoS",
    volume = "LATTICE2023",
    pages = "266",
    year = "2024"
}

@article{Jwa:2025fon,
    author = "Jwa, Seungyeob and Kim, Jeehun and Kim, Sunghee and Lee, Sunkyu and Lee, Weonjong and Park, Sungwoo",
    collaboration = "SWME",
    title = "{2025 update on $\varepsilon_K$ in the Standard Model with lattice QCD inputs}",
    eprint = "2503.00351",
    archivePrefix = "arXiv",
    primaryClass = "hep-lat",
    month = "2",
    year = "2025"
}

@article{deBlas:2025gyz,
    author = "de Blas, Jorge and others",
    title = "{Physics Briefing Book: Input for the 2026 update of the European Strategy for Particle Physics}",
    eprint = "2511.03883",
    archivePrefix = "arXiv",
    primaryClass = "hep-ex",
    reportNumber = "CERN--2025-008, CERN-ESU-2025-001",
    doi = "10.23731/CYRM-2025-008",
    month = "11",
    year = "2025"
}

@article{Belle:2017rcc,
    author = "Abdesselam, A. and others",
    collaboration = "Belle",
    title = "{Precise determination of the CKM matrix element $\left| V_{cb}\right|$ with $\bar B^0 \to D^{*\,+} \, \ell^- \, \bar \nu_\ell$ decays with hadronic tagging at Belle}",
    eprint = "1702.01521",
    archivePrefix = "arXiv",
    primaryClass = "hep-ex",
    reportNumber = "BELLE-CONF-1612",
    month = "2",
    year = "2017"
}

@article{DeSantis:2025qbb,
    author = "De Santis, Alessandro and others",
    title = "{Inclusive semileptonic decays of the $D_s$ meson: A first-principles lattice QCD calculation}",
    eprint = "2504.06063",
    archivePrefix = "arXiv",
    primaryClass = "hep-lat",
    reportNumber = "HIP-2025-9/TH",
    doi = "10.1103/3cxg-k322",
    journal = "Phys. Rev. D",
    volume = "112",
    number = "5",
    pages = "054503",
    year = "2025"
}

@article{DeSantis:2025yfm,
    author = "De Santis, Alessandro and others",
    title = "{Inclusive Semileptonic Decays of the $D_s$ Meson: Lattice QCD Confronts Experiments}",
    eprint = "2504.06064",
    archivePrefix = "arXiv",
    primaryClass = "hep-lat",
    reportNumber = "HIP-2025-10/TH",
    doi = "10.1103/snc6-cpz6",
    journal = "Phys. Rev. Lett.",
    volume = "135",
    number = "12",
    pages = "121901",
    year = "2025"
}

@article{Gambino:2020crt,
    author = "Gambino, Paolo and Hashimoto, Shoji",
    title = "{Inclusive Semileptonic Decays from Lattice QCD}",
    eprint = "2005.13730",
    archivePrefix = "arXiv",
    primaryClass = "hep-lat",
    reportNumber = "KEK-CP-376",
    doi = "10.1103/PhysRevLett.125.032001",
    journal = "Phys. Rev. Lett.",
    volume = "125",
    number = "3",
    pages = "032001",
    year = "2020"
}

@article{Bailas:2020qmv,
    author = "Bailas, Gabriela and Hashimoto, Shoji and Ishikawa, Tsutomu",
    title = "{Reconstruction of smeared spectral function from Euclidean correlation functions}",
    eprint = "2001.11779",
    archivePrefix = "arXiv",
    primaryClass = "hep-lat",
    reportNumber = "KEK-CP-374",
    doi = "10.1093/ptep/ptaa044",
    journal = "PTEP",
    volume = "2020",
    number = "4",
    pages = "043B07",
    year = "2020"
}

@article{Kellermann:2025pzt,
    author = {Kellermann, Ryan and Hu, Zhi and Barone, Alessandro and Elgaziari, Ahmed and Hashimoto, Shoji and Kaneko, Takashi and J{\"u}ttner, Andreas},
    title = "{Inclusive semileptonic decays from lattice QCD: Analysis of systematic effects}",
    eprint = "2504.03358",
    archivePrefix = "arXiv",
    primaryClass = "hep-lat",
    reportNumber = "KEK-CP-0407, CERN-TH-2025-059",
    doi = "10.1103/kltp-1p1f",
    journal = "Phys. Rev. D",
    volume = "112",
    number = "1",
    pages = "014501",
    year = "2025"
}

@article{Barone:2023tbl,
    author = {Barone, Alessandro and Hashimoto, Shoji and J{\"u}ttner, Andreas and Kaneko, Takashi and Kellermann, Ryan},
    title = "{Approaches to inclusive semileptonic B$_{(s)}$-meson decays from Lattice QCD}",
    eprint = "2305.14092",
    archivePrefix = "arXiv",
    primaryClass = "hep-lat",
    reportNumber = "KEK-CP-0394, CERN-TH-2023-087",
    doi = "10.1007/JHEP07(2023)145",
    journal = "JHEP",
    volume = "07",
    pages = "145",
    year = "2023"
}

@article{Fang:2026hru,
    author = "Fang, Wen-Sheng and Iguro, Syuhei and Li, Xin-Qiang and Sain, Ria and Watanabe, Ryoutaro and Zhang, Ben-Liang",
    title = "{$|V_{cb}|$ determinations from $\bar{B} \to D^{(*)} \ell \bar\nu$ decays within the SM and beyond}",
    eprint = "2606.17138",
    archivePrefix = "arXiv",
    primaryClass = "hep-ph",
    month = "6",
    year = "2026"
}

\end{document}